\documentclass[structabstract]{aa} 
\usepackage{amsmath}
\usepackage{amstext}
\usepackage{amsfonts}
\usepackage{amssymb}
\usepackage{graphicx}
\usepackage[varg]{txfonts}
\usepackage{xspace}
\usepackage{url}
\usepackage{natbib}
\usepackage{multirow}

\usepackage{siunitx}
\usepackage{longtable,lscape}	
\usepackage{hyperref}

\usepackage{color}
\usepackage[normalem]{ulem}

\begin{document} 

	\title{KMOS view of the Galactic centre} \subtitle{I. Young stars are centrally concentrated \thanks{Based on observations collected at the European Organisation for Astronomical Research in the Southern 	Hemisphere, Chile \mbox{(60.A-9450(A)}).}$^{,}$\thanks{Appendices are available in electronic form at http://www.aanda.org}$^{,}$\thanks{The extracted spectra are only available  at the CDS via anonymous ftp to cdsarc.u-strasbg.fr (130.79.128.5) or via http://cdsarc.u-strasbg.fr/viz-bin/qcat?J/A+A/584/A2}}

	\author{A.~Feldmeier-Krause 
			\inst{\ref{inst1}}
		\and N.~Neumayer
			\inst{\ref{inst4}}
		\and R.~Sch{\"o}del
			\inst{\ref{inst9}}			
		\and A.~Seth
			\inst{\ref{inst8}}	
		\and M.~Hilker
			\inst{\ref{inst1}}		
		\and P.~T.~de~Zeeuw
			\inst{\ref{inst1}}$^,$\inst{\ref{inst7}}	
		\and H.~Kuntschner
			\inst{\ref{inst1}}
		\and C.~J.~Walcher	
			\inst{\ref{inst11}}	
		\and N.~L{\"u}tzgendorf
			\inst{\ref{inst12}}
		\and M.~Kissler-Patig
			\inst{\ref{inst3}}
						} 

 \institute{European Southern Observatory (ESO), Karl-Schwarzschild-Stra{\ss}e 2, 85748 Garching, Germany 
 \email{afeldmei@eso.org}
 	\label{inst1}
	\and
          Max-Planck-Institut f{\"u}r Astronomie, K{\"o}nigsstuhl 17, 69117 Heidelberg, Germany
          	\label{inst4}
	    	 \and
         Instituto de Astrof\'{i}sica de Andaluc\'{i}a (CSIC), Glorieta de la Astronom\'{i}a s/n, 18008 Granada, Spain           
         \label{inst9}
          \and
           Department of Physics and Astronomy, University of Utah, Salt Lake City, UT 84112, USA
          	 \label{inst8}
	          \and 
          Sterrewacht Leiden, Leiden University, Postbus 9513, 2300 RA Leiden, The Netherlands 
          	\label{inst7}
	\and
	 Leibniz-Institut f{\"u}r Astrophysik Potsdam (AIP), An der Sternwarte 16, 14482 Potsdam, Germany
	 	\label{inst11}
      \and 
      ESTEC, Keplerlaan 1, 2201 AZ Noordwijk, The Netherlands 
      \label{inst12}
      	      \and
          Gemini Observatory, 670 N. A'ohoku Place, Hilo, Hawaii, 96720, USA 
          	\label{inst3}	
          }

\date{Received 16 April 2015; accepted 10 September 2015}

 \abstract
    {The Galactic centre hosts a  crowded, dense nuclear star cluster with a half-light radius of 4\,pc. Most of the stars in the Galactic centre are cool late-type stars, but there are also $\gtrsim$100 hot early-type stars in the central parsec of the Milky Way. These stars are only 3-8\,Myr old.  }
   {Our knowledge of the number and distribution of  early-type stars in the Galactic centre is incomplete.  Only
a few spectroscopic observations have been made beyond a projected distance of 0.5\,pc of the Galactic centre. The distribution and kinematics of early-type stars  are essential to understand the formation and growth of the nuclear star cluster.  }
   { We cover  the central \textgreater4\,pc$^2$ (0.75\,sq.arcmin) of the Galactic centre using the integral-field spectrograph KMOS (VLT). We extracted more than 1,000 spectra from individual stars and identified early-type stars based on their spectra. }
   {Our data set contains 114  bright early-type stars:  6 have narrow emission lines, 23 are Wolf-Rayet stars, 9 stars have featureless spectra, and 76 are O/B type stars. 
   Our wide-field spectroscopic data confirm that the distribution of young stars is compact, with 90\% of the young stars identified within 0.5\,pc of the nucleus. We identify 24 new O/B stars primarily at large radii.  We estimate photometric masses of the O/B stars and show that the total mass in the young population is $\gtrsim$12,000\,$M_\odot$.  The O/B stars all appear to be bound to the Milky Way nuclear star cluster, while less than  30\% belong to the clockwise rotating disk.  We add one new star to the sample of stars  affiliated with this disk.
   }
{The central concentration of the early-type stars is a strong argument that they have formed in situ. An alternative scenario, in which the stars formed outside the Galactic centre in a cluster that migrated to the centre, is refuted.  A large part of the young O/B stars is not on the disk, which
either means that the early-type  stars did not all form on the same disk or that the disk is dissolving rapidly.}
{}
\titlerunning{KMOS view of the Galactic Centre  I: Young Stars}
\authorrunning{A. Feldmeier-Krause et al.}
  \maketitle

\section{Introduction}

 Nuclear star clusters (NSCs) are a distinct component  at the centre of many galaxies. The central region of  $\sim$\:\!75\%$-$80\% of spiral galaxies \citep{carollo98,boker02,iskren14} and spheroidal galaxies \citep{cote06,mark14} contains a nuclear star cluster.  Nuclear star clusters are located at a distinguished spot in a galaxy: The centre of the galaxy's gravitational potential \citep{nadine11}. Galaxies grow by mergers and accretion, so that infalling gas and  stars can finally end up in the centre of a galaxy. Nuclear regions therefore have very high densities.  Many nuclear star clusters also contain a supermassive black hole \citep[e.g.][]{anil08,graham09}. The nuclear regions of galaxies are of special interest for galaxy formation and evolution studies because of the scaling correlations between the mass of the nuclear star cluster and other galaxy properties, such as the galaxy mass \citep[e.g.][]{wehner06,rossa06,ferrarese06,scott13}.

 The nuclear star cluster of the Milky Way (MW) is the best-studied case of a galaxy nucleus. The cluster was first detected by \cite{becklin} in the infrared. It has a half-light radius or effective radius $r_{\mathrm{eff}}$  of $\sim$110$-$127\arcsec\,\citep[4.2$-$5\,pc,][]{sb,Fritz14} and a mass of $\mathcal{M}_{MWNSC} = (2.5 \pm 0.4) \times 10^7$~$M_\odot$ \citep{sb}.   The central parsec of the Milky Way nuclear star cluster is extensively studied.  At a distance of only $\sim$8\,kpc \citep{ghez08,gillessen09s2,chatzopoulos}, it is possible to spatially resolve single stars. 
Monitoring of single stars over more than a decade led to an accurate measurement of the mass of the Milky Way central supermassive black hole: $\mathcal{M}_\bullet = 4.3\times 10^6$~$M_\odot$ \citep{eckart02,ghez05,ghez08,gillessen09}. The black hole is connected to the radio source Sagittarius A* (Sgr~A*).

Within the central $\sim$2\,pc around Sgr~A* lie ionised gas streamers, concentrated in three spiral arms \citep[e.g.][]{ekers83,herbst93}. They are called  the minispiral or Sgr~A West. The brightest features of the minispiral are the Northern Arm (NA), Eastern Arm (EA), Bar, and Western Arc  \citep[WA, e.g.][]{paumard04,zhao09,lau13}.

To understand the formation and growth of nuclear star clusters, it is important to study the stellar populations. 
Despite the complications from extinction and reddening, near-infrared  spectroscopy can be used to examine
stellar ages. For instance, studies of individual stars by \cite{blum03} and \cite{pfuhl11} suggested that  the dominant populations in the Milky Way nuclear star cluster are older than 5~Gyr.

Studies have shown that star formation in nuclear star clusters  continues until the present day \citep{jakob06}. Observations of nuclear clusters in edge-on spirals reveal  that young stars are located in flattened disks \citep{anil06,anilon4244_08}.  These younger components have a wide range of scales but most frequently appear to be centrally concentrated \citep{lauerm31_12,carson15}.

The Galactic centre likewise contains a young population of stars. Within the central parsec ($r$ \textless\,0.5\,pc) are $\gtrsim$100 hot early-type stars. These young stars are  O- and B-type  supergiants, giants, main-sequence stars, and post-main-sequence Wolf-Rayet (WR) stars \citep[e.g.][]{krabbe95,ghez03,paumard06,bartko10,do13}. The young  stars formed about 3$-8$~Myr ago \citep[e.g.][]{krabbe95,paumard06,lu13}. Dynamically, the young stars can be sorted into three different groups: (1)  stars within  $r$ \textless\,0.03\,pc (0.8\arcsec) are in an isotropic cluster, also known as S-star cluster.    Most of the $\gtrsim$20 stars are  B-type main-sequence stars. Then there are (2)  stars on a clockwise (CW) rotating disk with $r$\,$\approx\,$0.03$-$0.5\,pc (0.8$-$13\arcsec) distance to Sgr~A*, and (3) stars at the same radii as the stars in group (2), but not on the CW disk. It is under debate if there is a second, counter-clockwise rotating disk of stars within this  group \citep{genzel03,paumard06,bartko09,lu09,lu13,yelda14}. 
The stars in groups (2) and (3) have similar stellar populations \citep{paumard06}. It is unclear whether the stars of group (1) are the less massive members of the outer young population or if they were formed in one or  several distinct star formation events.

Most of the early-type stars are located within the central 1\,pc, but it is unclear if this is just an observational bias. Previous spectroscopic studies were mostly obtained within a radius of 0.5\,pc ($\sim$12\arcsec). \cite{bartko10} observed various fields with SINFONI and covered a surface  area of $\sim$500~sq.arcsec. However, the fields are asymmetrically distributed and mostly lie within 12\arcsec ($\lesssim$ 0.5\,pc) distance from the centre.  \cite{do13} observed an area of 113.7~sq.arcsec along the CW  disk. Their observations extend out to 0.58\,pc. \cite{stostad} mapped   an additional 80~sq.arcsec  out to 0.92\,pc  and found a break in the distribution of young stars at 0.52\,pc.
No previous spectroscopic study has fully sampled regions beyond the CW disk.

For this reason, we obtained $K$-band spectroscopy of the central 64\farcs9 $\times$ 43\farcs3  (2.51 $\times$ 1.68\,pc) of the Milky Way using the K-band Multi-Object-Spectrograph (KMOS, \citealt{kmos}) on the ESO/VLT. We covered an area of 2700~sq.arcsec  (0.75~sq.arcmin, \textgreater4\,pc$^2$), centred on Sgr~A* and symmetric in Galactic coordinates. From this data set we  extracted spectra for more than 1,000 individual stars and  obtained a map of the minispiral.  
We aim to classify the stars into late-type stars and early-type stars. For this purpose we use the CO absorption line as distinction. 
After the classification we investigate the  properties of the two different classes. 
Late-type stars will be treated separately in Feldmeier-Krause et al. (in prep.).

We here consider young populations of stars including O/B type stars, emission-line stars, and stars with featureless spectra. We also  present the intensity maps of ionised Brackett (Br)~$\gamma$ and He gas and of molecular H$_2$ gas. 
Over a nearly symmetric area of \textgreater 4\,pc$^2$ we investigate  the presence and spatial distribution of early-type stars. Furthermore, we derive photometric masses and collect the kinematics of the O/B stars. In addition, we   examine the spectral subclasses of the emission-line stars. 

This paper is organised as follows: In Sect.~\ref{sec:two} we describe the observations and data reduction. We outline the data analysis in Sect.~\ref{sec:analysis}. Our results are presented in Sect. \ref{sec:results}  and are discussed in Sect. \ref{sec:disc}. We  conclude with a summary in Sect. \ref{sec:con}.

\section{Observations and data reduction}
\label{sec:two}
\subsection{Spectroscopic observations}
\label{sec:obs}
\begin{figure*}
\centering
\includegraphics[width=18cm]{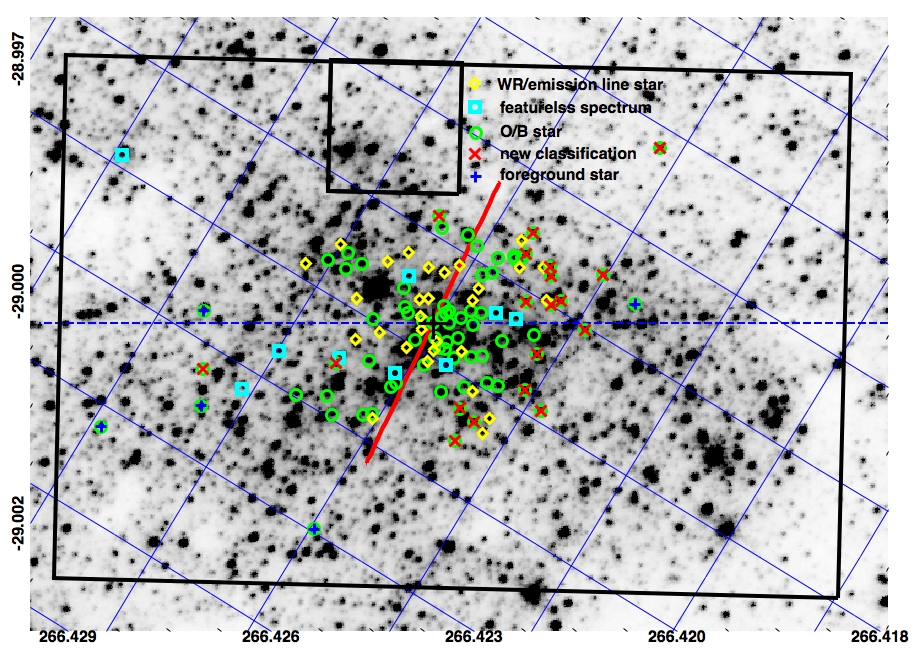}
\caption{Field of view and spatial distribution of early-type stars in the Milky Way nuclear star cluster. The black box shows the KMOS 64\farcs9 $\times$ 43\farcs3 field of view, i.e. 2.51 $\times$ 1.68\,pc, the small square in the upper middle was not observed due to the inactive IFU 13.  Blue lines are the equatorial coordinate grid with a spacing of 10\arcsec.  The dashed blue horizontal  line denotes the orientation of the  Galactic plane. The black cross  shows the position of Sgr~A*. The underlying image is from HST/NICMOS \citep{hst2}, aligned along Galactic coordinates. 
Yellow diamond symbols denote confirmed Wolf-Rayet (WR) and emission-line stars, cyan squares are stars with featureless spectra,  green circles denote O/B stars. Red x-symbols indicate new young star candidates, blue plus-symbols probable foreground stars. The red line denotes the line of nodes of the clockwise-rotating disk of young stars.}

\label{fig:cube}
\end{figure*}

Our spectroscopic observations were obtained with KMOS at VLT-UT1 (Antu)  on   September 23,  2013 during the KMOS science verification. KMOS consists of 24 IFUs with a field of view of 2\farcs8$\times$2\farcs8 each. We observed  in mosaic mode using the large configuration. This means that all 24 IFUs of KMOS are in a close arrangement,  and an area of  64\farcs9 $\times$ 43\farcs3 ($\sim$2880 sq.arcsec) is mapped with 16 dithers. 
There is a gap in the mosaic of 10\farcs8 $\times$ 10\farcs8 because one of the arms (IFU 13) was not working properly and had to be parked during the observations (see Fig.~\ref{fig:cube}). Therefore the total covered area is $\sim$2700\,sq.arcsec, corresponding to $\sim$4\,pc$^2$.
We observed two full mosaics of the same area with 16 dithers per mosaic. The mosaics are centred on $\alpha$=266\fdg4166 and $\delta$=$-$29\fdg0082 with a rotator offset angle at 120\degr. We chose the rotator offset angle  such that  the long side of a mosaic is almost aligned with the Galactic plane (31\fdg40 \,east of north, J2000.0 coordinates,  \citealt{2004reid}). The rotator angle only deviates by 1\fdg40 from the Galactic plane. Thus the covered  area is approximately point-symmetric with respect to Sgr~A*.

We used KMOS in the K-band ($\sim$1.934\,$\mu$m $-$ 2.460\,$\mu$m) with a spectral resolution $R = \frac{\lambda}{\Delta \lambda} \sim 4300$, which corresponds to a FWHM of  5.55~\AA\,measured on the sky lines. The pixel scale is $\sim$0.28\,nm/pixel in the spectral direction and 0.2\arcsec/pixel $\times$ 0.2\arcsec/pixel in the spatial direction. Each of the mosaic tiles consists of two 100\,s exposures. 
We observed one quarter of a mosaic on a dark cloud (G359.94+0.17, $\alpha \approx$ 266\fdg2, $\delta \approx$ -28\fdg9,   \citealt{dutra_darkcloud_01}) for sky subtraction.  B dwarfs were observed for telluric corrections. 

\subsection{Data reduction}
\label{sec:datared}

For  data reduction we used the KMOS pipeline Software Package for Astronomical Reduction with KMOS (SPARK, \citealt{spark}) in ESO Recipe Execution Tool (EsoRex). This package contains routines for processing dark frames, flat field exposures, arc frames obtained using argon and neon arc lamps, and standard star exposures. For the telluric spectra we used an IDL routine
that removes the Br~$\gamma$  absorption line from each telluric spectrum. The routine fits  the Br $\gamma$  line with a Lorentz profile and subtracts the fit from the telluric spectrum. It also divides the telluric spectrum by a blackbody spectrum to remove the stellar continuum.

We reduced both science and sky exposures by applying the following steps: flat fielding, wavelength calibration, cube construction, telluric correction, and spatial  illumination correction (using flat-field frames). The four sky exposures were average combined to a master sky, which we subtracted  from the object cubes. We used the method described by \cite{ohsky}, in which the sky cube is scaled to the object cube based on OH line strengths before subtraction.
Then we removed the cosmic rays from each object  cube with a 3D version of L.A.Cosmic \citep{lacosmic} provided by \cite{spark}. 

We extracted the spectra from the 736 data cubes using PampelMuse, a software package  written by \cite{pampel}.  PampelMuse was designed for extracting spectra from IFU observations of crowded stellar fields and enables clean extraction of stars even when their separation is smaller than the seeing. The program requires an accurate star catalogue. We used the
 catalogue provided by \citet[and in prep.]{rainer10}, which was obtained from NACO and HAWK-I observations. We ran PampelMuse
separately for every IFU because the astrometry of a mosaic cube is not accurate enough and because the point-spread function (PSF) of the observations varies in time.  As a consequence, the PSF in a mosaic varies between the individual 16 dither positions.

Within PampelMuse, the routine \textit{initfit} uses the source list and produces a simulated image with the spatial resolution of KMOS. This image is then used as a first guess for the position of the stars in the data cubes. Since the PSF varies with wavelength, \textit{cubefit} runs the PSF-fitting for each layer of the data cube. We restricted the PSF to a circular shape. This means that
the PSF is defined by  two  variables, the FWHM and the $\beta$-parameter of the Moffat profile.  The PSF variables, the coordinates, and the flux were fitted iteratively and for each layer of the data cube in the wavelength interval of [2.02 $-$ 2.42 $\mu$m]. We excluded wavelength regions with prominent gas emission lines (e.g. H$_2$, Br $\gamma$, He) from the  fit. 

The  coordinates and the PSF  vary only smoothly with wavelength, and the routine \textit{polyfit} fits a 1D polynomial to the coordinate and PSF parameters as a function of wavelength. The goodness of the PSF fit depends on the number of bright stars in the IFU. For IFUs  without bright stars in the field of view we  used the  PSF that was determined from IFUs with bright stars in the FOV. However, the PSF varies in time. 
Therefore we inspected the PSF fits for the 23 data cubes, where each data cube corresponds to a specific IFU. We did this separately for each exposure and selected the best PSF fits per exposure. We  combined the best  FWHM and best $\beta$  of one exposure to a mean FWHM and mean $\beta$, both as a function of wavelength. The FWHM lies between $\text{two }$\text{to three}\,pixels  for the 32 exposures because the seeing also changed from 0.7\arcsec\, to 1.3\arcsec\, during the night. With the knowledge of the PSF of each exposure  and the star coordinates,    the routine \textit{cubefit} was run again to determine the flux for each star on all layers of the data cubes. 

After extracting background-subtracted spectra with PampelMuse, we shifted each spectrum to the local standard of rest. PampelMuse extracted $\sim$12,000 spectra of more than 4,000 different stars with $K_S$\textless 17\,mag in the KMOS field of view. We discarded all spectra with a signal-to-noise ratio (S/N) below 10 or negative flux, leaving $\sim$3,000 spectra.
The  S/N   for each extracted spectrum was calculated by PampelMuse with Eq. 16 of \cite{pampel}. 
We combined the two spectra of each star from the two exposures. For $\sim$180 stars we had even more than two spectra from the two mosaics, since PampelMuse also extracted spectra from stars that were centred outside of the field of view of the IFU. 
  We combined the spectra with the best S/N to one spectrum per star by taking a noise-weighted mean. The S/N between the individual exposures typically differed by less than 10. We obtained spectra for more than 1,000 individual stars with a formal total S/N \textgreater 10.

We also constructed a mosaic using  the data cubes from all 32 exposures. This mosaic extends over 64\farcs9 $\times$ 43\farcs3, with a gap for the inactive IFU 13. To determine the astrometry of the mosaic, we used  the 1.9\,$\mu$m image of the HST/NICMOS Paschen-$\alpha$ Survey of the Galactic centre \citep{hst1,hst2} as a reference. This image has a pixel scale of 0\farcs1/pixel. We collapsed the KMOS mosaic data cube to an image and rebinned it to the HST pixel scale. The two images were iteratively cross correlated. Although the two images cover different wavelength regions, a large enough number of stars is detected in both images to line the frames up. Finally, we applied a correction to the local standard of rest.  This mosaic data cube  was used to measure the gas emission lines of the minispiral and circumnuclear ring.

\section{Data analysis}
\label{sec:analysis}
\subsection{Photometry}
\label{sec:photo}
To be able to determine the spectral classes and  colours of the stars, we complemented our spectroscopic data set with photometric measurements. 
\citet[and in prep.]{rainer10} observed the Milky Way nuclear star cluster with  NACO and HAWK-I and constructed a star catalogue. This catalogue provides $J$ (HAWK-I), $H$, and $K_S$ (HAWKI-I and NACO) photometry.  The NACO catalogue extends over the central $\sim$40\arcsec$\times$40\arcsec, HAWK-I data were used for  regions farther out.

The brightest stars are saturated in the HAWK-I and NACO images, and we complemented our photometry with other star catalogues. We used photometry from the SIRIUS catalogue \citep{shogo06} for eight stars,  and for three further bright stars without HAWK-I, NACO or SIRIUS photometry, we used photometry form the 2MASS catalogue  \citep{2mass}.
 For almost 1,000 
stars we have the $JHK_S$ photometry from either NACO/HAWK-I,  SIRIUS or 2MASS , for a further 100 
 stars we only have $HK_s$ photometry. For two stars we have no $K_S$ photometry, but $JH$ photometry. 
 
To obtain  clean photometry, we corrected for  interstellar extinction. In the Galactic centre, extinction varies on arcsecond scales \citep[e.g.][]{scoville03,rainer10,fritzextinktion}. The typical extinction values are about 2.5\,mag  in the $K_S$ band, 4.5\,mag in the $H$ band, and more than 7\,mag in the $J$ band.
We used the extinction map and the extinction law derived  from  \cite{rainer10}\footnote{We downloaded the extinction map from the CDS database. It turned out that the astrometry of the extinction map was wrong by a scale factor of 60. We reported this issue to CDS, and the astrometry was fixed on 26th March, 2015.} 
for the extinction correction of the photometry.  About 350 ($\sim$30\%) of the stars are outside the field of view of the extinction map. For these we assumed that the extinction is the mean value of the extinction map $A_{K_S}=2.70$\,mag. 

The extinction map was created after excluding foreground stars. Therefore, any foreground star will be strongly over-corrected to very negative colours. 
The intrinsic colours  of cluster members are in a very narrow range of about $-$0.13\,mag  \textless\,$(H\!-\!K_S)$ \textless\,0.38\,mag \citep[Table~7.6,  and 7.8]{rainer10,do13,rainer_review14,allen}. We  used this knowledge to identify foreground stars. Stars with a  bluer extinction-corrected $(H\!-\!K_S)_0$ colour than the intrinsic colour are   foreground stars.  
To account for uncertainties in the extinction correction, we used a larger colour interval and classified a foreground star when the extinction-corrected $(H-K_S)_0$ colour  was less than $-$0.5\,mag. 

Identifying background stars is less obvious. Very red stars may not be background stars, but be embedded in local dust features or have dusty envelopes. \cite{viehmann06} showed that several red sources in the Galactic centre are not background stars, but bow-shock sources. For red sources we have to consider the spectral type and the surroundings of the star to decide whether it is locally embedded or a background star.

\subsection{Completeness}
\label{sec:comp}
It is important to know how complete our spectroscopic data set is up to a given magnitude. Completeness is influenced by  various factors,  for example  the depth of the observation, the spatial resolution, but also the stellar number density of the observed field.  In a  dense environment, crowding becomes stronger, and fewer faint stars can be detected.

We used the photometric catalogue by \citet[and in prep.]{rainer10} to extract the stars, which means that our spectroscopic data set can at best be as complete as the photometric catalogue. Our data have a lower spatial resolution than the images used to produce the photometric catalogue, and therefore the completeness of our data set must be lower.  The photometric catalogue contains $\gtrsim$6,000 stars in the KMOS field of view. PampelMuse extracted spectra from more than 4,000 stars with $K_S$\textless 17\,mag. Only $\sim$1,000 of these have a spectrum with a S/N greater than 10.   We determined the completeness of the spectroscopic data set by comparing  our data set  with the photometric catalogue  in different magnitude bins. 
We  assumed that the photometric catalogue is complete to 100\% up to $K_S$=15\,mag, at least at a projected distance  $p$\,\textgreater10\arcsec\,from Sgr~A*.

\begin{figure}
        \resizebox{\hsize}{!}{\includegraphics{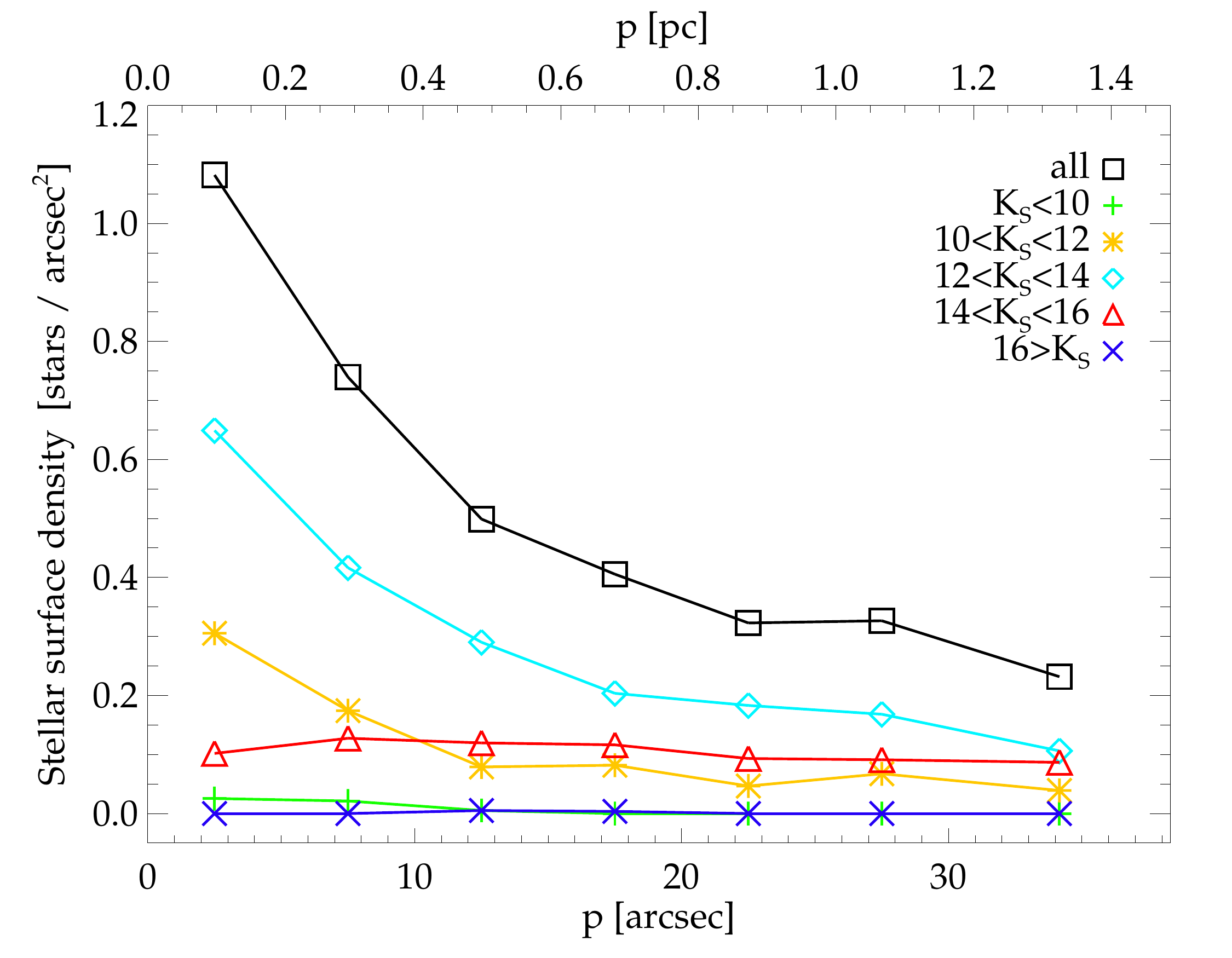}}
        \caption{Number density profile of the spectroscopic data set in different magnitude bins. }
        \label{fig:ndens}
        \end{figure}

The effect of crowding is illustrated in Fig. \ref{fig:ndens}.        
 We plot the number density profile of our spectroscopic data set as a function of the projected  distance $p$ to Sgr~A* in different magnitude bins. Most of the stars are in the magnitude bin of 12 $\le K_S \le$ 14. The number density of bright stars with 10 $\le K_S \le$ 14 decreases with increasing radius in the central 10\arcsec, while the number density of faint stars in the magnitude bin 14 $\le K_S \le$ 16 is nearly constant in the same radial range and even slightly increases. 

The reason for this is crowding:
There are more bright stars in the centre of the cluster, and they outshine the faint stars. Therefore we miss more faint stars in the centre than farther out.
This effect was shown in previous studies \citep[e.g.][]{rainer07,do09,bartko10}.

\begin{table}
\caption{Completeness limits of the spectroscopic data set for different radial bins. }
\label{tab:completeness}
\centering
\begin{tabular}{c c c}
\noalign{\smallskip}
\hline\hline
\noalign{\smallskip}
 distance& 80\% completeness at $K_S$&50\% completeness at $K_S$  \\
$\left [\text{arcsec}\right ]$  &$\left [\text{mag}\right ]$ &$\left [\text{mag}\right ]$\\
\noalign{\smallskip}
\hline
\noalign{\smallskip}
$p$ \textless\,5\arcsec& $  13.0\,\pm\, 0.3 $&$ 13.8\,\pm\,  0.1$ \\
5\arcsec$\leq$ $p$\,\textless 10\arcsec & $   13.3\,\pm\, 0.1 $&$  14.1\,\pm\,  0.1 $ \\
$p$\,$\geq$ 10\arcsec & $ 13.5\,\pm\, 0.2 $&$  14.1\,\pm\,  0.2 $ \\
\noalign{\smallskip}
\hline
\noalign{\smallskip}
\end{tabular}
\tablefoot{The limiting magnitude is given in $K_S$.}
\end{table}

 As a result of the higher  crowding in the centre, the completeness limits depend on the distance from the centre. For this reason we determined the completeness separately for stars located within $p$\textless5\arcsec\,from Sgr~A*, stars with 5\arcsec $\leq$$p$\,\textless 10\arcsec, and stars beyond 10\arcsec. 
  The spectroscopic completeness was then estimated by comparing the number of stars as a function of magnitude N($K_S$) in the spectroscopic data set with the total number of stars from the photometric catalogue. We calculated the fraction of stars that are missing in the spectroscopic data set for different magnitude bins to correct our number density results by that fraction. To derive the fraction of missing stars, we binned the stars  in magnitude bins of $\Delta K_S$.   We varied the size of the magnitude bins to test the effect of  the magnitude binning. We tried $\Delta K_S$=0.7\,mag, $\Delta K_S$=0.5\,mag, and $\Delta K_S$=0.3\,mag. The difference gives the uncertainties of the completeness limits. 
We list our 80\% and 50\% completeness limiting magnitudes in Table~\ref{tab:completeness} for the three different radial bins. At greater distances, the completeness limits are shifted to  fainter stars than in the centre as a result of crowding. 
 The completeness limits did not vary beyond their uncertainty when we  chose slightly different radial bins.

We investigated the effect of source confusion on our ability to classify stars. We conclude that crowding only has a minor effect on our completeness limit, and the S/N degradation does not severely affect our ability to classify stars brighter than our completeness limit.

\subsection{Spectral identification of late- and early-type stars}

We visually investigated the spectra and classified the stars into three categories: (a) late-type stars, (b) early-type stars,
and (c) uncertain type. Late-type stars are rather cool and have CO absorption lines.  Most of them are  of old to intermediate age \citep{pfuhl11}, although there are exceptions such as the red supergiant IRS~7, which is only $\sim$7\,Myr old \citep{carr10}. Late-type stars are in the majority with  $\sim$990 stars. They will be analysed in detail in Feldmeier-Krause et al. (in prep.).

Early-type stars can be separated into emission-line stars, O/B stars, and featureless spectra. 
The data set contains  29 stars with  emission lines,   23 of which are Wolf-Rayet (WR) stars and six stars have  narrow emission lines (see Sect.~\ref{sec:wr}). The  O/B star spectra have no CO lines but rather He and/or H (Br~$\gamma$) absorption lines. Our data set contains 76 O/B stars (see Sect.~\ref{sec:ob}). A further nine stars have featureless spectra without strong absorption  or emission features, but strongly increasing continuum (see Sect.~\ref{sec:fless}). They are associated with bow shocks. The remaining 40 spectra are in category (c) of uncertain type, mostly because the S/N was too low or because the spectra are contaminated by the light of nearby brighter stars.  

\begin{figure}
	\resizebox{\hsize}{!}{\includegraphics{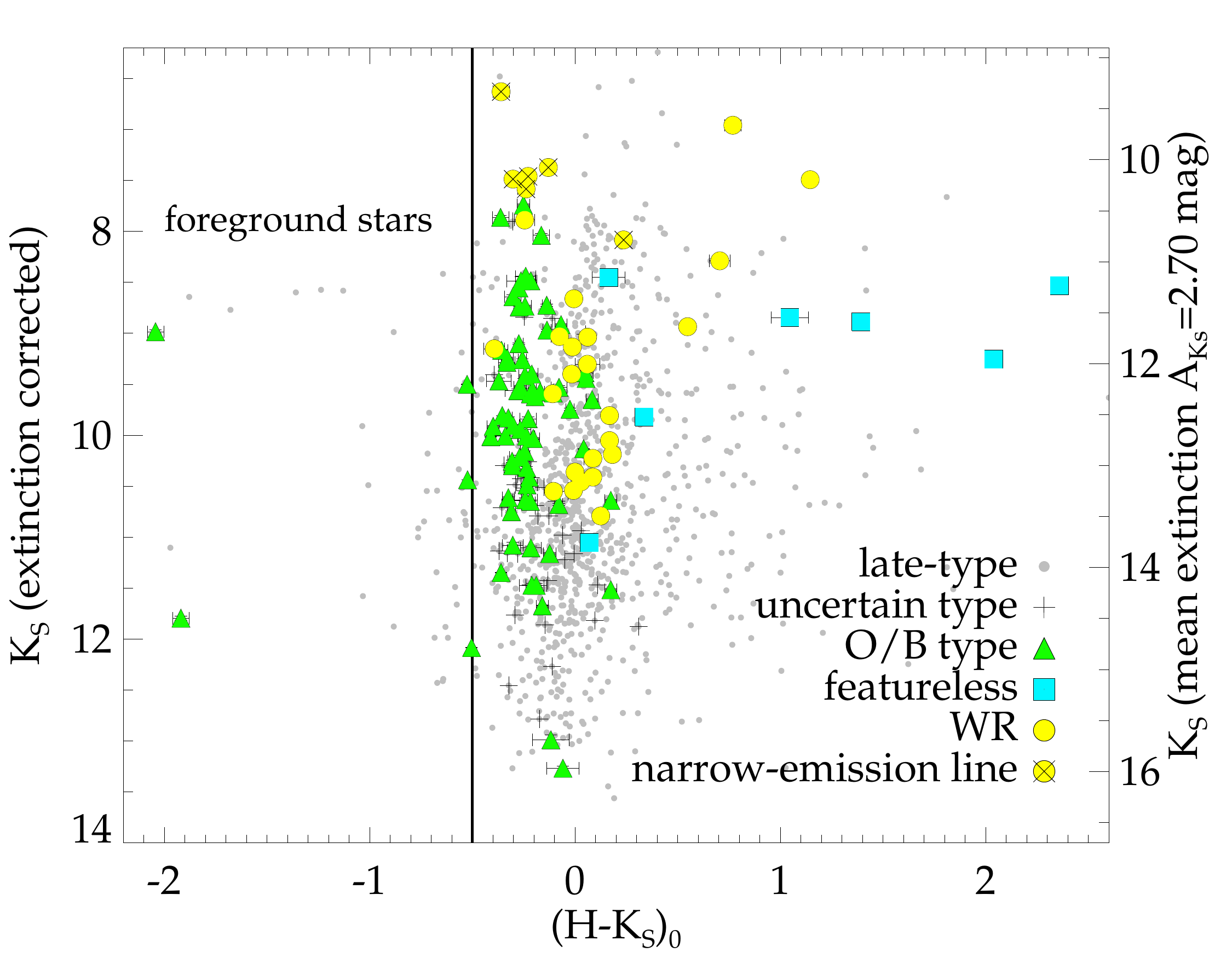} }
	\caption{Colour-magnitude diagram of the  stars within the KMOS field of view with extracted spectra and $H$ and $K_S$ photometry, after extinction correction. Stars with colours $(H\!-\!K_S)_0 <$ $-$0.5\,mag are most likely  foreground stars (left of the vertical line). Different symbols and colours denote different types of stars. Yellow circles denote emission-line/WR stars, cyan squares are sources with featureless spectra, green triangles are O/B type stars, grey dots are late-type stars, black plus-signs are  stars of uncertain type. The right $y-$axis denotes the $K_S$ magnitude after extinction correction,  with a mean extinction of $A_{K_S}$=2.70\,mag.}
		\label{fig:cmd}
	\end{figure}

 Figure~\ref{fig:cmd} shows a colour-magnitude diagram (CMD) using $H$ and $K_S$ after extinction correction.  The location of these stars is also indicated in Fig.~\ref{fig:cube} with the same colour coding. We would like to point out  that almost all WR stars are redder than the O/B stars. This is because they have evolved off the main sequence and may be producing dust. Therefore, the observed mean position of the WR stars  on the red side of the CMD supports our stellar classification and the accuracy of the CMD.

\subsection{Deriving stellar kinematics}
\label{sec:measurekin}
Stellar kinematics are useful to  study the origin  of the early-type stars. In this section we describe our routine to fit the radial velocities of O/B stars. 

The broad lines of the Wolf-Rayet stars make it difficult to determine their radial velocities. The lines are mostly a combination of several blended lines, and the stars have fast winds and outflows. For the featureless sources no spectral lines can be fitted. 
For the O/B stars we used the penalized pixel-fitting
(\textit{pPXF}) routine to fit the Br~$\gamma$ and He lines \citep{ppxf}. We used template spectra from three different libraries: \cite{wallace}, \cite{hanson05}, and  KMOS B-main-sequence stars. The KMOS B-main-sequence stars were observed in our program as telluric  standard stars.  From \cite{hanson05} we only used the O/B stars that were observed with  ISAAC/VLT ($R \sim 8,000$). We measured the radial velocities of the O/B templates by fitting  the Br\,$\gamma$ line and shifted the templates to the rest wavelength. 
The high-resolution templates were convolved with a Gaussian to match the spectral resolution of the KMOS data. Then we ran  \textit{pPXF}  on our data. 

The uncertainty of the radial velocity was measured using Monte
Carlo simulations. We added random noise to the spectra and fitted the radial velocities in 100 runs. The standard deviation of the 100 measurements was our uncertainty. The results are listed in Table~\ref{tab:etkin} and are analysed in Sect. \ref{sec:kin}.  
The wavelength region of the He~I and Br~$\gamma$ absorption lines also shows He~I and Br~$\gamma$ emission from ionised gas (see Sect. \ref{sec:gas}). The program PampelMuse subtracts the background when extracting the stellar spectra, but the surrounding gas emission increases the noise in this wavelength region. This induces  high uncertainties in our radial velocity measurements. For this reason, the median value of the radial velocity uncertainty is $\sigma_{median}\approx$ 60\,km~s$^{-1}$. 
We compared our radial velocity measurements with the data of \cite{bartko09} and \cite{yelda14}. There are nine  stars with independent radial velocity measurements from this work and the previous studies. Using these stars, we can estimate the so-called true external $\sigma$  of our measurement, meaning that we can test whether we over- or underestimate the uncertainties. The procedure  was described by \cite{reijns06}. First, we measured the mean velocity offsets $\langle v_i-v_j \rangle$ ($i=1,2,3; j=2,3,1$) between each pair of the three studies for the overlap stars and the respective standard deviation $\sigma^2_{v_i-v_j}$. Because $\sigma^2_{v_i-v_j} = \sigma^2_{v_i} + \sigma^2_{v_j}$, we can calculate the true $\sigma_{v_i}$  ($i=1,2,3$) from the three measurements of  $\sigma^2_{v_i-v_j}$.

A comparison of the external error $\sigma_{ext}$ with the mean error $\sigma_{mean}$ of the individual radial velocity measurements indicates whether  we over- or underestimate the uncertainty. 
The   external error $\sigma_{ext}$ = 45\,km~s$^{-1}$  for our radial velocities is smaller than the  mean error $\sigma_{mean}$  of the nine overlap stars. $\sigma_{ext}$ is approximately 0.7 times the mean error $\sigma_{mean}$. 
Of the nine stars with three independent radial velocity measurements, five stars  in our data set have a high S/N\textgreater 56 (Id 109, 205, 294, 331, 372), but  four stars (Id 707, 1123, 1238, 2233) have a low  S/N (\textless 30). 
The velocities of three of these four stars  with low S/N agree with the measurement of  \cite{bartko09} or \cite{yelda14} within the uncertainties. However, we  consider the radial velocity measurements of the five stars with the higher S/N more reliable. The external error calculated from the  five stars  with high S/N is $\sigma_{ext}$ = 27\,km~s$^{-1}$. This is 0.8 times the mean error $\sigma_{mean}$, thus our errors appear to be accurate to within 20\%.
Although nine independent radial velocity measurements are not enough for an accurate determination of $\sigma_{ext}$, our analysis indicates that we do not underestimate the radial velocity errors.

\section{Results}
\label{sec:results}

Here we first present  the O/B type stars, and we derive their masses from the photometry. We obtain maps of the emission line flux that is  generated by the minispiral and the circumnuclear ring. For stars with narrow emission lines and Wolf-Rayet stars we show  spectra and the spectral classification, followed by the spectra of featureless sources. We finally present the spatial distribution of the early-type stars. We also investigate the O/B star kinematics and  stellar orbits. 

\subsection{O/B type stars}
\label{sec:ob}
\subsubsection{Identifying O/B stars}
\label{sec:obid}

	\begin{figure*}
	\centering
	\includegraphics[width=15cm]{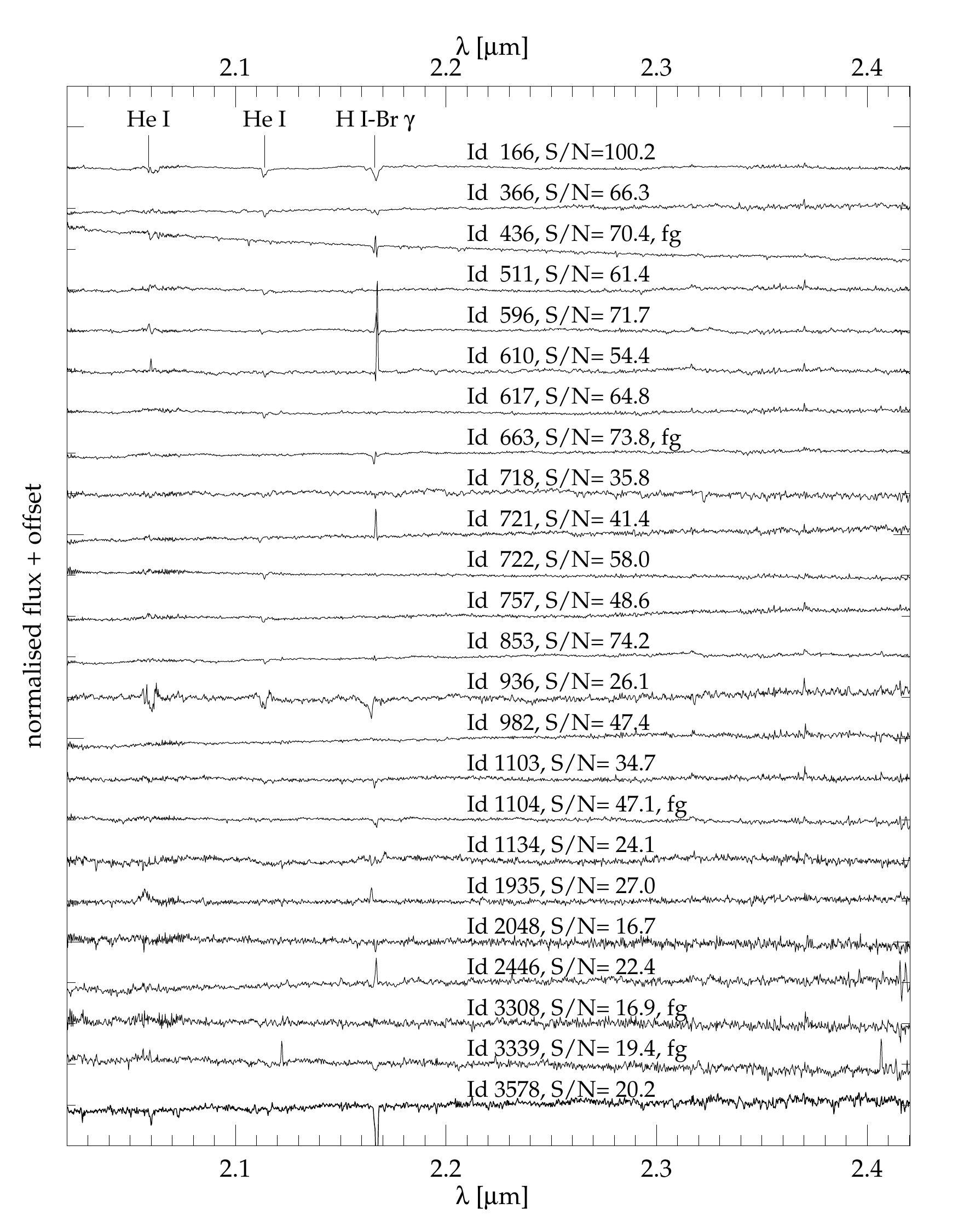}
	\caption{Spectra of the newly identified O/B type stars. The fluxes are normalised and an offset is added to the flux. The spectra are not shifted to rest wavelength. The numbers denote the identification numbers of the stars as listed in Tables \ref{tab:et}, \ref{tab:etmag}, and \ref{tab:etkin}. We also show the S/N and denote probable foreground stars with ``fg''.}
	\label{fig:obnew}
	\end{figure*}

O/B-stars have effective temperatures of $T_{\mathrm{eff}}>$10,000~$K$ \citep[e.g.][]{martins05,crowther06}.  The most prominent lines in O/B giant $K$-band spectra are the He~I (2.058~$\mu$m, 2.113~$\mu$m and $\sim$2.164~$\mu$m), H~I (4-7) Br~$\gamma$ (2.166~$\mu$m), and He~II  (2.1885~$\mu$m) lines \citep{hanson05}. The 2.113~$\mu$m complex is also partly generated by N~III. These lines appear mostly in absorption, but can also be in emission or absent, depending on the spectral type \citep{morris96}. 

Previous studies found $\sim$100 O/B supergiants, giants, and main-sequence stars in the innermost parsec of the Galaxy \cite[e.g.][]{paumard06,bartko09,do13,stostad}.  Our spectroscopic data set contains 76 O/B stars, 52 of which were reported in previous spectroscopic studies,  
but 24 sources appear not to have  been identified before, due primarily to the wider field of view of our observations relative to previous spectroscopic studies.

The spectra of the newly identified O/B stars are shown in Fig.~\ref{fig:obnew}. Five of them are probably foreground stars, as they have very blue  colours (Id 436, 663, 1104, 3308, and 3339, $(H-K_S)_0=-2.04, -0.53, -0.52, -0.50$, and $-$1.92\,mag). For one of the O/B stars (Id 982) we had to assume a mean extinction value of $A_{K_S}=2.70$\,mag
because this star is beyond the field of view of the  extinction map of \cite{rainer10}. This means there is a large uncertainty in the star's colour of $(H-K_S)_0=0.04$\,mag. If the local extinction is higher than the assumed mean extinction value of $A_{K_S}=2.70$\,mag, this could mean that this star also is a foreground star. For the star Id 2048 we have no colour information and cannot determine
whether this star is a foreground star.

We list our sample of O/B-type stars in Table~\ref{tab:et}. This table provides the star Id, right ascension R.A., declination Dec (in equatorial coordinates), the offset coordinates $\Delta$R.A. and $\Delta$Dec with respect to Sgr~A*, the magnitude $K_{S}$, remarks on the star colour, the star name and type (if available), a note to the respective reference, and the S/N.

The  O/B-type stars were identified by inspecting the spectra of all stars in our data set. To verify our visual classification, we measured the equivalent widths ($EW$) of the $^{12}$CO(2,0) line at 2.2935\,$\mu$m, and the Na~I doublet at 2.2062\,$\mu$m and 2.2090\,$\mu$m. We used the definitions of  band and continuum from \cite{frogel}.  For the late-type stars we obtain a mean value of $EW_{CO,LT}=18.30$ ($EW_{Na,LT}=4.60$) with a standard deviation of $\sigma_{CO,LT}=5.20$ ($\sigma_{Na,LT}=2.13$). The mean uncertainty  is only $\Delta_{EW_{CO,LT}}=0.39$ ($\Delta_{EW_{Na,LT}}=0.25$). For the O/B stars, the equivalent widths for CO and Na are lower, with a mean value of $EW_{CO,O/B}=-0.76$ and $\sigma_{CO,O/B}=3.25$ ($EW_{Na,O/B}=0.47$ and $\sigma_{Na,O/B}=1.75$). This means that the equivalent width of the CO  line of O/B stars is on average more than 3.67$\sigma$ smaller than for late-type stars, and  the equivalent width of Na is $\sim$1.97$\sigma$ smaller. We list the equivalent widths of CO and Na for the O/B stars in Table~\ref{tab:etmag}. 

O/B giants and supergiants have observed magnitudes of $K_S$=11$-$13\,mag at the Galactic centre, while O/B main-sequence stars have $K_S$=13$-$15\,mag \citep{eisenhauer05, paumard06}. To estimate the luminosity class of the O/B stars in our sample, we corrected  the  $K_S$ magnitude  using the extinction map provided by \cite{rainer10} and added a mean extinction of A$_{Ks}$=2.70\,mag to the $K_S$ magnitude. We chose A$_{Ks}$=2.70\,mag since this is the mean value of A$_{Ks}$  from \cite{rainer10} in our field of view. The resulting  values are  given in Table~\ref{tab:et} (see also the right y-axis in Fig.~\ref{fig:cmd}). With this rough magnitude cut, we estimate that about 70\% of the O/B stars in our data set are giants or supergiants, and 30\% are main-sequence stars.

\subsubsection{Mass estimates and dust extinction}
\label{sec:massdust}
To determine the spectral type of O/B stars in the $K$ band, the data quality has to be very high. The He~I line at 2.058~$\mu$m is in a region of high telluric absorption and low S/N. The minispiral emission  increases the noise at 2.058~$\mu$m and 2.166~$\mu$m. Since the gas emission is spatially highly variable, the background subtraction is imperfect. But even without these difficulties, a spectral classification is complicated. 
\cite{hanson05} collected $K$-band spectra of O and early-B stars of known spectral type. They found that for a determination of T$_{\mathrm{eff}}$ and log~$g$, the spectral resolution should be   $R\approx$~5,000 or higher. 
Furthermore, a S/N $>$ 100 is desirable. For  the stars in our data set, these conditions  are not fulfilled.  
Therefore we cannot place more constraints on the spectral types of our O/B star sample.  

Nevertheless, we can estimate the mass of the O/B stars under some assumptions from the photometry. The intrinsic colour of O/B stars  is in a very narrow range close to $(H-K_S)_0\approx-0.1$\,mag \citep[Table~7.6,  and 7.8]{allen}. Therefore the wide spread of the O/B stars over $\sim$1\,mag on the CMD (Fig. \ref{fig:cmd}) is mostly due to imperfect extinction correction. For the extinction correction we used the extinction map of \cite{rainer10}. It was derived  by averaging over several stars and is therefore only an approximation to the real local extinction.
However, because we spectroscopically selected the stars  and all O/B stars have intrinsic colours $(H-K_S)_0\approx-0.1$\,mag, we can use the photometric colours to obtain independent extinction  estimates.
We assumed an extinction law of $A_\lambda \propto \lambda^{-\alpha}$ to calculate the true extinction for each single O/B star and its true magnitude  $K_{S,0}$. We used the extinction law coefficient of $\alpha$=2.21 \citep{rainer10}. The results for $K_{S,0}$ and $A_{K_S}$ are listed in Cols. 4 and 5 of Table~\ref{tab:etmag} for the 73 O/B stars with $H$ and $K_S$ photometry. The    uncertainty  $\sigma_{Ks,0}$  contains the propagated  error of the measured photometry $\sigma_{H}$ and  $\sigma_{K_S}$, the error of the true intrinsic colour $\sigma_{H-K_S}$,  the extinction-law coefficient uncertainty $\sigma_{\alpha}$, and the Galactocentric distance uncertainty $\sigma_{R_0}$.

\begin{figure}
	\resizebox{\hsize}{!}{\includegraphics{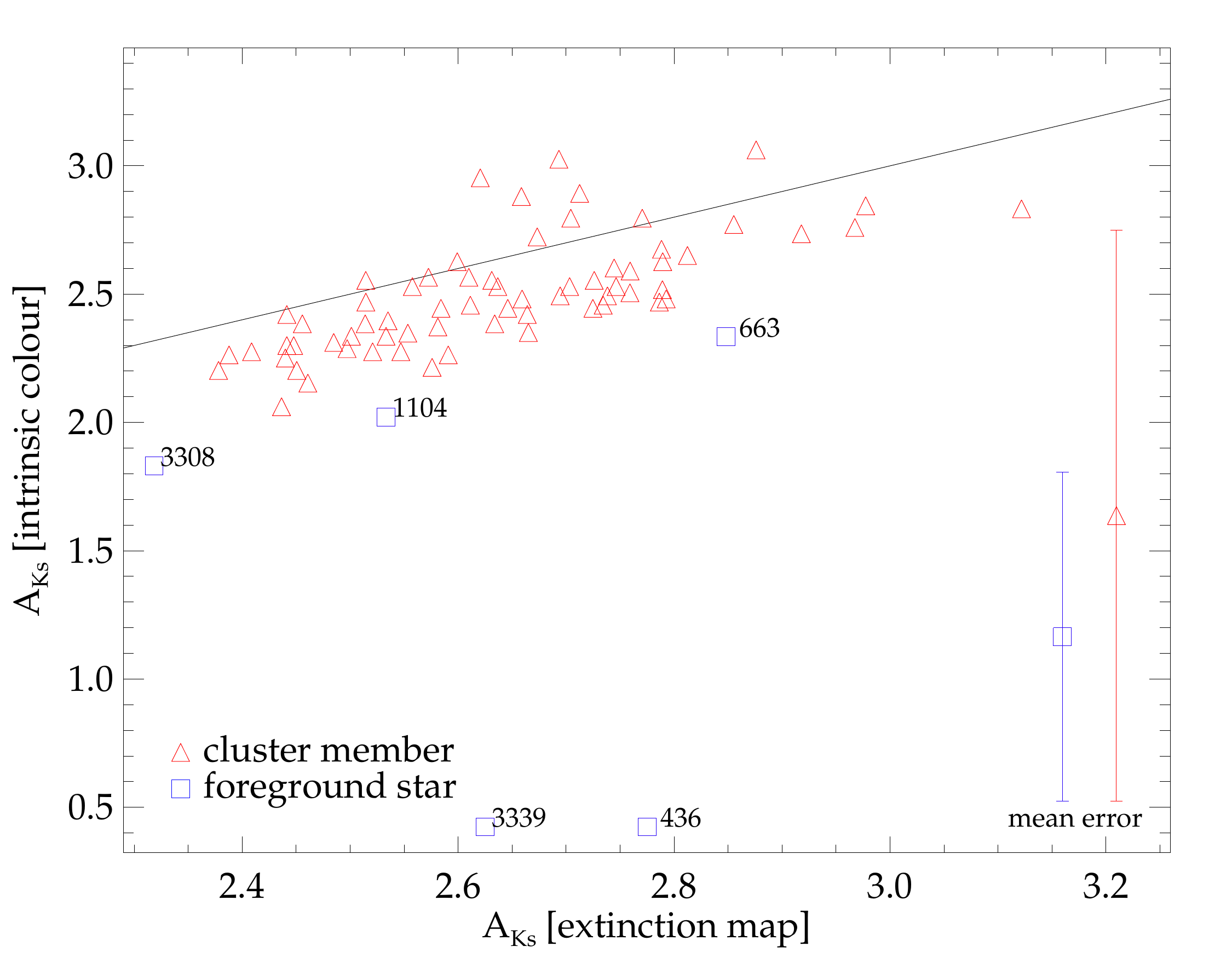}}
	\caption{Comparison of the extinction $A_{K_S}$ in magnitudes derived from the intrinsic colour with the extinction from the extinction map of \cite{rainer10} for O/B stars. The black line denotes the 1:1 line, blue squares are foreground stars, red triangles are cluster member stars. Typical error bars are shown in the lower right corner.  }	
	\label{fig:extinction}
	\end{figure}

The derived extinction values $A_{K_S}$ range from 0.42\,mag for probable foreground stars to 3.06\,mag. 
The median extinction value of O/B stars that are not flagged as foreground stars is 2.48\,mag with a standard deviation of 0.22\,mag. The extinction derived from the extinction map is mostly  higher, with a median of $A_{K_S}$=2.63\,mag and a standard deviation of 0.15\,mag. 
We plot the extinction derived from the intrinsic colours against the extinction from the extinction map of \cite{rainer10} in Fig.~\ref{fig:extinction}. For the two stars Id 436 and 3339 it is obvious that they are foreground stars, the extinction derived from the intrinsic colour is lower by more than 2\,mag than $A_{K_S}$ from the extinction map. We also classified the three stars Id 663, 1104, and 3308  as foreground stars. With the large uncertainty of the extinction, these stars might be  cluster member stars. 

There appears to be a systematic offset between the extinction:  $A_{K_S}$ derived from intrinsic colours is mostly lower by $\sim$0.2\,mag than the value of  $A_{K_S}$  from the extinction map. We varied different input parameters to test their effect on our result of $A_{K_S}$. 
A lower value of  $(H-K_S)_0$ than $-0.1$\,mag is  unlikely. However, when we changed the extinction law coefficient $\alpha$ from   2.21 \citep{rainer10} to 2.1, the offset of 0.2\,mag disappeared. Previous studies measured $\alpha$ in the range of 2.0 to 2.64 \citep{gosling09,stead09,shogo09,rainer10}. The value of $\alpha$ has the largest uncertainty and can therefore alone account for the offset.

We  also used isochrones to estimate the stellar mass given the position of the star in the CMD. We used the isochrones of  \cite{bressan12}, \cite{chen14} and \cite{tangiso14}  downloaded at \footnote{\url{http://stev.oapd.inaf.it/cmd}} with  solar metallicity. \cite{ramirez00}  found that the iron abundance Fe/H  of the Galactic centre stars is roughly solar. However, the $\alpha$-element abundance is super-solar \citep{cunha07,martins08}. \cite{paumard06} and \cite{lu13} showed that the young population in the Galactic centre is 3$-$8\,Myr old. We used isochrones in this age interval with a spacing of $\Delta (\log(age/yr))$ = 0.01. 
The isochrones are for 2MASS photometry, therefore we shifted the colours to our ESO photometry using the equations given by \citet[2003 version at \footnote{\url{http://www.astro.caltech.edu/~jmc/2mass/v3/transformations/}}]{colortransform}.

\begin{figure}
        \resizebox{\hsize}{!}{\includegraphics{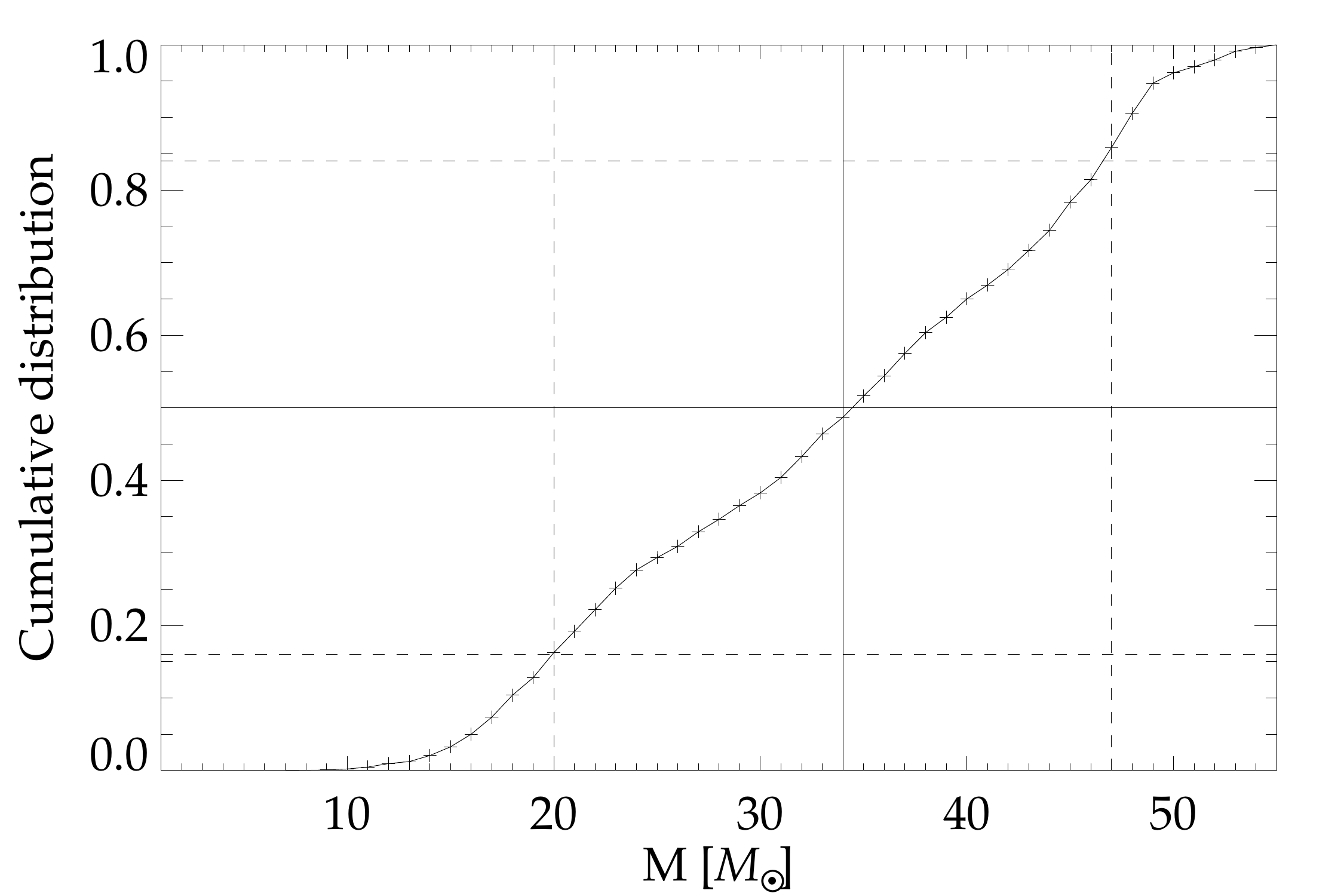}}
        \caption{Cumulative mass distribution of star Id 617. The horizontal lines denote 0.16, 0.5, and 0.86 percentiles, the vertical lines denote the corresponding masses. We derived a mass of $\mathcal{M}=34^{+13}_{-14}$\,$M_\odot$ for this star.}
        \label{fig:cum617}
        \end{figure}

For the O/B stars in our data set we computed the likelihood $\mathcal{L}$ of $(H-K_S)_0=(H-Ks)_{iso}$ and $K_{S,0}=K_{S,iso}$ 
\begin{eqnarray*}
\mathcal{L}=\frac{1}{\sqrt{{2\pi}}\sigma_{Ks,0}}\exp\left ( {-\frac{1}{2} \left( \frac{K_{S,0} - K_{S,iso}}{\sigma_{Ks,0}}\right)^2} \right ) \times
\\
\hspace{5mm}\frac{1}{\sqrt{{2\pi}}\sigma_{(H-Ks)_0}}\exp \left ({-\frac{1}{2} \left( \frac{(H-K_S)_{0} - (H-K_S)_{iso}}{\sigma_{(H-Ks)_0}}\right)^2} \right ),
\end{eqnarray*}
where $(H-Ks)_{iso}$ and $K_{S,iso}$ are the isochrone points from all isochrones in our age interval. 
To each isochrone point there is a corresponding stellar mass $\mathcal{M}$. Because we used various isochrones, there can be different stellar mass values for the same value of $(H-Ks)_{iso}$ and $K_{S,iso}$.  We have a  distribution of stellar masses, and we used the likelihood to calculate the probability mass function of the stellar mass for each O/B star separately.  In Table~\ref{tab:etmag} we list the median mass of each star in  the  probability function  (column $\mathcal{M}$),  the uncertainties are derived from the 0.16 and 0.84 percentiles. Figure~\ref{fig:cum617} shows the cumulative mass distribution  of star Id 617 as an example. 
The masses of our O/B star sample range from 43\,$M_\odot$ for the brightest stars to only 7\,$M_\odot$ for a probable foreground star.   When we used isochrones with a slightly higher metallicity, we obtained  lower stellar masses in most cases. However, the results agree within their uncertainties.

We  estimated the total mass of the young star cluster with some assumptions. In Sect. \ref{sec:photo} we have shown that the 80\% completeness limit is at $K_S$$\approx$13.2\,mag. When we  consider  only O/B stars with $K_S\leq$ 13.2\,mag and with $\mathcal{M}\geq$ 30\,$M_\odot$,  the mass function is approximately complete. The initial mass function (IMF) of  young stars in the Galactic centre is top-heavy \citep{bartko10,lu13}. 
We fitted  the  IMF $dN/dm = A \times m^{-\alpha}$ to the observed mass function in the mass interval $\left[ 30 M_\odot; 43 M_\odot \right ] $, where we have 51 stars. We used the software $mpfit$ \citep{mpfit}  to fit the coefficient $A$ and use $\alpha$ values from the literature. We refrained from fitting $\alpha$.  The covered mass interval and  the number of stars are too small to constrain the shape of the IMF.
Then we integrated the  IMF from $\mathcal{M}=\left[ 1 M_\odot; \mathcal{M}_{max} \right ]$ to obtain the total mass of the young star cluster. As our mass function contains only O/B stars and no emission-line stars, which are also young and in the same mass interval, we derived only a lower limit for the young star cluster mass.

Assuming an IMF with $\alpha$=1.7 \citep{lu13} and  $\mathcal{M}_{max}$=150\,$M_\odot$, we obtain $\mathcal{M}_{young, M\leq150 M_\odot}^{\alpha=1.7}$ = 21,000\,$M_\odot$, and with $\alpha$=0.45 \citep{bartko10}, we obtain $\mathcal{M}_{young, M\leq150 M_\odot}^{\alpha=0.45}$ = 32,000\,$M_\odot$.
With an upper integration limit of $\mathcal{M}_{max}$=80\,$M_\odot$, the young cluster mass is  $\mathcal{M}_{young, M\leq80 M_\odot}^{\alpha=1.7}$ = 16,000\,$M_\odot$  for $\alpha$=1.7 and $\mathcal{M}_{young, M\leq80 M_\odot}^{\alpha=0.45}$ = 12,000\,$M_\odot$  for $\alpha$=0.45. We thus give $\mathcal{M}_{total,young}\sim$12,000\,$M_\odot$ as a lower limit for the mass of the young star cluster. When we consider the lower mass limits of the stars, the total mass is decreased to $\mathcal{M}_{young, M\leq80 M_\odot}^{\alpha=1.7}$ = 6,000\,$M_\odot$ ($\mathcal{M}_{young, M\leq80 M_\odot}^{\alpha=0.45}$ = 10,000\,$M_\odot$). The binning uncertainty is also of the order $\sim$3,000\,$M_\odot$.

\subsection{Emission line sources}
\label{sec:emline}
There are three sources of emission lines in the Galactic centre: (a) Extended ionised gas streamers, the so-called  minispiral,  or Sgr~A East; (b) molecular gas; and (c) emission-line stars, which mostly are WR stars.

\subsubsection{Ionised gas streamers}
\label{sec:gas}

\begin{figure}
	\resizebox{\hsize}{!}{\includegraphics{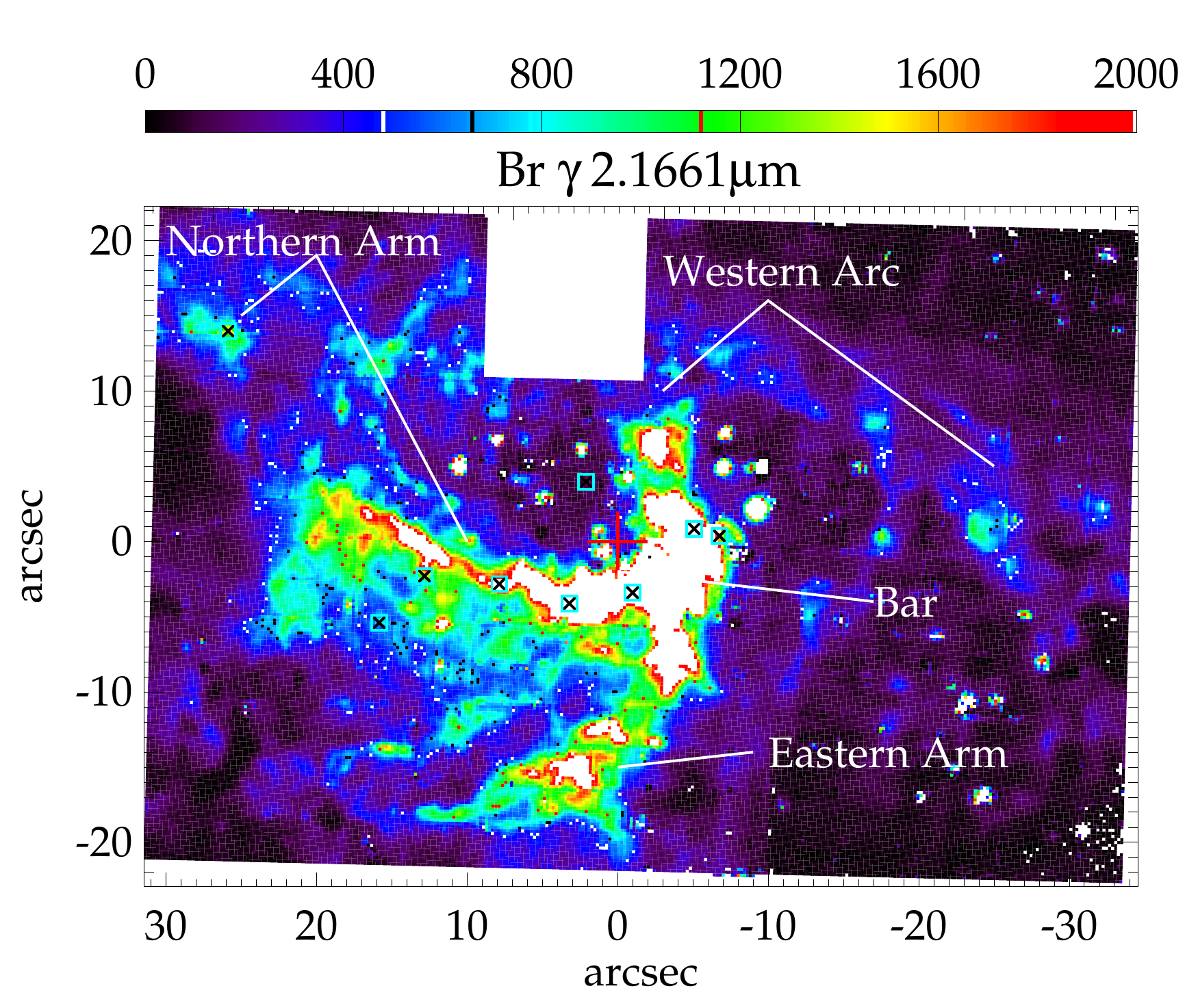}}
	\caption{Emission line map of Br~$\gamma$ gas at 2.1661\,$\mu$m of the full KMOS mosaic. 
        The axes  show the distance from Sgr~A* (red plus sign) in Galactic coordinates. Black crosses  with cyan surrounding square symbols denote the positions of the sources with featureless spectra (see Sect.~\ref{sec:fless}). The flux of Br~$\gamma$ emission is not saturated, but the scaling was set low in order to show the fainter, extended structure  of the minispiral.}	
        \label{fig:brgammaflux}
	\end{figure}

\begin{figure}
	\resizebox{\hsize}{!}{\includegraphics{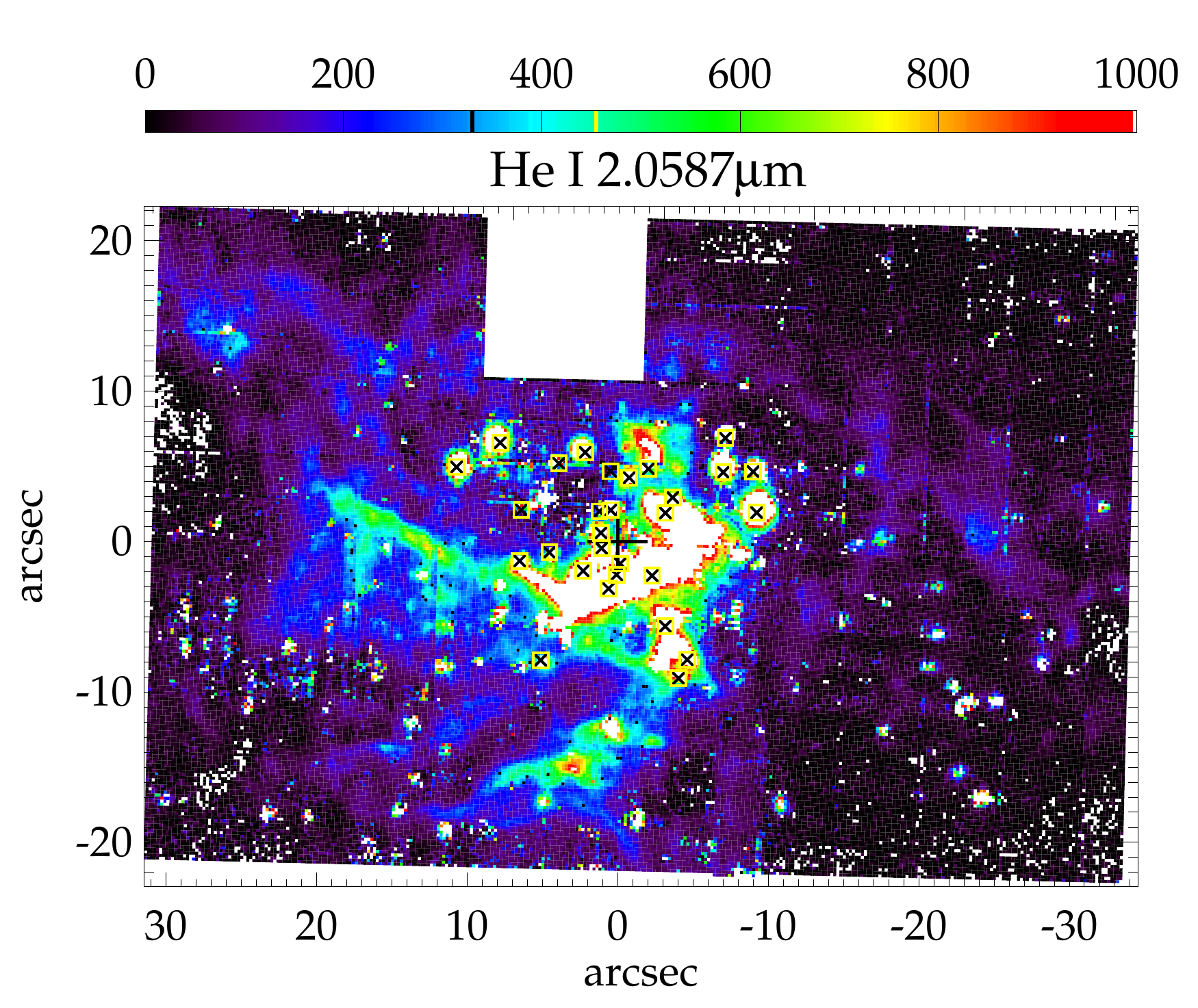}}
	\caption{Same as Fig.~\ref{fig:brgammaflux} for  He I  gas at 2.0587\,$\mu$m. The black plus sign denotes the position of Sgr~A*. The He I emission line is weaker than the Br~$\gamma$ line.  Black crosses with yellow surrounding square symbols denote the positions of the emission-line stars (see Sect.~\ref{sec:wr}).    The He~I line flux is not saturated in the data, but we set the scaling in this image low in order to show the extended structure  of the minispiral. }
	\label{fig:heflux}
	\end{figure}

The gas streamers of the minispiral can be seen  in our data in the   H~I~(4-7)~Br~$\gamma$ 2.166\,$\mu$m  and He~I 2.058\,$\mu$m ($2s$ $^1S$-$2p$ $^1P^O$)   lines. 
We fitted Gaussians to the  H~I Br~$\gamma$ and He~I 2.058\,$\mu$m emission lines using the KMOS mosaic. The resulting flux maps are shown in Fig.~\ref{fig:brgammaflux} for Br~$\gamma$ and in Fig.~\ref{fig:heflux} for He~I emission. The images are oriented in the Galactic coordinate system 
and are centred on Sgr~A*, which is shown as a red or black cross. 
We chose the applied  flux scaling in the Figs.~\ref{fig:brgammaflux} and  \ref{fig:heflux} to show the extended minispiral structure, but the flux of  the emission lines is not saturated in the data. 
The  H~I Br~$\gamma$ emission is stronger than the He~I emission, therefore the He~I map is noisier. 

The gas emission is very bright and complicates the measurement of equivalent widths of the  
He~I and H~I Br~$\gamma$ absorption features of O/B-type stars. Since the gas emission is also highly variable on small spatial scales, we refrained from modelling the gas emission. PampelMuse subtracted the surrounding background  from the spectra, but residuals remain in our data. 
Subtracting the gas emission close to the star can be  complicated even for high-angular resolution data \citep[see][]{paumard06}. 
However, as the gas emission lines are very narrow compared to emission lines from Wolf-Rayet stars and because most emission-line stars have additional C or N lines, we can distinguish between the different emission line sources. 

	
\subsubsection{	Molecular gas }
The molecular gas in the Galactic centre is concentrated in a circumnuclear ring (CNR). This clumpy gas ring  extends over a projected distance of $\sim$1.6 to 7\,pc \citep[41\arcsec$-$3\arcmin, e.g.][]{yusefh2,Lee08,smithcnd14}   and rotates with $\sim$\:\!110\,km/s \citep{christopherhcn,isaacanja}.  The gas ring consists of two prominent symmetric lobes north-east and south-west of Sgr~A*.

\begin{figure}
	\resizebox{\hsize}{!}{\includegraphics{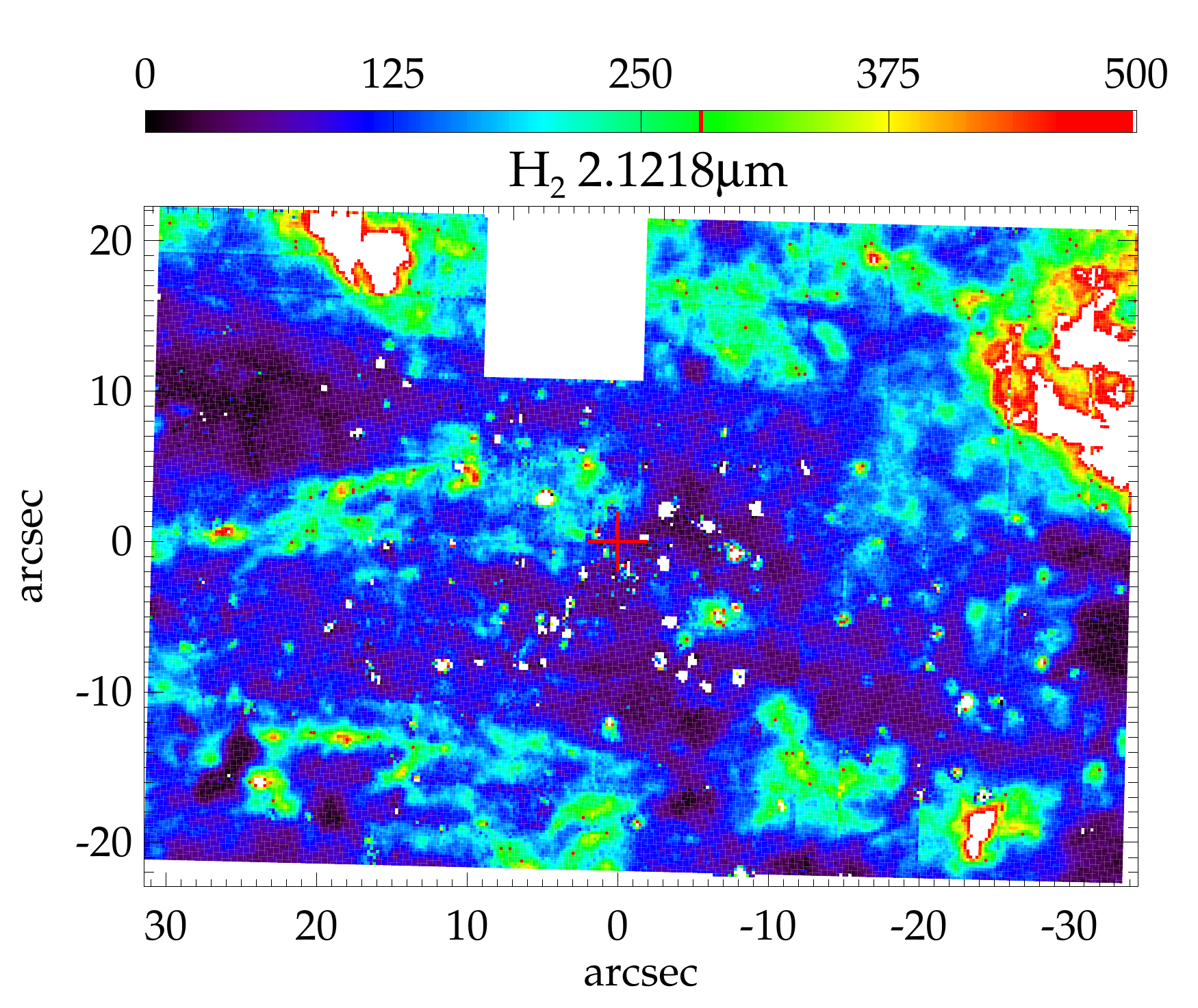}}
	\caption{Same as Fig.~\ref{fig:brgammaflux} for  H$_2$  gas at 2.1218\,$\mu$m. The red plus sign denotes the position of Sgr~A*.  The H$_2$ line flux is not saturated in the data, but we set the scaling in this image low in order to show the extended structure of the gas. }
	\label{fig:h2flux}
	\end{figure}

Our data set maps only the inner edge of the circumnuclear ring. We fitted Gaussians to the  H$_2$ emission line  at 2.1218\,$\mu$m (1-0 S(1)) using the KMOS mosaic. Figure~\ref{fig:h2flux} shows the H$_2$ flux map in the Galactic coordinate system. There are several gas streamers and clumpy structures within the projected distance of the circumnuclear ring.

\subsubsection{Emission-line stars: Spectral classification}
 \label{sec:wr}
	\begin{figure*}
	\centering
	\includegraphics[width=17cm]{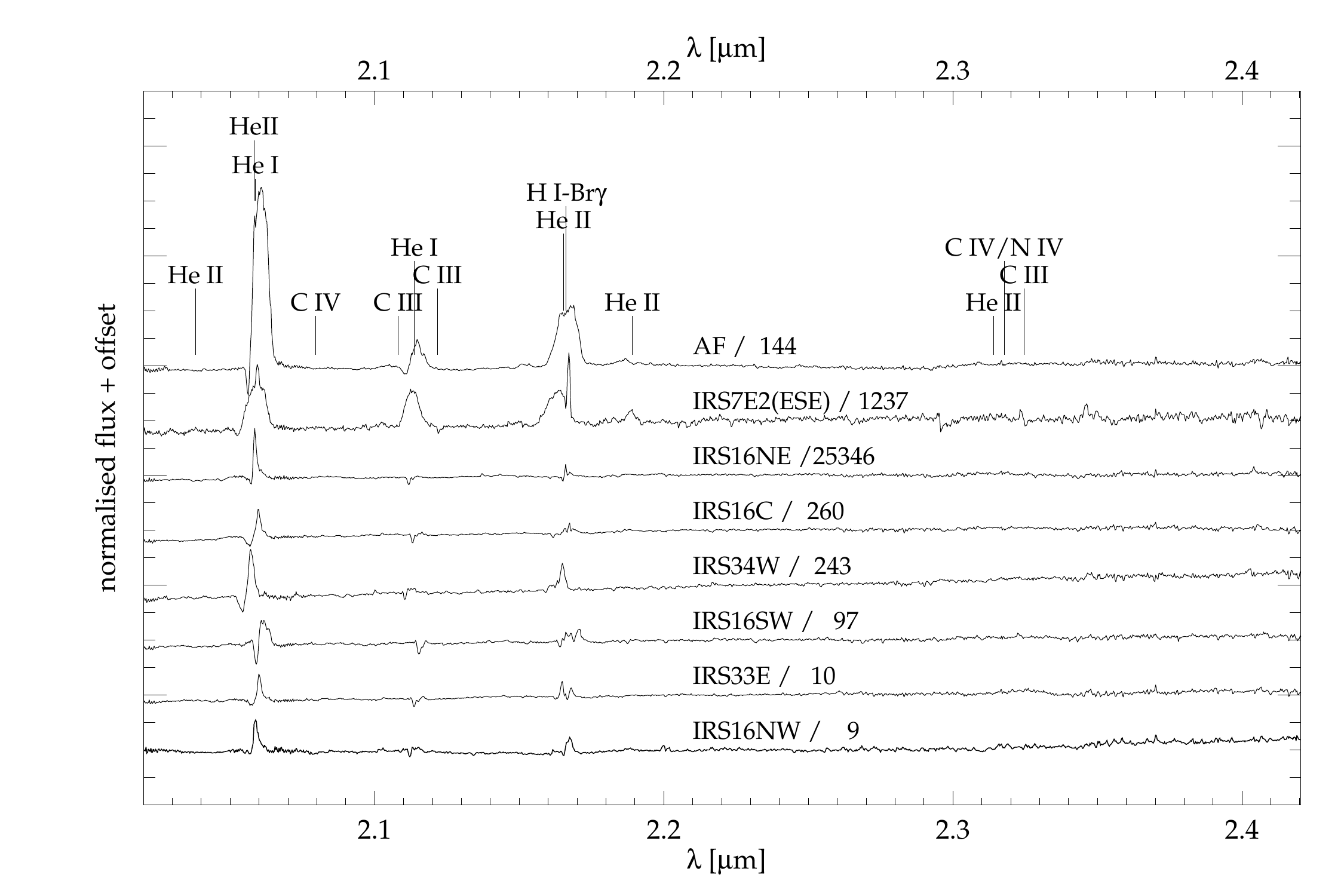}
	\caption{Spectra of stars with  narrow emission lines
and Wolf-Rayet stars. The lower six spectra are Ofpe/WN9 stars with  narrow emission lines (FWHM$\sim$200~km~s$^{-1}$).   
        Fluxes are normalised and an offset is added to the flux. The spectra are not shifted to rest wavelength. The narrow emission  line  in the  2.167~$\mu$m emission line of spectrum Id~1237 is a residual of poorly subtracted minispiral emission. 
	}
	\label{fig:ofpe}
	\end{figure*}
	  
	\begin{figure*}
	\centering
	\includegraphics[width=17cm]{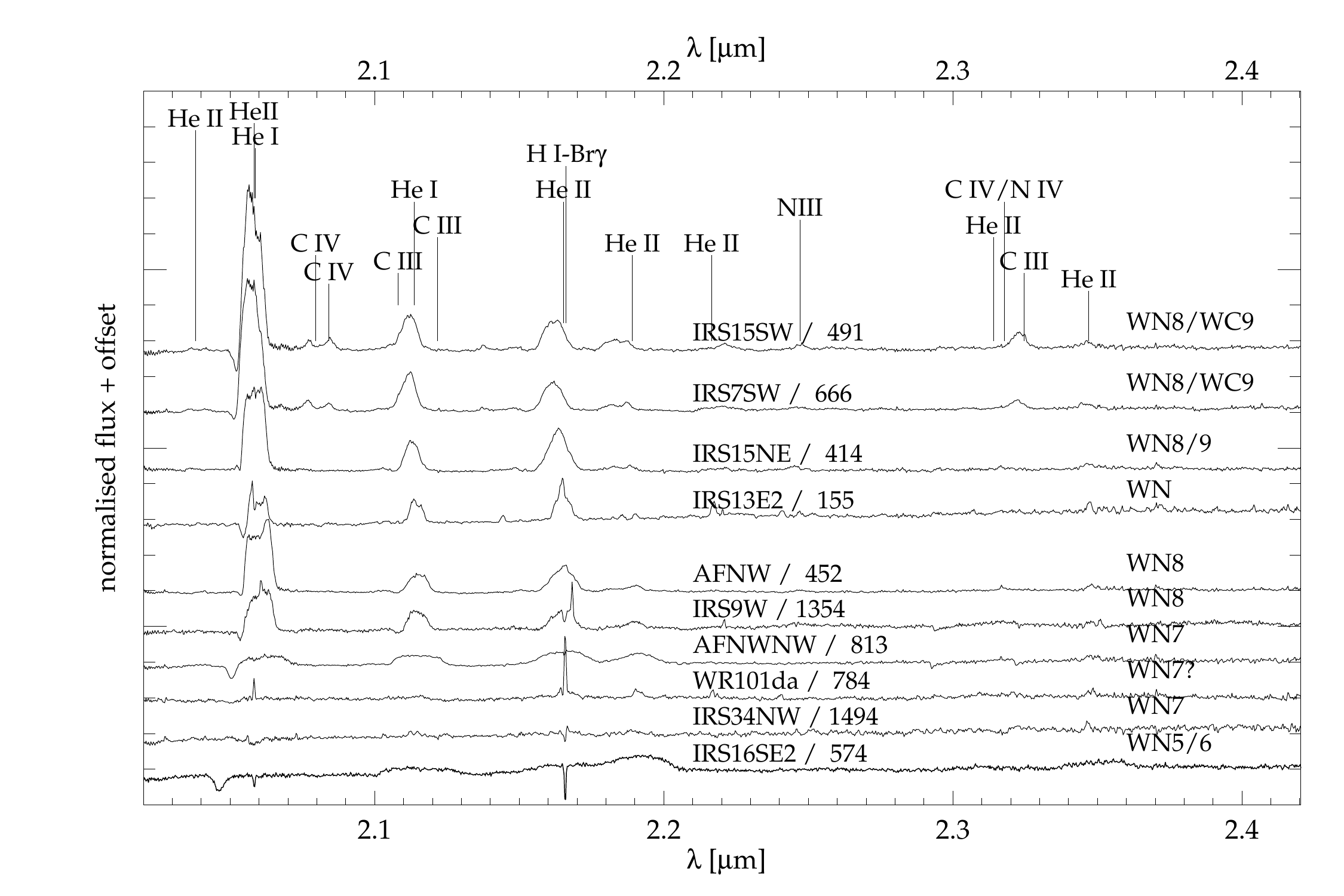} 
	\caption{Spectra of Wolf-Rayet stars of type WN and WN/WC, ordered by increasing WN-type from bottom to top.  	The fluxes are normalised and an offset is added to the flux. The spectra are not shifted to rest wavelength. }
	\label{fig:wn}
	\end{figure*}

	\begin{figure*}
	\centering
	\includegraphics[width=17cm]{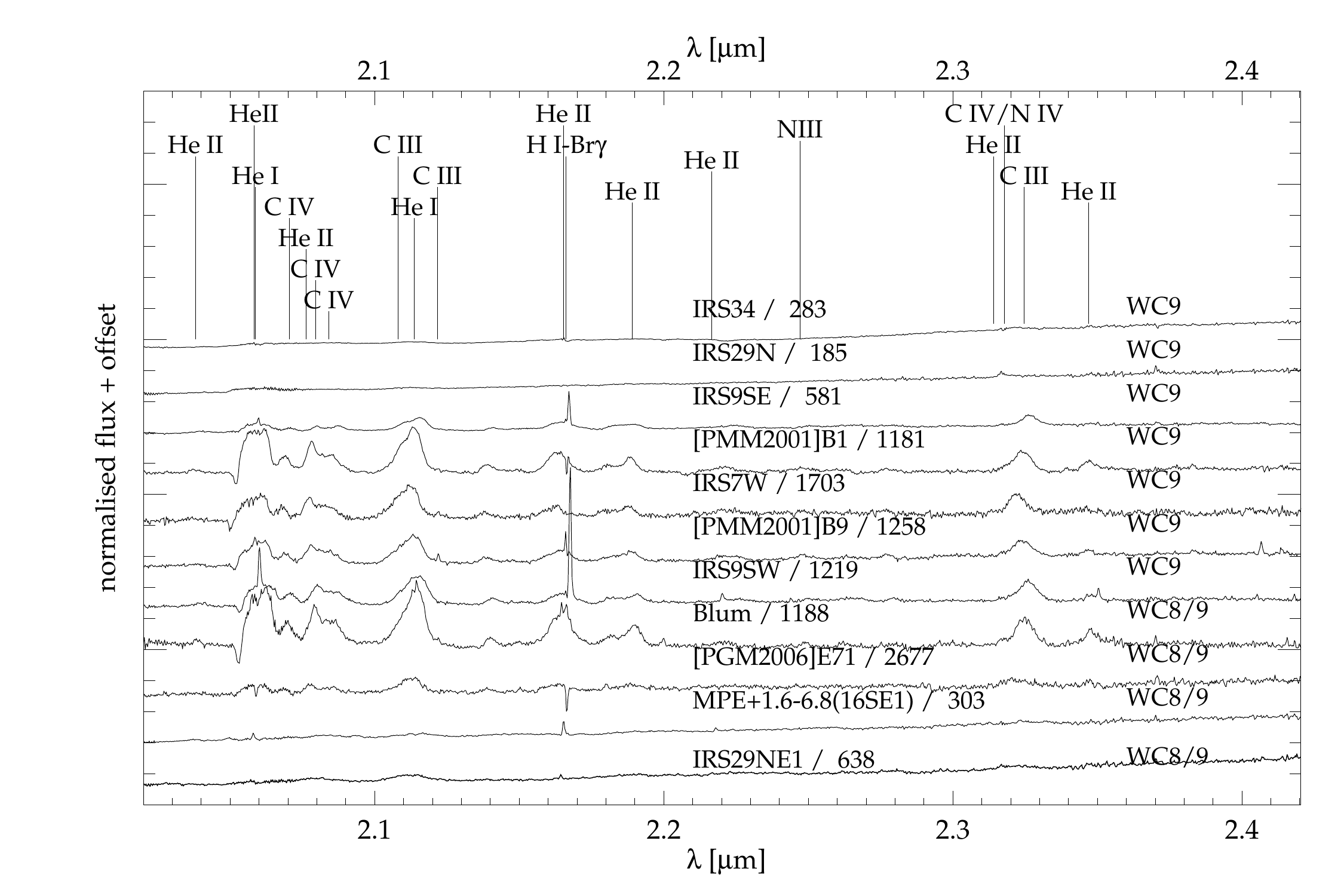} 
	\caption{Spectra of Wolf-Rayet stars of type WC. The classification from \citet[PGM2006]{paumard06} is written for  all spectra.  The narrow emission line at 2.167~$\mu$m of  spectrum Id 784 is a residual of the subtracted minispiral emission. 
	The fluxes are normalised and an offset is added to the flux. The spectra are not shifted to rest wavelength. }
	\label{fig:wc}
	\end{figure*}

Stars with a He~I 2.058~$\mu$m emission line can belong to many different types such as WR stars, intermediate types such as Ofpe/WN9 (O-type spectra with additional  H, He, and N emission lines, and other peculiarities), and luminous blue variable (LBV) stars. \cite{paumard01}  suggested two classes of He~I 2.058~$\mu$m emission-line stars in the Galactic centre: Stars with narrow emission lines (FWHM$\sim$200~km s$^{-1}$) and stars with very broad emission lines (FWHM$\sim$1000~km~s$^{-1}$). \cite{paumard01} roughly sorted  narrow-line stars into LBV-type stars, with temperatures of 10,000$-$20,000~K, and broad-line stars to WR-type stars, with higher temperatures of \textgreater30,000~K.  In broad-line star spectra the lines have a higher peak value above the continuum than in narrow-line star spectra. 

Wolf-Rayet  stars are evolved, massive stars (\textgreater 20~$M_\odot$ while on the main sequence, \citealt{sanderwc12}). 
 Their spectra show strong  emission lines because these stars are losing mass. \cite{figerwr97} provided a list of WR emission lines in the $K$ band; among them the He~I, He~II, H~I, N~III, C~III, and C~IV transitions. WR stars can be sorted into WN and in WC types. WN-type  spectra are dominated by nitrogen lines
and WC-type  spectra are dominated by carbon and oxygen.

We have  29 spectra with a He~I 2.058~$\mu$m emission line/WR stars. These stars are already known, for instance from \cite{krabbe95}, \cite{blum03} and \cite{paumard06}.
We list these  stars  in Table~\ref{tab:wrlist}, and their spatial distribution is shown in Fig.~\ref{fig:cube} with  yellow symbols. The spectra  are shown in Figs.~\ref{fig:ofpe}, \ref{fig:wn}, and \ref{fig:wc}. In some spectra the residual from the minispiral gas emission after the subtraction is still visible, for example in Id.~1237/IRS~7E2~(ESE) at $\sim$2.167~$\mu$m. The brightest WR stars are also visible in the emission line maps in Figs. \ref{fig:brgammaflux} and \ref{fig:heflux} as bright point sources.~  
As a result of their large FWHM,  the emission lines  are blends of several lines. Therefore radial velocity measurements of WR stars are highly uncertain with our data.  \cite{tanner06} obtained high-resolution spectra and measured the radial velocities of emission lines stars in the Galactic centre.

\cite{paumard06} listed eight stars in their Table~2 as Ofpe/WN9 stars because they showed   narrow emission lines and a He~I complex at 2.113~$\mu$m.  
The KMOS spectra of  these stars  are shown in Fig.~\ref{fig:ofpe}. All spectra have   P Cygni profiles at 2.058~$\mu$m, at the He~I line. This indicates that  these stars are a source of strong stellar winds ($\sim$200\,km s$^{-1}$). However, two of the stars (Id 144/AF and Id 1237/IRS~7E2~(ESE))  look different in our data from the other Ofpe/WN9 stars. They have significantly broader lines, with FWHM$\sim$700\,km s$^{-1}$ instead of $\sim$200~km s$^{-1}$. The 2.113~$\mu$m feature is mostly in emission and not in absorption, in contrast to the other six as Ofpe/WN9 identified stars. Furthermore, a feature at He~II 2.1891~$\mu$m appears in emission.   \cite{figerwr97}  showed that the ratio between the  2.1891~$\mu$m feature and the 2.11~$\mu$m feature is strongly correlated with subtypes for WN stars and increases with earlier subtype. This 2.1891~$\mu$m feature is also present in  the other WN stars of our data (see Fig.~\ref{fig:wn}). Therefore we conclude that the stars Id 144/AF and Id 1237/IRS~7E2~(ESE) are not Ofpe/WN9 stars but are hotter stars, such as WN8 or WN9.  \cite{tanner06} also classified star AF (Id 1206) as a broad emission-line star.

The spectra of stars classified as WN stars in \cite{paumard06} are shown in Fig.~\ref{fig:wn}. 
WN stars can be separated into an early (WN2 to WN6) and a late
group (WN6 to WN9). 
The only early WN star in our data, Id 574/IRS~16SE2, is a WN5/6 star \citep{horrobin04}. The spectrum of Id 155/IRS~13E2 is classified as that of an WN star by \cite{paumard06} without further specification. We find that this spectrum resembles the late WN8 spectra of Id 452/AFNW and Id 1354/IRS~9W. Stars Id 784/WR101da and Id 1494/IRS34~NW were classified as WN7 stars. Their spectra have only weak emission lines, for example at 2.189~$\mu$m (He II) and 2.347~$\mu$m (He II). 

Star Id 491/IRS~15SW was classified as a transition-type  WN8/WC9 star by \cite{paumard06}. In addition to the aforementioned He~I and He~II emission lines, the spectrum shows the C~IV doublet at 2.0796 and 2.0842~$\mu$m, and C~III at $\sim$2.325~$\mu$m in emission.  These features are much weaker than the He and H lines. 
The spectrum of Id 666/IRS~7SW has the same C~IV and C~III lines, although it was classified as WN8  by \cite{paumard06}. Therefore we suggest that Id 666/IRS~7SW  is a WN/WC transition-type star like Id  491/IRS~15SW.

WC stars  have  C~III and C~IV emission lines that are  about
as strong as 
the He lines. Figure~\ref{fig:wc} shows the spectra of WC stars in our data set. The classifications are adopted from  \cite{paumard06}. We find that for stars Id 185/IRS~29N,  Id 283/IRS~34, Id 303, and Id 638/IRS~29NE1 the emission lines are rather weak. This cannot be caused by  the S/N, which is higher than 55 for all of the four spectra. The continua of these four spectra show a  steep rise with  wavelength, and these stars are also very red ($(H-K_S)_0$ \textgreater 0.54\,mag).  This suggests that these stars are embedded in dust \citep{geballe06}. The continuum emission from the surrounding dust dilutes the stellar spectral lines (for a discussion see Appendix \ref{sec:discem}). 

In summary, we confirm  that 29 stars  are emission-line stars. We classify the stars Id 144/AF and Id 1237/IRS~7E2 as broad
emission-line stars and the star Id 666/IRS~7SW as a WN8/WC9 star, in contrast to \cite{paumard06}. Four of the stars (Id 185/IRS~29N,  Id 283/IRS~34, Id 303, and Id 638/IRS~29NE1) have only weak emission lines, which can be explained by  bright surrounding  dust.  Despite their red colours, we do not consider them  to
be background stars. We discuss these findings in Appenix~\ref{sec:discem}.
\begin{table*}
\caption{Emission-line and Wolf-Rayet stars}
 \label{tab:wrlist}
 \centering
\begin{tabular}{rccllllr}
\noalign{\smallskip}
\hline\hline
\noalign{\smallskip}

Id&RA&Dec&colour&name&type&PGM2006\tablefootmark{d}&S/N\\
&$ \left[ ^{\circ} \right]$&$ \left[ ^{\circ} \right]$\\
\noalign{\smallskip}
\hline
\noalign{\smallskip}
$     9$&$  266.41684 $&$   -29.007483 $&$...$&$IRS16NW$&$Ofpe/WN9$\tablefootmark{b}&$E19$&$  95.8$\\
$    10$&$  266.41705 $&$   -29.008696 $&$...$&$IRS33E$&$Ofpe/WN9$\tablefootmark{b}&$E41$&$  91.1$\\
$    97$&$  266.41718 $&$   -29.008080 $&$...$&$IRS16SW$&$Ofpe/WN9$\tablefootmark{b}&$E23$&$  81.7$\\
$   243$&$  266.41553 $&$   -29.007401 $&$ red$&$IRS34W$&$Ofpe/WN9$\tablefootmark{b}&$E56$&$  71.1$\\
$   260$&$  266.41711 $&$   -29.007637 $&$...$&$IRS16C$&$Ofpe/WN9$\tablefootmark{b}&$E20$&$ 102.2$\\
$ 25346$&$  266.41772 $&$   -29.007565 $&$...$&$IRS16NE$&$Ofpe/WN9$\tablefootmark{b}&$E39$&$  94.5$\\
\hline
$   144$&$  266.41476 $&$   -29.009741 $&$...$&$AF$&$WN$\tablefootmark{a}&$E79$&$  78.2$\\
$   155$&$  266.41577 $&$   -29.008307 $&$...$&$IRS13E2$&$WN$\tablefootmark{a}&$E51$&$  45.8$\\
$   414$&$  266.41724 $&$   -29.004581 $&$...$&$IRS15NE$&$WN8/9$\tablefootmark{b}&$E88$&$  54.8$\\
$   452$&$  266.41440 $&$   -29.008829 $&$...$&$AFNW$&$WN8$\tablefootmark{b}&$E74$&$  68.8$\\
$   491$&$  266.41632 $&$   -29.005037 $&$...$&$IRS15SW$&$WN8/WC9$\tablefootmark{b}&$E83$&$  50.3$\\
$   574$&$  266.41776 $&$   -29.008135 $&$...$&$IRS16SE2$&$WN5/6$\tablefootmark{b}&$E40$&$  33.8$\\
$   666$&$  266.41556 $&$   -29.006466 $&$...$&$IRS7SW$&$WN8/WC9$\tablefootmark{a}&$E66$&$  59.9$\\
$   784$&$  266.41541 $&$   -29.008274 $&$...$&$WR101da$&$WN7?$\tablefootmark{b}&$E60$&$  39.3$\\
$   813$&$  266.41376 $&$   -29.008535 $&$...$&$AFNWNW$&$WN7$\tablefootmark{b}&$E81$&$  57.3$\\
$  1237$&$  266.41824 $&$   -29.006472 $&$...$&$IRS7E2(ESE)$&$WN$\tablefootmark{a}&$E70$&$  35.6$\\
$  1354$&$  266.41782 $&$   -29.009386 $&$...$&$IRS9W$&$WN8$\tablefootmark{b}&$E65$&$  36.4$\\
$  1494$&$  266.41562 $&$   -29.007044 $&$...$&$IRS34NW$&$WN7$\tablefootmark{b}&$E61$&$  33.8$\\
\hline
$   185$&$  266.41632 $&$   -29.007420 $&$ red$&$IRS29N$&$WC9$\tablefootmark{b}&$E31$&$  82.0$\\
$   283$&$  266.41516 $&$   -29.007618 $&$ red$&$IRS34$&$WC9$\tablefootmark{c}&$...$&$ 167.1$\\
$   303$&$  266.41742 $&$   -29.008127 $&$ red$&$MPE$+$1.6$-$6.8(16SE1)$&$WC8/9$\tablefootmark{b}&$E32$&$  68.6$\\
$   581$&$  266.41861 $&$   -29.010094 $&$...$&$IRS9SE$&$WC9$\tablefootmark{b}&$E80$&$  91.1$\\
$   638$&$  266.41647 $&$   -29.007254 $&$ red$&$IRS29NE1$&$WC8/9$\tablefootmark{b}&$E35$&$  55.1$\\
$  1181$&$  266.41980 $&$   -29.007748 $&$...$&$[PMM2001]B1$&$WC9$\tablefootmark{b}&$E78$&$  38.6$\\
$  1188$&$  266.41406 $&$   -29.009296 $&$...$&$Blum$&$WC8/9$\tablefootmark{b}&$E82$&$  21.1$\\
$  1219$&$  266.41818 $&$   -29.010050 $&$...$&$IRS9SW$&$WC9$\tablefootmark{b}&$E76$&$  41.9$\\
$  1258$&$  266.41776 $&$   -29.006857 $&$...$&$[PMM2001]B9$&$WC9$\tablefootmark{b}&$E59$&$  39.7$\\
$  1703$&$  266.41605 $&$   -29.006159 $&$...$&$IRS7W$&$WC9$\tablefootmark{b}&$E68$&$  20.4$\\
$  2677$&$  266.41730 $&$   -29.006008 $&$...$&$...$&$WC8/9$\tablefootmark{b}&$E71$&$  26.5$\\
\noalign{\smallskip}
\hline
\noalign{\smallskip}
\end{tabular}
\tablefoot{
\tablefoottext{a}{Spectral classification from this work}
\tablefoottext{b}{Spectral type from \cite{paumard06}}
\tablefoottext{c}{Spectral type from \cite{blum03}}
\tablefoottext{d}{PGM2006 refers to the nomenclature of  \cite{paumard06}}}
\end{table*}

\subsection{Featureless spectra}
\label{sec:fless}

\begin{figure*}
\centering
\includegraphics[width=17cm]{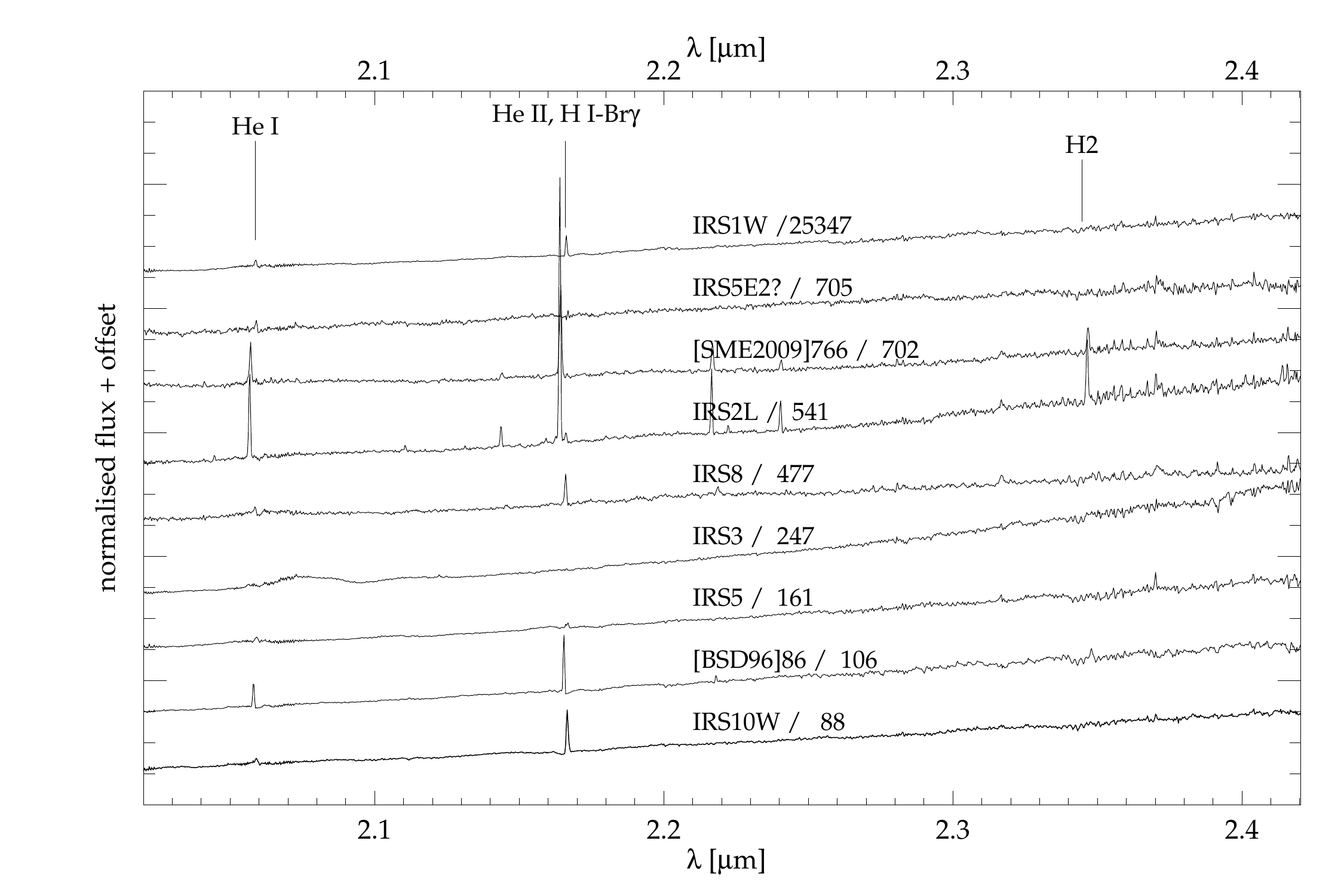}
\caption{Featureless spectra. Emission lines are due to imperfect background subtraction and caused by He~I, H~I, and H$_2$ gas emission. The fluxes are normalised and an offset is added to the flux. The spectra are not shifted to rest wavelength. }
\label{fig:fless}
\end{figure*}  

\begin{table*}
\caption{Stars with featureless spectra}
 \label{tab:fless}
 \centering
\begin{tabular}{rccrlllrl}
\noalign{\smallskip}
\hline\hline
\noalign{\smallskip}
Id&RA&Dec&K$_S$\tablefootmark{a}&colour&name&type&S/N&location\\
&$ \left[ ^{\circ} \right]$&$ \left[ ^{\circ} \right]$&$ \left[ \text{mag} \right]$\\
\noalign{\smallskip}
\hline
\noalign{\smallskip}

$    88$&$  266.41888 $&$   -29.006372 $&$          11.170 $&$...$&$IRS10W$&$...$&$ 110.4$&$NA$\\
$   106$&$  266.41757 $&$   -29.008551 $&$          11.680 $&$ red$&$[BSD96]86$&$...$&$ 115.7$&$Bar$\\
$   161$&$  266.41956 $&$   -29.005119 $&$          11.250 $&$ ?$&$IRS5$&$WR$\tablefootmark{f}&$ 102.1$&$NA$\\
$   247$&$  266.41608 $&$   -29.006762 $&$          11.220 $&$ red$&$IRS3$&$WC5/6$\tablefootmark{c}/$AGB$\tablefootmark{d}&$  93.6$&$-$\\
$   477$&$  266.41730 $&$   -28.999689 $&$          11.956 $&$ red$&$IRS8$&$O5$-$6$\tablefootmark{e}&$  58.4$&$NA-edge$\\
$   541$&$  266.41574 $&$   -29.008911 $&$          11.730 $&$ red$&$IRS2L$&$...$&$  50.5$&$Bar$\\
$   702$&$  266.41559 $&$   -29.009377 $&$          12.403 $&$...$&$[SME2009]766$&$...$&$  51.2$&$Bar$\\
$   705$&$  266.42090 $&$   -29.004850 $&$          13.510 $&$...$&$IRS5NE?$&$G8\,$III\tablefootmark{g}&$  47.9$&$NA$\\
$ 25347$&$  266.41846 $&$   -29.007660 $&$    ...    $&$ ?$&$IRS1W$&$Be?$\tablefootmark{b}/WR\tablefootmark{f}&$ 124.5$&$NA$\\

\noalign{\smallskip}
\hline
\noalign{\smallskip}
\end{tabular}
\tablefoot{
\tablefoottext{a}{$K_S$ magnitudes from \cite{rainer10}, if available} 
\tablefoottext{b}{Spectral type from \cite{paumard06}}
\tablefoottext{c}{Spectral type from \cite{horrobin04}}
\tablefoottext{d}{Spectral type from \cite{pott05}}
\tablefoottext{e}{Spectral type from \cite{geballe06}}
\tablefoottext{f}{Spectral type from \cite{tanner05}, \cite{joel14}}
\tablefoottext{g}{Spectral type from \cite{perger08}}
}
\end{table*}

Previous studies pointed out that several  sources    apparently
have featureless, steep   $K$-band spectra in the Galactic centre. For example, the spectra of   IRS~3 and IRS~1W show no  detectable emission or absorption features \citep[e.g.][]{krabbe95,blum03}. 
 These sources are often extended in mid-infrared images, and it was shown that they are bow shocks. Bow shocks are  caused by bright emission-line stars that either have strong winds or move through the minispiral \citep[e.g.][]{tanner05,tanner06,geballe06,viehmann06,perger08,popyoungphot,joel14}. 
 
We detected several featureless sources in our KMOS data. They are listed in Table~\ref{tab:fless}, and their spectra are shown in Fig.~\ref{fig:fless}. 
The first column of Table~\ref{tab:fless} denotes the Id, R.A., and Dec  from our catalogue.
Most of the sources with featureless spectra are located close to the minispiral.  The last column of Table~\ref{tab:fless} gives their location within the minispiral. We also indicate their positions in Fig.~\ref{fig:brgammaflux}. Many of the stars are either connected with the Northern Arm (NA) or the Bar. 
Only star Id 247/IRS~3  is in a region of low ionised gas emission. Nevertheless, it is the most reddened of these sources. 

Bow shocks arise through the interaction of the interstellar medium (like the minispiral gas) with  the material expelled from mass-losing stars. The central sources of Id 161/IRS~5 and Id 25347/IRS~1W are probably WR stars \citep{tanner05,joel14}. The source Id 247/IRS~3 was classified as WC5/6 \citep{horrobin04} and as an AGB star \citep{pott05}.  The spectrum of Id 247/IRS~3 shows a broad emission bump at 2.078~$\mu$m, but this  could be caused by the close WN5/6 star IRS~3E. This star is rather faint ($K_S$ = 14.1\,mag), however, compared to star Id 247 ($K_S$ = 11.2\,mag), and therefore the spectrum has a too low S/N and is missing from our list of WR stars.

 The spectrum of Id 477/IRS~8 in our data is  nearly featureless, but \cite{geballe06} was able to separate the spectrum of IRS~8 into the contribution of the bow shock and the actual star, IRS~8*. They showed that star IRS~8*  has several weak emission and absorption lines and classified IRS~8* as O star. 
One star in our sample of  featureless sources (Id 705) was classified as  a late-type star by \cite{perger08}.

These featureless sources are also very red in $H-K_S$, as we
show in Fig.~\ref{fig:cmd}.
We find that of all of the sources, the spectra of Id  247/IRS~3 and Id 541/IRS~2L have the steepest continuum rise to longer wavelengths (slope m=$\Delta flux$/$\Delta \lambda$=4.2 and 3.3, respectively). These sources are probably not background stars, but surrounding dust causes the reddening. 

In brief,  many stars with featureless spectra were either classified as young emission-line or O-type star, or their red colour and continuum shape suggest that they are young, embedded stars.  Therefore we also consider these stars as young early-type stars of the Milky Way nuclear star cluster. 

\subsection{Spatial distribution of  early-type stars}
\label{sec:spatial}
Our wide-field study of early-type stars confirms the results of  previous studies in smaller regions \citep[e.g.][]{stostad}.  Young stars are mostly concentrated at $p<$0.5\,pc (see Fig.~\ref{fig:cube}).
Previous spectroscopic data sets were spatially asymmetric   with respect to Sgr~A* and therefore were  potentially biased. For example, \cite{do13} observed the Galactic centre along the projected disk of young stars. Our data set is completely symmetric with respect to Sgr~A* out to $p$=12\arcsec\,($\sim$0.48\,pc). In the radial range to $p$=21\arcsec\,(0.84\,pc) we only miss a small field of 10\farcs8$\times$10\farcs8, therefore the area is complete to 91\% out to $p$=21\arcsec\,(0.84\,pc). The spatially nearly full coverage  allows us to study the spatial distribution of early-type stars. 

\begin{figure}
        \resizebox{\hsize}{!}{\includegraphics{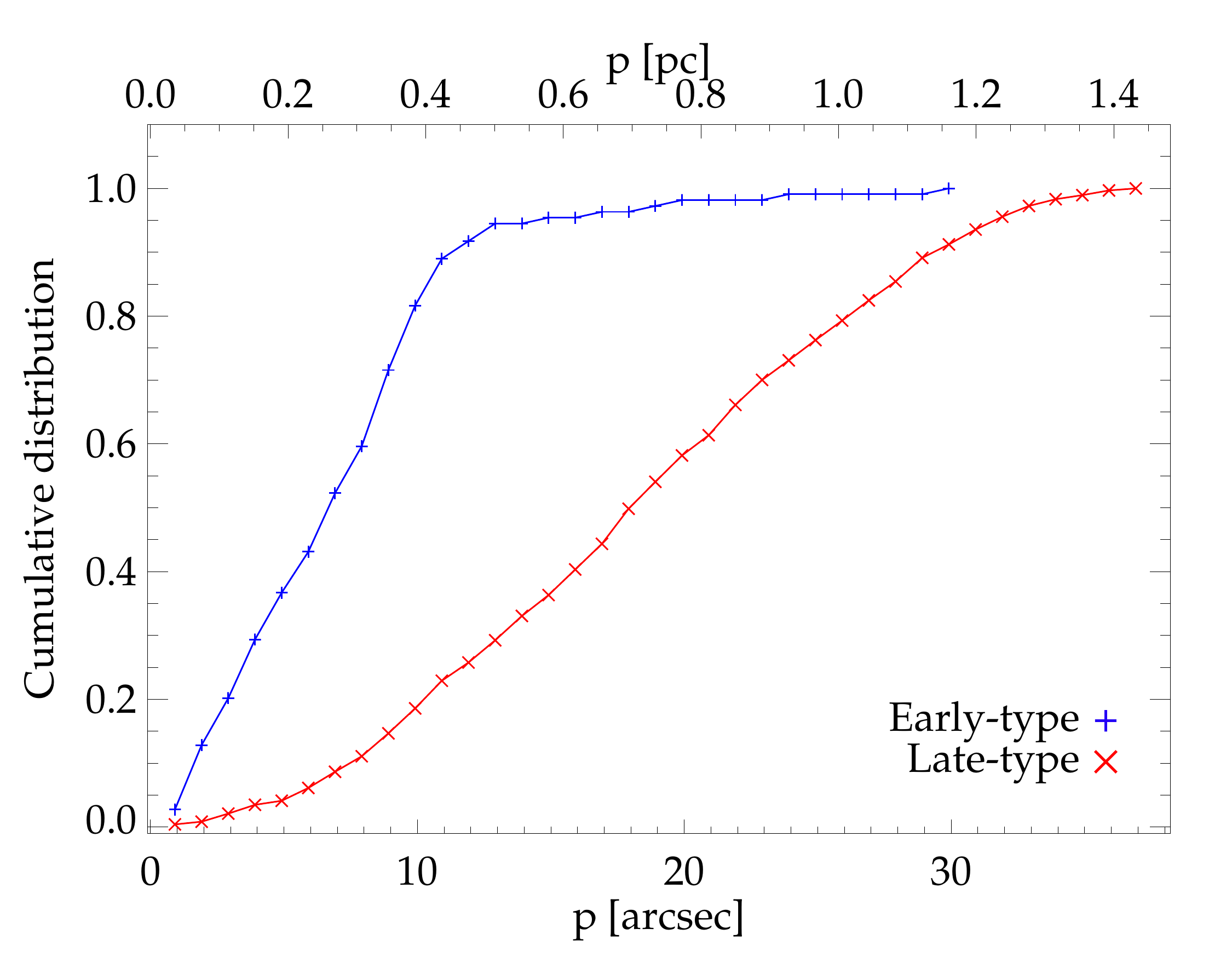}}
        \caption{Cumulative number counts of early-type stars (blue   plus signs) and late-type stars (red crosses) as a function of projected distance $p$ from Sgr~A*, normalised to one. Foreground stars were excluded. }
        \label{fig:cumfrequ}
        \end{figure}

Figure \ref{fig:cumfrequ} shows the cumulative number counts of our observed early-type stars and late-type stars normalised to one, as a function of projected distance $p$ to Sgr~A*. Most of the early-type stars lie within the central parsec and reach a cumulative frequency of 0.9 at $p$=12\arcsec\,(0.47\,pc), whereas the late-type stars are distributed throughout the entire cluster range.   For this plot we did not correct for completeness. Including a completeness correction would steepen the  lines in the innermost regions even more. The median projected  distance to the centre  is only 6.6\arcsec\,($\sim$0.26\,pc) for the early-type stars, but 19\arcsec\,(0.74\,pc) for the late-type stars. 
 We list the projected distance $p$ to Sgr~A* for the O/B stars in Table~\ref{tab:etkin}. The outermost O/B star  that is not a foreground star is Id 982 with $p$=23.6\arcsec\,(0.92\,pc). Only the featureless source Id 477/IRS~8 has a larger distance $p$=29.4\arcsec among the early-type stars.
  
\begin{figure}
	\resizebox{\hsize}{!}{\includegraphics{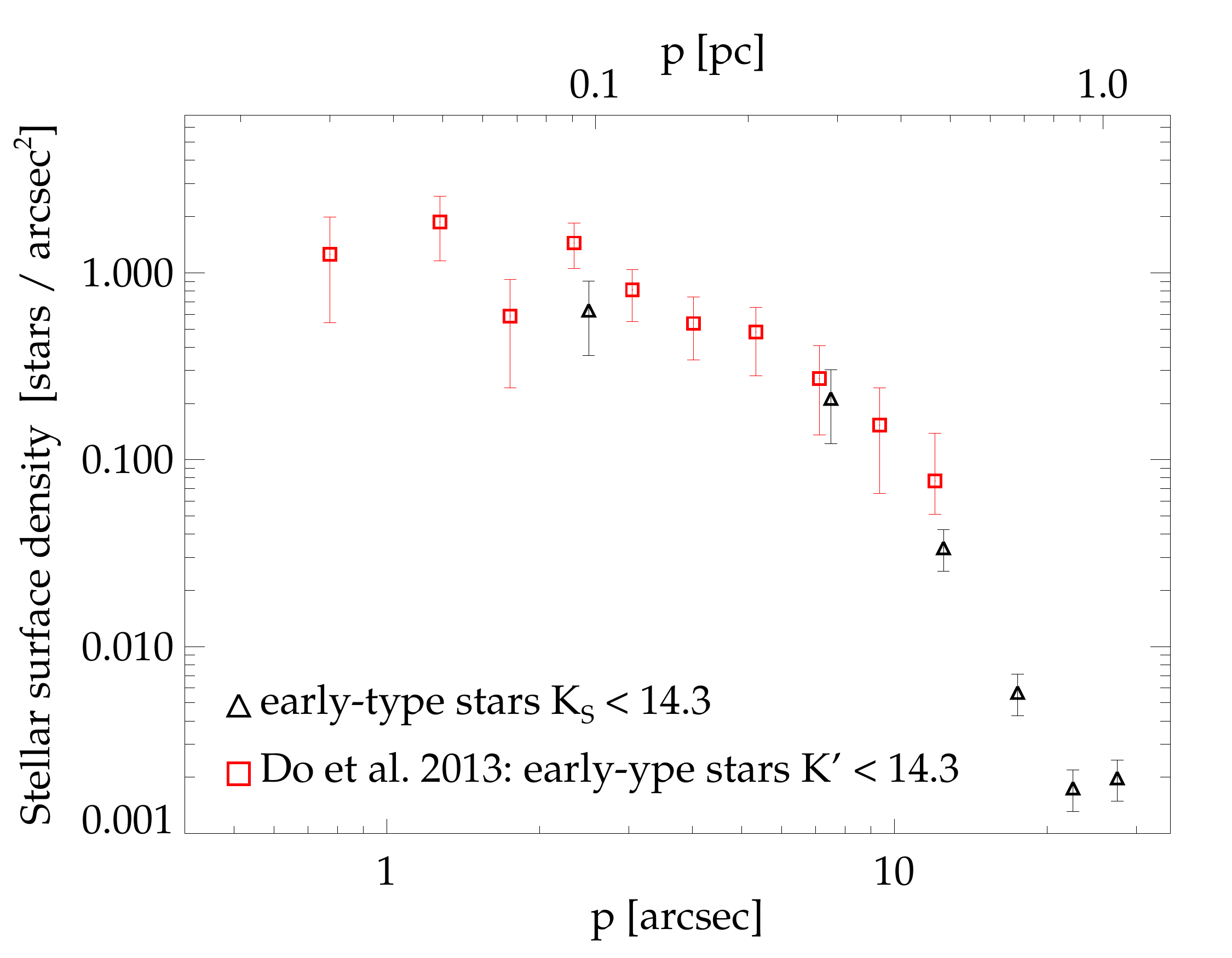}}
	\caption{Stellar surface density profile for all early-type stars (O/B, emission line stars, and sources with featureless spectra). We exclude possible foreground stars and apply a completeness correction (see Sect. \ref{sec:photo}).  We consider only stars brighter than $K_S$=14.3. Black triangles denote this study, red squares the results of \cite{do13}.}	
	\label{fig:sdens}
	\end{figure}

While we benefit from the large spatial coverage, our data set lacks  the spatial resolution and the higher completeness of other studies \citep[e.g.][]{bartko10,do13}.   In Sect.~\ref{sec:comp} we calculated the fraction of stars that we missed in different radial and magnitude bins. We used three radial bins ($p$\textless 5\arcsec, 5\arcsec$\leq$$p$\textless10\arcsec, and $p\geq$10\arcsec) and magnitude bins with a width of $\Delta K_S$=0.5\,mag. We corrected our number counts of early-type stars in the different magnitude and radial bins by including the fraction of missed stars. Then we computed a   completeness-corrected stellar number density of bright stars with $K_S$\textless 14.3.  We find excellent agreement with the results of \citet[$K'$\textless 14.3]{do13}, as shown in Fig. \ref{fig:sdens}. Our data set extends to larger radii beyond 10\arcsec. There are only a few stars in this region, and the number density of bright early-type stars decreases by more than two orders of magnitude from the centre to a projected distance of $p$=1\,pc.

Inspection of Fig. \ref{fig:cube}  shows that the distribution of early-type stars  (i.e. O/B stars, emission-line stars, and sources with featureless spectra)  appears  elongated, primarily along the Galactic plane.  
 However, there is a slight misalignment of the  distribution of early-type stars with respect to the Galactic plane. Most early-type stars beyond 0.5\,pc ($\sim$12.8\arcsec) are either in the Galactic north-west (NW, top right), or  south-east (SE, bottom left) quadrant. 
  We note that on larger scales the rotation axis also seems offset from the Galactic plane in a similar direction \citep{isaacanja}.  
The early-type stars are  more centrally concentrated in the north-east (NE) and south-west (SW) fields  than in the SE and NW fields. The median projected distances $\hat{p}$ are $\hat{p}_{NE}$=0.19\,pc (5.0\arcsec) and $\hat{p}_{SW}$=0.23\,pc  (5.8\arcsec) in the NE and SW field, but $\hat{p}_{SE}$=0.26\,pc (6.6\arcsec) and $\hat{p}_{NW}$=0.30\,pc (7.8\arcsec) in the SE and NW fields.

To quantify a possible asymmetric distribution,  we compared the number of early-type stars    in the different quadrants Galactic  NE, SE, SW, and NW. The centre is the position of Sgr~A*. We corrected for the slightly asymmetrically covered area and compare the number of stars $N_{field}$ in different fields.  Probable foreground stars were not taken into account.
We find that there are about the same number of early-type stars in the NE, SE and SW fields, but  $\approx$1.4 times more early-type stars in the NW quadrant (corresponding to more than ten stars). This is in contrast to the distribution of late-type stars, for which there are the fewest stars in the  Galactic NW. 

As some of the early-type stars in the central $\sim$0.5\,pc are on a disk, asymmetry is not unexpected. However, Fig.\ref{fig:cube}
shows that the line of nodes of the disk is $\sim$60\degr\;offset from the Galactic plane. The early-type stars also appear offset from the Galactic plane, but not by as much.   An important observational bias is introduced by the spatially variable extinction. This is also shown in Fig.~\ref{fig:cube}. In the underlying 1.90\,$\mu$m image there are some patchy regions with less flux, for instance in the SW corner of the image. We detect fewer stars in these regions and find an asymmetric spatial distribution of early-type stars. As our extinction map does not extend to this region, we cannot quantify the effect of the variable extinction.  Thus we cannot conclude whether dust alone can explain the asymmetry.

\subsection{Kinematics  of  early-type stars}
\label{sec:kin}
The early-type stars in the Galactic centre can be distributed into different groups based on their kinematics. In the central $p$\textless 0.03\,pc ($\sim$0.8\arcsec) is the S-star cluster. This group of $\gtrsim$20 stars has high orbital eccentricities $e$ \citep[$\bar{e}$ = 0.8,][]{gillessen09}. Most of the stars are B-type main-sequence stars ($K_s\gtrsim$14\,mag). These stars are mostly too faint and too crowded to be in our data set. The only exception is S2 (Id 2314), which is one of the brightest S-stars with $K_S$=14.1\,mag \citep[for a Table of 51 sources in the S-star cluster see][]{sabha10}. 

 At greater distances, 0.03\,pc\textless$p$\textless0.5\,pc (0.8\arcsec$-$13\arcsec), there is a clockwise (CW) rotating disk of young stars with moderate orbital eccentricities ($e\sim0.3$). This disk contains WR, O, and B stars \citep{yelda14}. Not all stars in this radial range lie on the disk, there is also a more isotropic off-disk population. The disk and off-disk populations are very similar and probably coeval \citep{paumard06}.  It is not yet settled whether there is a second, counterclockwise rotating disk.  To assess whether a star belongs to the disk or not and if the star is on a bound orbit, we have to know the stellar kinematics. 

We measured the radial velocities of O/B stars as described in Sect. \ref{sec:measurekin}, and Table~\ref{tab:etkin} lists the radial velocities $v_z$ of the O/B stars. When no good radial velocity measurement was possible with our spectra, we list the radial velocity of \cite{bartko09}, \cite{yelda14}, or an error-weighted mean of their measurements.
Furthermore, we match the O/B stars of our data set with the proper motions of   \cite{yelda14} and \cite{Rainerpm09}. These measurements are also listed in Table~\ref{tab:etkin} as $v_{RA}$ and $v_{Dec}$. To transfer the proper motion velocities into km\,s$^{-1}$, we assumed a Galactocentric distance of $R_0=$~8\,kpc.

 About 20 young stars are on the CW disk \citep{yelda14}. The CW disk has the orbital parameters inclination $i=130^\circ$ and ascending node $\Omega=96^\circ$ \citep[e.g.][]{yelda14,bartko09,lu09,paumard06}. The stars on the CW disk are approaching (negative radial velocity,  $v_z$\textless0 ) in the equatorial North-West, and receding (positive radial velocity,  $v_z$\textgreater 0) in the equatorial South-East. Based on this simple criterion, we can exclude the membership of 23 stars of our O/B star sample, 7 of which are newly identified O/B stars. We list in  the second last  column of Table~\ref{tab:etkin} whether $v_{z}$ agrees with the  rotation of the CW disk or not. 
If the entry  in the second last column of Table~\ref{tab:etkin} is 0, a  membership to the CW disk is not possible  according to $v_z$, given the   longitude of the ascending node $\Omega$ is 96$^\circ$. If the disk is warped, as found by  \cite{bartko09},  the value of  $\Omega$ would change with  the distance to Sgr~A*. Then the radial velocity criterion would exclude one star less. 
 
\begin{figure}
	\resizebox{\hsize}{!}{\includegraphics{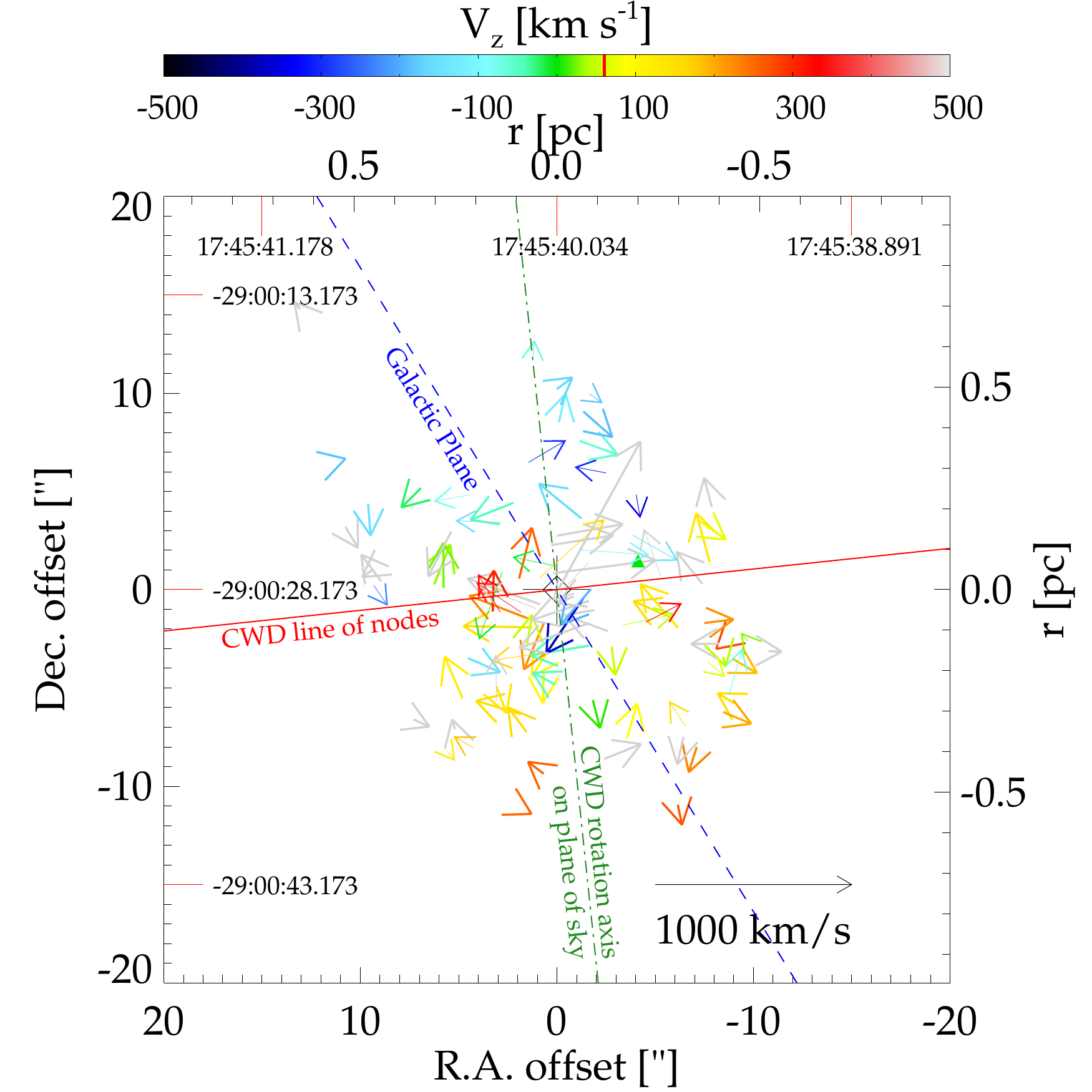}}	\caption{  Three-dimensional stellar kinematics of the O/B stars and emission line stars (smaller arrows). The arrows denote the proper motions, colours signify different radial velocities $v_z$. The black cross  indicates the position of Sgr~A*.         The coordinates show the offset to Sgr~A* in equatorial coordinates, the numbers at the top and left denote  absolute equatorial coordinates.  The Galactic plane is plotted as a blue dashed line, and the line of nodes of the clockwise disk with $\Omega$=96$^\circ$ is shown as a red line. The green dot-dashed line illustrates the projected rotation axis of the clockwise disk.}
		\label{fig:3dkin}
	\end{figure}

The stellar kinematics are illustrated   in Fig. \ref{fig:3dkin}. 
For 45 O/B stars we have the radial velocity and proper motions,  and for 22 stars proper motions alone.  
The directions of the arrows denote the proper motion direction, the lengths of the arrows denote the proper motion velocity $v_{pm}$ assuming a distance of 8\,kpc. Additionally, we overplot the kinematics of the emission-line stars with 27 radial velocities adopted from \cite{tanner06}, and 28 proper motion measurements  as slightly smaller arrows.   Proper motions are taken from \cite{yelda14} if available, and from  \cite{Rainerpm09}  otherwise. Because we used the disk parameters of  \cite{yelda14}  in our analysis, we give preference to the proper motions derived by this study. 

\subsection{O/B star orbits}
To identify stars on radial or tangential orbits, the angular momentum $j_z=xv_y-yv_x$ can be used, or as suggested by \cite{madigan14}, $j_z$ normalised to the maximum angular momentum at projected radius $p$
\begin{equation}
h=\frac{xv_y-yv_x}{\sqrt{G\mathcal{M}_\bullet p}}.
\end{equation}
$x$ and $y$ denote the distance to Sgr~A* in equatorial coordinates, $v_x$ and $v_y$ are the proper motions in the same coordinate system, $\mathcal{M}_\bullet$=4.3$\times$10$^6 M_\odot$ \citep{ghez08,gillessen09} is the mass of the supermassive black hole, and $G$ is the gravitational constant. 

The $h$-value constrains the stellar orbital eccentricity and
shows whether the star is on a projected orbit that is clockwise (CW) tangential ($h\sim1$)  or counterclockwise tangential ($h\sim-1$). We also list $h$  in Table~\ref{tab:etkin}. If $h$ is negative, this star is probably not a member of the CW disk, although the  radial velocity $v_z$ may agree with the CW disk. A value of $h \approx 0$ does in principle mean the star is on a radial projected orbit. But this can have different reasons: Either the star has a high orbital eccentricity ($e\gtrsim$0.8), a highly inclined orbit  ($i\gtrsim$70$^{\circ}$, with 90$^{\circ}$ meaning edge-on), or both.  If we have both  proper motion and radial velocity for a star, we can compare the magnitude of  the proper motion velocity $v_{pm}=(v_{RA}^{2}+v_{Dec}^{2})^{1/2}$ to the total three-dimensional velocity $v_{tot}$. 
If the proper motion velocity is much lower than the radial velocity, that is, the three-dimensional velocity vector is mainly pointing along our line of sight, the star is on a close to edge-on orbit.   For example, a value of  $v_{pm}/v_{tot}\leq$ 0.2 indicates a high inclination of the orbit. Then a low value of  $\vert h \vert$ tells us nothing about the eccentricity of the stellar orbit. 

Twenty-four stars have $\vert h \vert \leq$ 0.2, suggesting a high eccentricity $e$, a high inclination $i$, or both. For 18 of these stars we have kinematics in three dimensions, thus we can infer for three stars that they have orbits with high inclination, they are marked with a footnote in Table~\ref{tab:etkin} (Id 483, 728, and 853). 
On the other hand, we have 11 stars with $\vert h \vert \leq$ 0.2, for which the ratio  $\vert v_{pm}\vert/v_{tot}\geq$ 0.6 indicates a rather low inclination. Therefore the orbits of these stars have truly high eccentricities.

Although a low value of $\vert h \vert$ does not necessarily mean a radial orbit, a value of $h$ \textgreater  0.6 is an indication that a star is on the CW disk. 
Our data set contains  14 stars ($\sim$20\%) with $h$ \textgreater 0.6, for eight ($\sim$12\%) of them $v_{z}$ also agrees with the CW disk, but for four of them it does not. Only one star of the  new O/B stars is a good candidate for being on the CW disk: Id 596 has $h$=0.82 and is at a distance of $p$=7.35\arcsec\,($\sim$0.3\,pc) from Sgr~A*. 

To determine the full orbit of a star and thereby constrain the disk membership, it is also necessary to consider the distance of the star along the line of sight. 
  \cite{lu09}  and \cite{yelda14}  included measurements of the plane-of-sky  acceleration  to constrain the stars'  line-of-sight distances to Sgr~A*.  To better constrain the orbital parameters, previous studies  \citep[e.g.][]{lu09,bartko09,yelda14,joel14} computed density maps of the orbital planes and ran Monte Carlo simulations.

 For one star, Id 96, the value of $\vert h \vert$ is even higher than 1: $h=-$1.05. According to \cite{madigan14}, this means the star is still on a bound orbit, as $\vert h \vert \leq \sqrt{2}$. But it requires the semi-major axis of the stellar orbit  to be  larger than $p$, and the star is closer to pericenter than apocenter.

\begin{figure}
	\resizebox{\hsize}{!}{\includegraphics{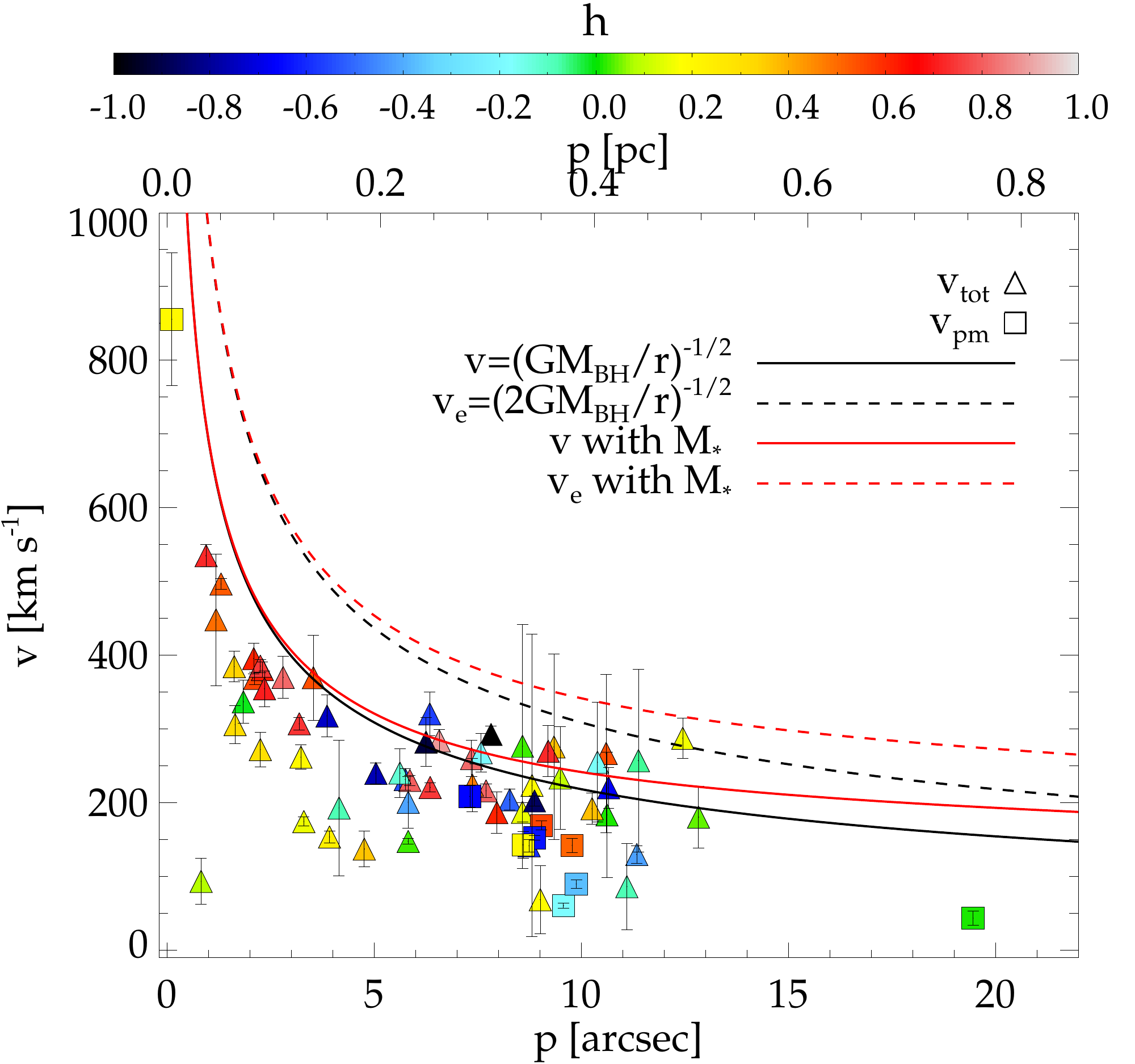}}
	\caption{Velocity profile for the O/B stars. The total velocity is plotted against the projected distance $p$ to Sgr~A*. Triangles denote fully known kinematics in three dimensions, squares denote only two-dimensional projected proper motion measurements and are therefore only lower limits of the total velocity. The colour-coding illustrates the normalised angular momentum $h=(xv_{y}-yv_{x})/(\sqrt{G \mathcal{M_{\bullet}}p})$.  The full black line denotes the velocity profile for a central point mass with $\mathcal{M_\bullet}$=4.3$\times10^6 M_\odot$, the dashed line denotes the escape velocity $v_e$ for such a point mass. Red lines mean that we also consider a stellar mass $\mathcal{M_{*}}$.}
		\label{fig:rv}
	\end{figure}

\cite{isaacanja} detected two high-velocity stars at $p$=3\,pc (80\arcsec) and $p$=5\,pc (130\arcsec) with $v_z$=292\,km~s$^{-1}$ and $-$266\,km~s$^{-1}$.  Our data set also contains stars with high velocities.  To check if the O/B stars are bound to the nuclear star cluster, we plot the total velocity $v_{tot}$ against the projected distance $p$ to Sgr~A* in Fig.~\ref{fig:rv}. For stars without a radial velocity measurement we plot  $v_{pm}$  , which is only a lower limit of $v_{tot}$. The colour-coding denotes the value of $h$. The full black line denotes the Keplerian velocity with a single point mass in the centre  with mass $\mathcal{M_{\bullet}}$ = 4.3$\times10^6 M_\odot$.  When we also take the stellar mass into account, the velocity increases. To illustrate this, we plot the velocity as a red line when we assume the stellar mass profile from \cite{isaacanja}.  Most stars lie below this line and must therefore be bound to the Milky Way nuclear star cluster.  

The dashed lines show the escape velocity $v_e$. 
Only one star  has a velocity close to the escape velocity: Id 722. It is at a projected distance of 12.45\arcsec\,with $v_{tot}$=(287$\pm$27)\,km\,s$^{-1}$. The proper motion of this star also points away from Sgr~A*. For Fig.~\ref{fig:rv} we plot only the projected distances of the stars, which are only lower limits. The true distance of star Id 722 might well be larger. But when we consider the  stellar  mass distribution, the star's velocity is lower than the escape velocity.  
 The normalised angular momentum is $h$=0.16, and if one takes  the stellar mass into account for $h$ as well, $\vert h \vert$ becomes even smaller. This indicates that this star with $\mathcal{M}=32\pm14\,M_\odot$ is on a radial orbit, but the star is still gravitationally bound to the Milky Way nuclear star cluster.
 
Some of the  stars  have large velocity uncertainties. Six  stars may have velocities above the escape velocity (Id 366, 511, 610, 617, 728, and 853), four of them (610, 617, 728, 853) also have a low value of $\vert h \vert$\textless 0.2, suggesting either a high eccentricity or inclination in their orbits.  To better
constrain the stellar orbits, a more accurate radial velocity measurement is required.  

\section{Discussion}
\label{sec:disc}
\subsection{Detection of 19 new O/B stars}

\label{sec:discob}
Our sample of 76 O/B stars mostly consists of previously known O/B stars.  However, 24 O/B stars have not been reported before, and 19 of these O/B stars are probably also cluster member stars. Three stars (Id 663, 1104, and 3308) are possibly foreground stars, while  two stars (Id 436, 3339) are definitely foreground stars. 

To verify our classification as O/B type stars, we measured the equivalent widths of the CO line at 2.2935\,$\mu$m and the Na~I doublet at $\sim$2.206\,$\mu$m in Sect.~\ref{sec:ob} and compared the result to the mean value of the late-type stars. For most of the O/B stars, the equivalent widths deviate by $\sim$3$\sigma$ from the mean value of late-type stars for the CO line and by $\sim$2$\sigma$ from the mean late-type star value for  Na~I. 

Only ten O/B stars have  $EW_{CO}$ \textgreater 2.7\,\AA, i.e. within  3$\sigma$ of the late-type stars' mean value ($EW_{CO,LT}$ = 18.3). However,  six of these ten stars have been classified as O/B stars in previous studies. 
For the other four stars (Id 718, 2446, 3308, and 3578), the significance of either the CO non-detection or of the Na non-detection is at least 2.7$\sigma$. 
The S/N of the spectra is  in the range of 16.9 to 35.8, this is rather low compared to the O/B star median S/N of 46. It is possible that the low S/N or poorly subtracted light from surrounding late-type stars produces a weak CO line signal.  If these stars indeed have  weak CO lines, the low values of  $EW_{CO}$ (3.2\,\AA$-$8.3\,\AA) would suggest  effective temperatures $T_{\mathrm{eff}}$\textgreater 4,500\,$K$ \citep{pfuhl11}. Then these stars could be of intermediate age ($\sim$100\,Myr).

\subsection{O/B star mass estimates}
We estimated the masses of the O/B stars in Sect.~\ref{sec:massdust} based on the assumptions that the intrinsic colours  $(H-K_S)_0$ are in a narrow range, close to $-$0.10 \citep{intrinsiccolor}, their metallicity is roughly solar \citep{ramirez00}, and their ages are in the range of 3$-$8\,Myr \citep{paumard06,lu13}. This means that we assumed the same age for the new O/B star detections as for the previously known O/B star population, for which this age estimate was derived. This may be an oversimplification. The new O/B-type stars are at larger projected distances from Sgr~A* and may have formed in a different star formation event. This means that their age and metallicity may be different.  Some of these stars have a mass of $\sim$10\,$M_\odot$. \cite{renzini9mstar} showed that a star with a mass $\mathcal{M}$=9\,$M_\odot$ and solar metallicity spends $\sim$20\,Myr on the main sequence. This means that these stars may be much older than 3$-$8\,Myr.  On the other hand, among the new classified O/B stars are also several  bright,  massive stars ($\mathcal{M}$\textgreater 20\,$M_\odot$). These stars  must be younger and close to the age of 3$-$8\,Myr. The spectra of the newly identified O/B stars with  S/N $\gtrsim$ 50 contain a He~I absorption line at 2.113~$\mu $m. \cite{hanson96}  showed that this line disappears in early O stars.  Therefore the  stars  with 2.113~$\mu $m absorption are later than  O7~V,  O7~III, or O9~I. This suggests that the new O/B stars belong to the same population as  the already classified O/B stars, none of which is earlier than O7 \citep{paumard06}. 

The estimated mass in Table~\ref{tab:etmag} is the median of the distribution of possible stellar masses weighted by the likelihood of the star position in the colour magnitude diagram.  We considered the uncertainty of the Galactocentic distance $\sigma_{R_0}$, the extinction law coefficient $\sigma_\alpha$, the photometry ($\sigma_H$ and $\sigma_{K_S}$), and the intrinsic colour $\sigma_{(H-K_S)_0}$ in the propagated uncertainty. 
For the brightest of these stars we derive median masses of  more than 40\,$M_\odot$. Such a high mass was observed for O6 I  or O6 V stars \citep[Table~15.8]{allen}. But  the stars with reported spectral type in our sample are of  type O8 and later. For example, we derive  a mass of $\mathcal{M}$=(41$^{+9}_{-13})$\,$M_\odot$ for the O8$-$9.5 star Id 331. 
 Stars of this spectral type have masses of $\sim$28\,$M_\odot$ (O8 I star) and lower \citep[Table~15.8]{allen}. This suggests that we rather overestimate the   stellar masses. 
 
We showed that the derived value for the extinction $A_{K_S}$ from intrinsic colours is lower than the value of $A_{K_S}$ from the extinction map adopted from \citealt{rainer10} (see Fig.~\ref{fig:extinction}). We would find better agreement by assuming an extinction law exponent of $\alpha$=2.1 instead of $\alpha$=2.21 \citep{rainer10}. Then the derived magnitudes $K_{S,0}$ of the stars would be higher, that is, the stars would be fainter. We tested if this would lower the stellar masses, but found that it is only a minor effect. For most of the stars the lower limit of our mass estimate agrees with the  mass expected for the spectral type. 

\subsection{Total mass of young stars}
We calculated a lower limit for the total mass of  young stars to $\mathcal{M}_{young, M\leq150 M_\odot}^{\alpha=1.7}$ = 21,000\,$M_\odot$ assuming a top-heavy initial mass function (IMF) with slope $\alpha$=1.7 \citep{lu13} and a maximum stellar mass of 150\,$M_\odot$. This result agrees with the cluster total mass found by \cite{lu13} of 14,000\,$M_\odot$ to 37,000\,$M_\odot$ in the same integration range $\left[ 1 M_\odot; 150 M_\odot \right]$.    With an extremely top-heavy IMF \citep[$\alpha$=0.45,][]{bartko10}, the mass is even higher  $\mathcal{M}_{young, M\leq150 M_\odot}^{\alpha=0.45}$ = 32,000\,$M_\odot$, but still in agreement with \cite{lu13}. 
We integrated up to stars with $\mathcal{M}$=150\,$M_\odot$, but the most massive star in the Galactic centre has  only $\mathcal{M}$=80\,$M_\odot$ \citep{martins07}. When we integrate in the interval $\left[ 1 M_\odot; 80 M_\odot \right]$, the cluster mass is only $\mathcal{M}_{young, M\leq80 M_\odot}^{\alpha=1.7}$ = 16,000\,$M_\odot$, for $\alpha$=1.7, and $\mathcal{M}_{young, M\leq80 M_\odot}^{\alpha=0.45}$ = 12,000\,$M_\odot$  for $\alpha$=0.45.
We thus give $\mathcal{M}_{total,young}\sim$12,000\,$M_\odot$ as a lower limit for the mass of the young star cluster. 

For these calculations  we assumed that the present-day mass function slope is the same as the IMF slope, but this  is only valid for a simple stellar population \citep{elmegreen06}. This is a reasonable assumption, since  we used stars  in the mass range $\left[ 30 M_\odot; 45 M_\odot \right]$ to fit the IMF, and these stars should all be younger than 8\,Myr. 
As we scaled the IMF only to the observed mass function of O/B stars and did not consider the  massive and  young emission-line stars, the derived total cluster mass  is only a lower limit.

\subsection{Disk membership}  
\label{sec:discdisk}

We estimated if an O/B star can be a member of the clockwise (CW) disk based on the stellar angular momentum $h$ and on the stellar radial velocity $v_z$.  As the disk is receding in the SE and approaching in the NW, we can exclude the membership of stars with  $v_z$\,\textgreater\,0 that are located in the  NW and of  stars with $v_z$\,\textless\,0 that are located in the SE. We can also exclude stars with an angular momentum $h$\,$\lesssim$\,0.5 if we can show that the orbit is not edge-on. 
 This allowed us to exclude the CW disk membership for 53 stars ($>$69\%) in  our O/B star sample, of which 16 stars are newly classified  O/B stars.  When we assume that the CW disk is warped, as found by  \cite{bartko09},  these numbers change by only one star (54 stars, $>${72}\%).
 
A disk fraction of  $\lesssim$30\% agrees with the results of \cite{yelda14}, who studied the kinematics of O/WR stars and found a disk fraction of $\sim$20\%. \cite{yelda14} also showed that the significance of the disk decreases with distance to Sgr~A*. Our  sample of new O/B stars mostly lies at $\sim$10\arcsec\,(0.4\,pc), that is, close to the assumed outer edge of the CW disk at $\sim$13\arcsec\,(0.5\,pc). It might be that the outer edge of the disk is even closer to Sgr~A* \citep{stostad}.

  On the other hand, we found ten stars (one new) for which a membership to the CW disk is possible based on their proper motions, projected location, and radial velocity. However, this does not mean that these stars are necessarily  members of the CW disk, as the   three-dimensional location with respect to Sgr~A* was not taken into account. We were unable to constrain whether the   remaining 13 O/B stars might be  members of the CW disk or not. The radial velocity uncertainty allows both a receding and an approaching motion, there is no proper motion available, or the inclination $i$ is too high to determine the angular momentum $h$. Three of these undetermined stars are probably foreground stars. 

For a better determination of the stellar orbits,  a more sophisticated analysis such as that reported by  \cite{lu09}, \cite{bartko09}, \cite{yelda14}, and \cite{joel14}, is necessary.
  In the future, the missing proper motions and radial velocities
that are missing so far will probably also be available (Pfuhl et al. in prep.).

\subsection{Origin of the early-type stars}
Our data set covers the central 2.51 $\times$ 1.68\,pc (\textgreater 4\,pc$^2$) of the Galactic centre. No previous study covered such a large region with a comparable spatial resolution. We were able to extract stars as faint as $K_S$\,=\,15\,mag  with a completeness of 80\% at $K_S$\,$\approx$\,13.5\,mag. For the bright supergiants and giant stars with $K_S$\,\textless\,13\,mag, we can  assume that our data set is roughly complete out to $p$\,=\,0.84\,pc (21\arcsec). 

Bright O/B stars with $K_S$\,\textless\,13\,mag have a well-determined age of 3$-$8\,Myr. We can add the red supergiant IRS~7, the emission-line stars, and sources with featureless spectra and $K_S$\,\textless\,13\,mag, which are in the same age range. Then we find that 90\% of these 79 massive  stars are located within $p$\,=\,0.44\,pc (11.4\arcsec). This confirms the finding of \cite{stostad} that the cluster of young stars has an outer edge at $\sim$13\arcsec\,(0.52\,pc). This central confinement can help to constrain the origin and formation scenarios for the young stars.

It was suggested that young stars in the Galactic centre were formed in a massive young star cluster that fell towards the centre from $r\gtrsim$10\,pc  \citep{gerhard01,mcmillan03}. In this scenario  the infalling cluster is stripped and disrupted. But then we would expect a higher number of young  stars  beyond $p$\,=\,0.5\,pc \citep{fujii10,perets10}. Our data set contains only three bright early-type stars (and three faint early-type stars)  beyond $p$\,=\,0.5\,pc, but 76 (23)  within $p$\,=\,0.5\,pc. An infalling cluster would leave a trail of early-type stars, but we find no evidence for such a structure. It might be possible that the infalling cluster had been mass segregated and left  a trail of fainter early-type stars that we could not detect. But other studies with a  smaller spatial coverage but higher completeness for fainter stars were likewise unable to detect any signs of a trail \citep{bartko10,stostad}.

The late-type stars are much less concentrated than the early-type stars. This agrees with the findings in other nuclear star clusters \citep[e.g.][]{anil10404,iskren14,carson15}.  The old component of the nuclear star cluster is often spheroidal and more extended than the  disk of young stars. One counterexample is NGC 4244, where the blue disk is more extended than the older spheroidal component \citep{anilon4244_08}. 
\cite{anil06} argued that  young stellar disks in nuclear star clusters have a lifetime of  $\lesssim$\,1\,Gyr before being disrupted. 
In other nuclear star clusters, the young disks are often aligned with the host galaxy disk \citep[e.g. NGC 404, NGC 4244, and NGC 4449;][]{anil06,anilon4244_08,iskren14}. This is not the case for the CW disk in the Galactic centre. The projected distribution of young stars beyond $p$\,=\,0.47\,pc (12\arcsec) appears to be elongated along the Galactic plane, but slightly misaligned to it. It is unclear if this effect is only caused by the variable extinction. 

The CW disk of young stars can be explained by  \textit{in situ} star formation in a dense disk or stream around Sgr~A* \citep{levin03,paumard06}. The material would come  from infalling  molecular clumps and gas clouds \citep[e.g.][]{wardle08,gualandris12,eric15}.  
As the stars are very concentrated in the centre, the star-forming region must have had a size of $r$\,$\lesssim$\,0.5\,pc. 
However, the majority of the early-type stars in the Galactic centre are not on the CW disk (see Sect.~\ref{sec:discdisk} and \citealt{yelda14}).  As the stars are only 3$-$8\,Myr old, the young stars either  did not all form in a disk or the disk is dissolving more rapidly than expected. 

One possible disruption scenario is the infall of another molecular cloud to the Galactic centre \citep{mapelli13}. This cloud is disrupted in the supermassive black hole potential and forms an irregular, dense gas disk. Perturbations induced from this gas disk  might be able to dismember the CW disk of young stars \citep{mapelli13}. This could explain the isotropic cluster of young stars in the same radial range as the CW disk. 
However, other reasons that cause an  instability of the CW disk are also possible \citep[e.g.][]{hobbs09,chensstar14}.  More simulations and theoretical work are needed to  explore the possibilities.

\subsection{Early-type stars beyond the central 0.5\,pc}
There are only six stars with projected distances $p$\,\textgreater\,0.5\,pc. These are three O/B stars (Id 982, 2048, and 2446) and three featureless sources (Id 161, 477, and 705). However, the classification of two of the O/B stars as cluster member stars is uncertain. Id 982 is located outside the coverage of  the extinction map by \cite{rainer10}, therefore the colour has a large uncertainty.  Id 2048 lacks full colour information. 

Id 2446 at $p$\,=\,0.75\,pc (19.5\arcsec) is the only outer O/B star with available proper motions. The proper motion vector points away from Sgr~A*, but the velocity is low enough for the star to be  bound to the cluster ($v_{pm}$\,=\,(43.5$\pm$9.6)\,km~s$^{-1}$). The angular momentum is  $h$\,=\,0.01, which means that the star is on a radial projected orbit.

Two of the three featureless sources beyond 0.5\,pc (Id 161/IRS~5 and 705/IRS~5NE) are part of the IRS~5 complex \citep{perger08}. \cite{viehmann06} pointed out that the IRS~5 sources are remarkably bright in the mid-infrared, but less prominent in the near-infrared. The O/B star Id 483/IRS~5SE also belongs to this group. IRS~5SE consists of two components, IRS~5SE1 and IRS~5SE2 \citep{viehmann06}, which cannot be resolved in our data set. In this region we have two additional spectra that we were unable to classify (IRS~5S and IRS~5E); the remaining stars in this area have  late-type signatures.

Id 477/IRS~8 is a special case. This featureless source has the largest distance to Sgr~A* of all early-type stars in our field of view. \cite{geballe06} classified IRS~8 as an O5$-$O6 giant or supergiant. This makes IRS~8 the earliest O/B star in the Galactic centre. All the O/B stars within $p$\,=\,0.5\,pc are of type O8 and later. This would make IRS~8 the youngest  known star in the Galactic centre. However, \cite{geballe06} suggested that IRS~8  originally was a  member of a close binary and was   rejuvenated.

\section{Summary}
\label{sec:con}
We observed  the central \textgreater4\,pc$^2$ of the Galactic centre with the integral field spectrograph KMOS. Among more than 1,000 spectra from single stars were 114 early-type star spectra. We analysed these early-type spectra, and found the following:

\begin{enumerate}
\item
We detected 24 previously unknown O/B-type stars. Of these, 19 stars are probable cluster members. The new O/B stars are at projected distances of 0.3\,pc$-$0.92\,pc and cover masses from $\sim$10$-$40\,$M_\odot$. 

\item 
We derived a lower mass limit for the young cluster mass $\mathcal{M}_{total,young}$=12,000\,$M_\odot$.  We used different initial mass function slopes from the literature and  integrated  in the range $\mathcal{M}~=~\left[ 1\,M_\odot; 80\,M_\odot \right] $.

\item
With our spatially extended and nearly symmetric coverage, we studied the spatial distribution  of early-type stars. 
We found that the early-type stars are strongly concentrated in the projected central $p$\,=\,0.4\,pc and that only a few stars lie beyond 0.5\,pc. This contradicts a scenario where the early-type stars formed outside the Galactic centre in a massive cluster that fell towards the centre and depleted the stars at their current location. This formation scenario would leave behind a trail of early-type stars at projected distances of $p$\,\textgreater\,0.5\,pc, which we did not detect. This is a strong argument for the \textit{in situ} formation of the early-type stars.

\item
We studied the kinematics of the O/B stars and  showed that    one of the new O/B stars is a good candidate to be a member of the clockwise rotating disk. However, the majority ($\gtrsim$69\%) of the O/B stars is not on the disk. This means that either these stars have not formed on the clockwise disk or that the disk is already strongly disrupted. 
We found no stars that are unbound to the Milky Way nuclear star cluster. 
\end{enumerate}

  \begin{acknowledgements} 
 This research was supported by the DFG cluster of excellence Origin and Structure of the Universe (www.universe-cluster.de). C.~J.~W. acknowledges support through the Marie Curie Career Integration Grant 303912.
This publication makes use of data products from the Two Micron All Sky Survey, which is a joint project of the University of Massachusetts and the Infrared Processing and Analysis Center/California Institute of Technology, funded by the National Aeronautics and Space Administration and the National Science Foundation. 
This research made use of the SIMBAD database (operated at CDS, Strasbourg, France). 
\\
We would  like to thank the ESO staff who helped us to prepare our observations and obtain the data. 
Special thanks go to Alex Agudo Berbel, Yves Jung, Ric Davis, and Lodovico Coccato for advice and assistance in the data reduction process.
We are also grateful to Sebastian Kamann for providing us with his code PampelMuse. We  thank Morgan Fouesneau,  Iskren Georgiev, and Paco Najarro for  discussions and suggestions. We finally thank the anonymous referee for   comments and suggestions.
 \end{acknowledgements}
 \bibliographystyle{aa}		

\bibliography{bibs2}		

\newpage
\begin{appendix}
\section{Spectral  classification of emission-line stars}
\label{sec:discem}
We have spectra of 29 emission line stars. These stars were classified in previous studies as either Ofpe/WN9 type, WN, or WC type stars. Ofpe/WN9 types have narrower lines and are cooler (T$_{\mathrm{eff}}$=10,000$-$20,000\,K) than WN and WC stars (T$_{\mathrm{eff}}$\textgreater 30,000\,K).  We found some disagreement with previously reported spectral classifications for some of the emission-line stars.

\textit{Id 144/AF and Id 1237/IRS 7E2(ESE):}
We find that the two stars Id 144/AF and Id 1237/IRS 7E2(ESE), which were listed as Ofpe/WN9 by \cite{paumard06}, have broad emission lines and are rather WN8 or WN9 stars.   \cite{paumard06} stated that their spectra  of these stars    are of high quality. But we also have a high S/N, 78.2 for Id 144/AF and 35.6 for Id 1237/IRS 7E2. Furthermore,  the high-resolution spectra of \cite{tanner06} agree with Id 144/AF being a broad emission-line star, but they have no data for  Id 1237/IRS 7E2(ESE). \cite{paumard03} also found a broad He I line in the spectrum of  Id 144/AF. The resolving power $R$ reported by \cite{tanner06} was   14,000 and    23,300, but only 1,500 and 4,000  in the data used by  \cite{paumard06}. Our data set with $R$=4,300 agrees with  the high-resolution results from \cite{tanner06}. Because of their broad emission lines (FWHM $\sim$ 700\,km s$^{-1}$) and the resemblance of the spectra of Id 144/AF and Id 1237/IRS 7E2 with the WN8 and WN9 spectra in our data set,  we classify Id 144/AF and Id 1237/IRS 7E2 as broad emission-line stars, probably WN8 or WN9.  Id 1237/IRS 7E2 is also classified as a WN8 star by \cite{martins07}. 
\cite{martins07} used non-LTE atmosphere models to derive the properties of Galactic centre stars. For Id 144/AF they found a degeneracy between the effective temperature $T_{\mathrm{eff}}$ and the helium abundance He/H. In addition,  the wind of this star could be stronger than the wind of the  Ofpe/WN9 stars, and Id 144/AF may be more evolved. 
\cite{martins07} suggested that Ofpe/WN9 stars evolve to WN8 stars. 

\textit{Id 666/IRS 7SW:}
We reclassify  the star Id 666/IRS 7SW as WN8/WC9. This star was classified as WN8 in \cite{paumard06}. The  C~III and C~IV lines  distinguish a WN8/WC9 star from a WN8 star. These lines are weaker than the He and H lines. Therefore a low S/N can lead to a misidentification as a WN8 star, but \cite{paumard06} state that their spectrum of Id. 666/IRS 7SW is of high quality. Our spectrum of this star has a  S/N of 59.9, and we can clearly identify the C~IV doublet  at 2.0796\,$\mu$m,  and  2.0842\,$\mu$m  C~III at 2.325\,$\mu$m (see Fig. \ref{fig:wn}). The spectrum is very similar to the spectrum of the WN8/WC9 star  Id 491/IRS~15SW, therefore we conclude that Id 666/IRS 7SW is also a  WN8/WC9 type star. The two stars Id 491/IRS~15SW and Id 666/IRS 7SW are also in a similar location at the colour-magnitude diagram ($(H-K_S)_0$ = $-$0.02 and $-$0.11, $K_{S,0}$=9.40 and 9.59, Fig. \ref{fig:cmd}), which confirms their similarity. This classification agrees with
that of \cite{martins07}.

\textit{Id 185/IRS~29N,  Id 283/IRS~34, Id 303, and Id 638/IRS~29NE1:}
Four of the eleven WC stars have only shallow emission lines in our data set. These stars are Id 185/IRS~29N,  Id 283/IRS~34, Id 303, and Id 638/IRS~29NE1. The lines are very broad, but only weakly pronounced. Previous studies \citep{paumard01,tanner06} did not detect any distinct He~I emission for Id 185/IRS~29N, while \cite{paumard03} reported a broad He~I emission line for the same star. 
\cite{tanner06} suggested that  Id 638/IRS~29NE1 is variable and  that the spectral features changed with time. \cite{rafelski07} studied the light curve  of IRS~29N over a time line of ten years and found photometric variability. They suggested that these sources could be a wind-colliding binary system. \cite{gamen12} showed that stars can change their spectra within months.

Apart from the weak but broad lines, all the four sources 
are also very red  ($(H-K_S)_0$ \textgreater 0.55). 
 Their continua rise steeply with wavelength. This is an indication that these sources are embedded in dust (see e.g. \citealt{geballe06} for IRS~8). 
 The continuum in the spectra might not be the stellar continuum, but the continuum of the circumstellar dust, which dominates the lines  \citep{figer99,chiar01}.  Therefore the emission lines appear only as weak, broad bumps in the spectrum.
 
 Circumstellar dust is common for WC9 stars such as  Id 185/IRS~29N and   Id 283/IRS~34. For earlier types such as WC8, dust formation is rather uncommon and might indicate colliding winds \citep{sanderwc12}. The two stars Id 303 and Id 638/IRS~29NE1  are WC8/9 stars. Id 303 is located close to the minispiral, at least in projection. So it might  be a bow-shock that causes the reddening of Id 303. The bow-shock sources Id 161/IRS~5 and Id 25347/IRS~1W  are probably also embedded WR stars \citep{tanner05,joel14}, but their emission lines are outshone by the bow-shock continuum.
  Id 638/IRS~29NE1, however,  is not located inside the minispiral. But as mentioned earlier, the spectral features seem to change with time. This could be explained by circumstellar dust.
 
 \textit{Id 243/IRS~34W:}
The star  with a narrow emission line,  Id 243/IRS~34W,  has moderate reddening ($(H-K_S)_0$ = 0.24)  and a steeply rising continuum.
 As \cite{paumard03} already pointed out, this star  is fainter than the other stars with narrow emission lines (see also the
colour-magnitude diagram, yellow circle with black cross, Fig. \ref{fig:cmd}). The star Id 243/IRS~34W  shows a long-term photometric variability  \citep{ott99,paumard03}.  Therefore this star may also be dust embedded. 
   Id 243/IRS~34W is a LBV candidate, and \cite{humphreysetacar} showed for the case $\eta$ Carinae that LBV eruptions are accompanied with circumstellar dust obscurations.

\section{O/B star tables}

\clearpage

\clearpage
\onecolumn
\small
\begin{longtable}{rccrrllllcl}
\caption{\label{tab:et}O/B stars.}\\
\hline\hline 
Id&RA&Dec&$\Delta$RA&$\Delta$Dec&$K_S^{a}$&Colour&Name&Type&Note$^{b}$&S/N\\
&$ \left[ ^{\circ} \right]$&$ \left[ ^{\circ} \right]$&$ \left[ \arcsec \right]$&$ \left[ \arcsec \right]$&$ \left[ \text{mag} \right]$\\
\hline
  \endfirsthead \caption{continued.}\\ \hline\hline
Id&RA&Dec&$\Delta$RA&$\Delta$Dec&$K_S^{a}$&Colour&Name&Type&Note$^{b}$&S/N\\
&$ \left[ ^{\circ} \right]$&$ \left[ ^{\circ} \right]$&$ \left[ \arcsec \right]$&$ \left[ \arcsec \right]$&$ \left[ \text{mag} \right]$\\
\hline \\
\endhead 
\hline \\
\endfoot
 \hline  \\
\endlastfoot  \\
$    64$&$  266.41745 $&$   -29.007652 $&$  -1.632 $&$   0.570 $&$           10.74 $&...&IRS16CC&O9.5-B0.5&$4,5$&$ 88.4$\\
$    96$&$  266.41437 $&$   -29.007425 $&$   6.613 $&$   1.387 $&$           10.44 $&...&...&...&$3$&$ 82.3$\\
$   109$&$  266.41724 $&$   -29.008343 $&$  -1.060 $&$  -1.916 $&$           10.57 $&...&MPE+1.0-7.4(16S)&B0.5-1&$4,5$&$ 92.7$\\
$   166$&$  266.41397 $&$   -29.009418 $&$   7.660 $&$  -5.788 $&$           11.26 $&...&...&...&$1,5,7$&$100.2$\\
$   205$&$  266.41882 $&$   -29.007736 $&$  -5.305 $&$   0.268 $&$           11.18 $&...&IRS1E&B1-3&$4$&$ 78.0$\\
$   209$&$  266.41571 $&$   -29.009912 $&$   3.014 $&$  -7.567 $&$           11.19 $&...&...&...&$3?$&$ 38.8$\\
$   227$&$  266.41742 $&$   -29.009571 $&$  -1.548 $&$  -6.338 $&$           11.43 $&...&...&?&$2$&$ 84.5$\\
$   230$&$  266.41681 $&$   -29.005444 $&$   0.082 $&$   8.521 $&$           11.19 $&...&...&O9-B&$2,5$&$ 81.3$\\
$   273$&$  266.41684 $&$   -29.004976 $&$  -0.000 $&$  10.204 $&$           11.44 $&...&...&O9-B&$2,5$&$ 95.8$\\
$   294$&$  266.41681 $&$   -29.008450 $&$   0.082 $&$  -2.300 $&$           11.14 $&...&IRS33N&B0.5-1&$4,5$&$ 85.4$\\
$   331$&$  266.41705 $&$   -29.008289 $&$  -0.571 $&$  -1.723 $&$           11.34 $&...&IRS16SSW&O8-9.5&$4$&$ 56.4$\\
$   366$&$  266.41705 $&$   -29.010412 $&$  -0.570 $&$  -9.366 $&$           11.62 $&...&...&...&$1,5$&$ 66.3$\\
$   372$&$  266.41827 $&$   -29.008778 $&$  -3.832 $&$  -3.481 $&$           11.81 $&...&...&...&$4$&$ 71.0$\\
$   436$&$  266.41968 $&$   -29.003155 $&$  -7.622 $&$  16.761 $&$           11.69 $&fg&...&...&$1,7$&$ 70.4$\\
$   443$&$  266.41730 $&$   -29.008221 $&$  -1.224 $&$  -1.476 $&$           12.09 $&...&IRS16SSE1&O8.5-9.5&$4$&$ 47.1$\\
$   445$&$  266.41647 $&$   -29.005707 $&$   0.981 $&$   7.574 $&$           11.67 $&...&...&O9-B&$2,5$&$ 54.5$\\
$   483$&$  266.42029 $&$   -29.005968 $&$  -9.237 $&$   6.633 $&$           11.96 $&...&IRS 5SE&B3&$8$&$ 78.8$\\
$   507$&$  266.41632 $&$   -29.008602 $&$   1.386 $&$  -2.850 $&$           11.95 $&...&...&O8.5-9.5&$2$&$ 66.2$\\
$   508$&$  266.41867 $&$   -29.007784 $&$  -4.897 $&$   0.096 $&$           12.23 $&...&...&O9.5-B2II&$4$&$ 62.6$\\
$   511$&$  266.41458 $&$   -29.010035 $&$   6.027 $&$  -8.006 $&$           12.10 $&...&...&...&$1,5$&$ 61.4$\\
$   516$&$  266.41754 $&$   -29.006582 $&$  -1.879 $&$   4.422 $&$           11.44 $&...&...&B0-3&$2$&$ 70.4$\\
$   562$&$  266.41690 $&$   -29.007578 $&$  -0.163 $&$   0.838 $&$           12.26 $&...&S1-3&?&$4$&$ 22.8$\\
$   567$&$  266.41855 $&$   -29.006989 $&$  -4.574 $&$   2.959 $&$           11.87 $&...&...&...&$3,5$&$ 75.4$\\
$   596$&$  266.41837 $&$   -29.009346 $&$  -4.075 $&$  -5.527 $&$           12.29 $&...&...&...&$1$&$ 71.7$\\
$   610$&$  266.41849 $&$   -29.009783 $&$  -4.399 $&$  -7.100 $&$           11.99 $&...&...&...&$1$&$ 54.4$\\
$   617$&$  266.41415 $&$   -29.009539 $&$   7.171 $&$  -6.221 $&$           12.51 $&...&...&...&$1$&$ 64.8$\\
$   663$&$  266.42368 $&$   -29.002535 $&$ -18.368 $&$  18.993 $&$           12.20 $&fg&...&...&$1$&$ 73.8$\\
$   668$&$  266.41711 $&$   -29.008015 $&$  -0.734 $&$  -0.735 $&$           12.54 $&...&...&...&$4$&$ 40.3$\\
$   707$&$  266.41733 $&$   -29.008621 $&$  -1.305 $&$  -2.918 $&$           12.32 $&...&...&B0-3&$4,5$&$ 28.3$\\
$   716$&$  266.41733 $&$   -29.009823 $&$  -1.303 $&$  -7.244 $&$           12.62 $&...&...&...&$3$&$ 48.4$\\
$   718$&$  266.41608 $&$   -29.010204 $&$   2.036 $&$  -8.617 $&$           12.71 $&...&...&...&$1$&$ 35.8$\\
$   721$&$  266.41904 $&$   -29.006376 $&$  -5.884 $&$   5.164 $&$           12.23 $&...&...&...&$1$&$ 41.4$\\
$   722$&$  266.41489 $&$   -29.010849 $&$   5.209 $&$ -10.938 $&$           12.72 $&...&...&...&$1$&$ 58.0$\\
$   725$&$  266.41602 $&$   -29.008396 $&$   2.202 $&$  -2.108 $&$           12.27 $&...&...&O9-B0&$2$&$ 41.2$\\
$   728$&$  266.41425 $&$   -29.008633 $&$   6.932 $&$  -2.959 $&$           12.35 $&...&...&...&$3$&$ 26.8$\\
$   757$&$  266.41400 $&$   -29.008787 $&$   7.583 $&$  -3.516 $&$           12.12 $&...&...&...&$1,5$&$ 48.6$\\
$   762$&$  266.41788 $&$   -29.008120 $&$  -2.774 $&$  -1.112 $&$           12.17 $&...&IRS16SE3&O8.5-9.5&$4$&$ 25.5$\\
$   785$&$  266.41705 $&$   -29.008959 $&$  -0.571 $&$  -4.134 $&$           12.30 $&...&...&B0-1&$4$&$ 38.8$\\
$   838$&$  266.41455 $&$   -29.006830 $&$   6.126 $&$   3.529 $&$           12.28 $&...&...&...&$3$&$ 68.4$\\
$   847$&$  266.41595 $&$   -29.008617 $&$   2.365 $&$  -2.905 $&$           12.15 $&...&...&B0-1&$2$&$ 46.2$\\
$   853$&$  266.41730 $&$   -29.010958 $&$  -1.221 $&$ -11.330 $&$           12.45 $&...&...&...&$1$&$ 74.2$\\
$   890$&$  266.42026 $&$   -29.006920 $&$  -9.148 $&$   3.207 $&$           12.86 $&...&...&...&$6$&$ 54.2$\\
$   900$&$  266.41632 $&$   -29.007969 $&$   1.387 $&$  -0.570 $&$           12.65 $&...&[GEO97]W14&O8.5-9.5&$4$&$ 30.0$\\
$   936$&$  266.41479 $&$   -29.009893 $&$   5.458 $&$  -7.498 $&$           12.73 $&...&...&...&$1$&$ 26.1$\\
$   941$&$  266.41693 $&$   -29.007181 $&$  -0.245 $&$   2.266 $&$           12.53 $&...&...&O9-B0&$4$&$ 45.4$\\
$   951$&$  266.41989 $&$   -29.006598 $&$  -8.170 $&$   4.367 $&$           13.00 $&...&...&OB&$2$&$ 59.7$\\
$   958$&$  266.41415 $&$   -29.008760 $&$   7.176 $&$  -3.419 $&$           12.61 $&...&...&...&$3$&$ 24.7$\\
$   973$&$  266.41489 $&$   -29.008347 $&$   5.221 $&$  -1.929 $&$           14.17 $&...&...&...&$3?$&$ 50.9$\\
$   974$&$  266.41782 $&$   -29.009180 $&$  -2.608 $&$  -4.930 $&$           12.96 $&...&...&...&$3$&$ 45.0$\\
$   982$&$  266.40979 $&$   -29.010162 $&$  18.812 $&$  -8.466 $&$           12.84 $&?&...&...&$1$&$ 47.4$\\
$  1048$&$  266.41983 $&$   -29.007536 $&$  -8.000 $&$   0.989 $&$           13.11 $&...&...&...&$4$&$ 50.6$\\
$  1103$&$  266.41342 $&$   -29.008680 $&$   9.133 $&$  -3.131 $&$           15.97 $&...&...&...&$1$&$ 34.7$\\
$  1104$&$  266.42184 $&$   -29.004236 $&$ -13.427 $&$  12.868 $&$           13.14 $&fg&...&...&$1$&$ 47.1$\\
$  1123$&$  266.41641 $&$   -29.007929 $&$   1.142 $&$  -0.426 $&$           12.92 $&...&[GEO97]W10&O8-9.5&$4$&$ 26.2$\\
$  1134$&$  266.41507 $&$   -29.009359 $&$   4.727 $&$  -5.576 $&$           15.69 $&...&...&...&$1$&$ 24.1$\\
$  1238$&$  266.41629 $&$   -29.007799 $&$   1.469 $&$   0.041 $&$           13.34 $&...&[GEO97]W7&O9-9.5&$2$&$ 24.9$\\
$  1245$&$  266.41968 $&$   -29.007692 $&$  -7.591 $&$   0.426 $&$           13.35 $&...&...&...&$4$&$ 32.6$\\
$  1327$&$  266.41507 $&$   -29.008177 $&$   4.732 $&$  -1.318 $&$           12.71 $&...&...&...&$3$&$ 44.9$\\
$  1350$&$  266.41626 $&$   -29.009369 $&$   1.548 $&$  -5.610 $&$           13.32 $&...&...&...&$3$&$ 34.2$\\
$  1474$&$  266.41684 $&$   -29.007048 $&$  -0.000 $&$   2.747 $&$           13.02 $&...&...&O8-9&$4$&$ 28.0$\\
$  1534$&$  266.41632 $&$   -29.008284 $&$   1.387 $&$  -1.703 $&$           13.34 $&...&[RGH2007] GEN-1.70-1.65&O-B&$4$&$ 29.0$\\
$  1554$&$  266.41446 $&$   -29.008282 $&$   6.363 $&$  -1.696 $&$           13.19 $&...&...&...&$3?$&$ 50.8$\\
$  1619$&$  266.41660 $&$   -29.008102 $&$   0.653 $&$  -1.051 $&$           14.18 $&...&S1-8&...&$4$&$ 19.7$\\
$  1643$&$  266.41641 $&$   -29.005293 $&$   1.145 $&$   9.064 $&$           13.78 $&...&...&O9-B0&$2$&$ 32.9$\\
$  1892$&$  266.41644 $&$   -29.006804 $&$   1.062 $&$   3.625 $&$           13.45 $&...&...&O8-9&$2$&$ 16.3$\\
$  1935$&$  266.41434 $&$   -29.006641 $&$   6.699 $&$   4.209 $&$           13.38 $&...&...&...&$1$&$ 27.0$\\
$  2048$&$  266.41342 $&$   -29.010553 $&$   9.118 $&$  -9.874 $&$           15.70 $&?&...&...&$1$&$ 16.7$\\
$  2233$&$  266.41656 $&$   -29.007904 $&$   0.734 $&$  -0.336 $&$           13.81 $&...&S0-14&O9.5-B2&$4$&$ 23.1$\\
$  2314$&$  266.41681 $&$   -29.007795 $&$   0.082 $&$   0.055 $&$           14.38 $&...&S2,S0-2&B0-2&$4$&$ 25.4$\\
$  2420$&$  266.41687 $&$   -29.009148 $&$  -0.082 $&$  -4.813 $&$           13.86 $&...&...&...&$4$&$ 19.9$\\
$  2446$&$  266.42099 $&$   -29.003845 $&$ -11.139 $&$  14.275 $&$           14.05 $&...&...&...&$1$&$ 22.4$\\
$  3308$&$  266.41364 $&$   -29.011532 $&$   8.540 $&$ -13.396 $&$           14.79 $&fg&...&...&$1$&$ 16.9$\\
$  3339$&$  266.42307 $&$   -29.007935 $&$ -16.647 $&$  -0.446 $&$           14.50 $&fg&...&...&$1$&$ 19.4$\\
$  3578$&$  266.41916 $&$   -29.009645 $&$  -6.192 $&$  -6.606 $&$           14.22 $&...&...&...&$1$&$ 20.2$\\
$  3773$&$  266.41422 $&$   -29.008575 $&$   7.014 $&$  -2.753 $&$           14.92 $&?&[RGH2007] GEN-1.70-1.65&O-B&$2$&$ 34.4$\\
$ 11652$&$  266.41449 $&$   -29.007727 $&$   6.285 $&$   0.302 $&$ ... $&?&...&...&$3?$&$ 15.2$\\
\end{longtable}
\tablefoot{The table lists the Stellar identification number Id, the coordinates in RA and Dec, and the offset coordinates from Sgr~A* $\Delta$RA and $\Delta$Dec in arcseconds (R.A.$_{SgrA*}$=266.41684\degr, Dec$_{SgrA*}$=$-$29.00781056\degr).  The $K_S$ magnitude was extinction corrected and shifted to a common extinction of $A_{Ks}$=2.70\,mag. We list the five probable foreground stars in the colour column. If the star war previously listed and classified, we denote the name and type. Column ``Note'' lists the reference to the stellar identification as an early-type star. Column ``S/N'' denotes the signal-to-noise ratio.
$^{(a)}$ $K_S$ magnitudes from \cite{rainer10}, extinction corrected and shifted to a  common extinction of $A_{Ks}$ = 2.70 mag;
$^{(b)}$ (1) First spectroscopic classification  reported in this work; (2) Spectral type from \cite{paumard06};  
 (3) Spectral type from \cite{bartko09}; (4) Spectral type from \cite{do13};
 (5) Photometric early-type candidate from \cite{shogo12}; (6) Classified as early-type star by \cite{stostad};
(7) Early-type star candidate from \cite{isaacanja}; (8) Classified as early-type star by \cite{perger08}.
``fg'' denotes a likely foreground star.
}
\clearpage

\clearpage
\onecolumn
\normalsize
\begin{longtable}{r|rr|rrr|rr}
\caption{\label{tab:etmag}O/B stars II.}\\
\hline \hline 
&\multicolumn{2}{c}{Extinction map}&\multicolumn{2}{c}{Intrinsic colour}&\\
Id&$K_{S,0} $&$A_{K_S} $&$K_{S,0}\hspace{5mm}$&$A_{K_S}\hspace{5mm}$&$\mathcal{M}\hspace{4mm}$&$EW_{CO}$&$EW_{NA}$\\
&$ \left[ \text{mag} \right]$&$ \left[ \text{mag} \right]$&$ \left[ \text{mag} \right]\hspace{4mm}$&$ \left[ \text{mag} \right]\hspace{4mm}$&$\left [  M_\odot \right ] \hspace{2mm}$\\
\hline
  \endfirsthead \caption{continued.}\\ \hline\hline
&\multicolumn{2}{c}{Extinction map}&\multicolumn{2}{c}{Intrinsic colour}&\\
Id&$K_{S,0} $&$A_{K_S} $&$K_{S,0}\hspace{5mm}$&$A_{K_S}\hspace{5mm}$&$\mathcal{M}\hspace{4mm}$&$EW_{CO}$&$EW_{NA}$\\
&$ \left[ \text{mag} \right]$&$ \left[ \text{mag} \right]$&$ \left[ \text{mag} \right]\hspace{4mm}$&$ \left[ \text{mag} \right]\hspace{4mm}$&$\left [  M_\odot \right ] \hspace{2mm}$\\
\hline \\
\endhead 

\hline
\endfoot
 \hline
\endlastfoot  \\
$    64$&$ 8.04 $&$2.63$&$  8.12\pm  1.15$&$  2.55\pm  1.14$&$ 42^{+ 10}$\hspace{1mm}$_{- 11}$&-3.0& 1.0\\
$    96$&$ 7.74 $&$2.92$&$  7.92\pm  1.23$&$  2.74\pm  1.21$&$ 43^{+  9}$\hspace{1mm}$_{- 12}$&-5.3&-0.5\\
$   109$&$ 7.87 $&$2.79$&$  8.18\pm  1.11$&$  2.47\pm  1.10$&$ 42^{+ 10}$\hspace{1mm}$_{- 11}$&-3.0& 0.3\\
$   166$&$ 8.56 $&$2.58$&$  8.77\pm  1.07$&$  2.37\pm  1.05$&$ 41^{+ 10}$\hspace{1mm}$_{- 12}$&-1.8& 0.4\\
$   205$&$ 8.48 $&$2.66$&$  8.66\pm  1.12$&$  2.48\pm  1.10$&$ 41^{+ 10}$\hspace{1mm}$_{- 12}$& 0.9&-0.6\\
$   209$&$ 8.49 $&$2.53$&$  8.68\pm  1.07$&$  2.34\pm  1.04$&$ 41^{+ 10}$\hspace{1mm}$_{- 12}$&-2.0& 3.1\\
$   227$&$ 8.73 $&$2.51$&$  8.77\pm  1.11$&$  2.47\pm  1.10$&$ 41^{+ 10}$\hspace{1mm}$_{- 12}$& 0.7&-0.9\\
$   230$&$ 8.49 $&$2.58$&$  8.63\pm  1.10$&$  2.44\pm  1.09$&$ 41^{+ 10}$\hspace{1mm}$_{- 11}$&-0.6& 0.9\\
$   273$&$ 8.74 $&$2.48$&$  8.91\pm  1.04$&$  2.31\pm  1.03$&$ 41^{+  9}$\hspace{1mm}$_{- 13}$&-0.3&-0.0\\
$   294$&$ 8.44 $&$2.76$&$  8.61\pm  1.17$&$  2.59\pm  1.15$&$ 41^{+ 10}$\hspace{1mm}$_{- 12}$& 0.3& 3.6\\
$   331$&$ 8.64 $&$2.74$&$  8.89\pm  1.13$&$  2.49\pm  1.11$&$ 41^{+  9}$\hspace{1mm}$_{- 13}$&-0.4& 0.8\\
$   366$&$ 8.92 $&$2.51$&$  8.88\pm  1.15$&$  2.55\pm  1.13$&$ 41^{+  9}$\hspace{1mm}$_{- 13}$&-2.1&-1.5\\
$   372$&$ 9.11 $&$2.50$&$  9.31\pm  1.03$&$  2.29\pm  1.02$&$ 39^{+ 10}$\hspace{1mm}$_{- 13}$&-3.6& 0.8\\
$   436$&$ 8.99 $&$2.78$&$ 11.34\pm  0.29$&$  0.42\pm  0.23$&$ 19^{+ 18}$\hspace{1mm}$_{-  3}$& 1.3& 1.6\\
$   443$&$ 9.39 $&$2.71$&$  9.21\pm  1.30$&$  2.89\pm  1.29$&$ 39^{+ 11}$\hspace{1mm}$_{- 13}$& 3.8&-1.0\\
$   445$&$ 8.97 $&$2.61$&$  9.01\pm  1.15$&$  2.57\pm  1.14$&$ 40^{+ 10}$\hspace{1mm}$_{- 13}$&-0.2& 0.6\\
$   483$&$ 9.26 $&$2.44$&$  9.45\pm  1.02$&$  2.25\pm  1.00$&$ 38^{+ 10}$\hspace{1mm}$_{- 14}$& 1.4& 0.4\\
$   507$&$ 9.25 $&$2.73$&$  9.52\pm  1.11$&$  2.46\pm  1.09$&$ 38^{+ 10}$\hspace{1mm}$_{- 14}$&-2.2& 2.5\\
$   508$&$ 9.53 $&$2.60$&$  9.50\pm  1.18$&$  2.63\pm  1.17$&$ 38^{+ 11}$\hspace{1mm}$_{- 14}$&-1.6& 1.2\\
$   511$&$ 9.40 $&$2.41$&$  9.53\pm  1.03$&$  2.28\pm  1.01$&$ 38^{+ 10}$\hspace{1mm}$_{- 14}$&-2.1&-0.3\\
$   516$&$ 8.74 $&$2.97$&$  8.95\pm  1.24$&$  2.76\pm  1.23$&$ 40^{+ 10}$\hspace{1mm}$_{- 12}$& 4.6& 1.6\\
$   562$&$ 9.56 $&$2.51$&$  9.69\pm  1.07$&$  2.38\pm  1.06$&$ 37^{+ 11}$\hspace{1mm}$_{- 14}$&-4.0& 1.0\\
$   567$&$ 9.17 $&$2.79$&$  9.48\pm  1.12$&$  2.48\pm  1.10$&$ 38^{+ 10}$\hspace{1mm}$_{- 14}$&-0.6& 0.9\\
$   596$&$ 9.59 $&$2.57$&$  9.59\pm  1.15$&$  2.57\pm  1.14$&$ 38^{+ 10}$\hspace{1mm}$_{- 15}$&-1.3&-0.9\\
$   610$&$ 9.29 $&$2.72$&$  9.57\pm  1.10$&$  2.44\pm  1.09$&$ 38^{+ 10}$\hspace{1mm}$_{- 15}$&-5.6&-1.3\\
$   617$&$ 9.81 $&$2.46$&$ 10.12\pm  0.97$&$  2.15\pm  0.96$&$ 34^{+ 13}$\hspace{1mm}$_{- 14}$&-0.2& 0.5\\
$   663$&$ 9.50 $&$2.85$&$ 10.02\pm  1.05$&$  2.33\pm  1.04$&$ 35^{+ 12}$\hspace{1mm}$_{- 15}$& 1.9& 0.4\\
$   668$&$ 9.84 $&$2.61$&$  9.99\pm  1.11$&$  2.46\pm  1.09$&$ 35^{+ 12}$\hspace{1mm}$_{- 15}$& 1.3&-1.0\\
$   707$&$ 9.62 $&$2.79$&$  9.74\pm  1.20$&$  2.67\pm  1.19$&$ 37^{+ 11}$\hspace{1mm}$_{- 15}$& 1.1&-2.2\\
$   716$&$ 9.92 $&$2.58$&$ 10.28\pm  1.00$&$  2.21\pm  0.99$&$ 33^{+ 13}$\hspace{1mm}$_{- 14}$& 0.4& 0.0\\
$   718$&$10.01 $&$2.50$&$ 10.17\pm  1.06$&$  2.34\pm  1.04$&$ 34^{+ 13}$\hspace{1mm}$_{- 15}$& 3.5& 3.9\\
$   721$&$ 9.53 $&$2.69$&$  9.73\pm  1.12$&$  2.49\pm  1.11$&$ 37^{+ 11}$\hspace{1mm}$_{- 15}$&-1.8& 0.1\\
$   722$&$10.02 $&$2.44$&$ 10.39\pm  0.93$&$  2.06\pm  0.92$&$ 32^{+ 14}$\hspace{1mm}$_{- 14}$&-1.2&-0.6\\
$   725$&$ 9.57 $&$2.75$&$  9.78\pm  1.14$&$  2.53\pm  1.13$&$ 37^{+ 11}$\hspace{1mm}$_{- 15}$&-8.7&-0.4\\
$   728$&$ 9.65 $&$2.66$&$  9.43\pm  1.29$&$  2.88\pm  1.28$&$ 39^{+ 10}$\hspace{1mm}$_{- 15}$&-4.4&-1.5\\
$   757$&$ 9.42 $&$2.70$&$  9.59\pm  1.14$&$  2.53\pm  1.12$&$ 38^{+ 10}$\hspace{1mm}$_{- 15}$&-3.1&-1.2\\
$   762$&$ 9.47 $&$2.59$&$  9.80\pm  1.03$&$  2.26\pm  1.01$&$ 37^{+ 10}$\hspace{1mm}$_{- 15}$& 1.2& 2.8\\
$   785$&$ 9.60 $&$2.74$&$  9.74\pm  1.17$&$  2.60\pm  1.16$&$ 37^{+ 11}$\hspace{1mm}$_{- 15}$&-1.0&-2.3\\
$   838$&$ 9.58 $&$2.86$&$  9.66\pm  1.24$&$  2.77\pm  1.23$&$ 37^{+ 11}$\hspace{1mm}$_{- 15}$&-1.5& 1.0\\
$   847$&$ 9.45 $&$2.88$&$  9.26\pm  1.37$&$  3.06\pm  1.36$&$ 39^{+ 11}$\hspace{1mm}$_{- 14}$&-4.8& 0.7\\
$   853$&$ 9.75 $&$2.70$&$  9.65\pm  1.25$&$  2.80\pm  1.24$&$ 37^{+ 11}$\hspace{1mm}$_{- 15}$&-0.3& 1.1\\
$   890$&$10.16 $&$2.38$&$ 10.34\pm  0.99$&$  2.20\pm  0.98$&$ 33^{+ 13}$\hspace{1mm}$_{- 15}$&-4.0& 1.5\\
$   900$&$ 9.95 $&$2.65$&$ 10.15\pm  1.10$&$  2.44\pm  1.09$&$ 34^{+ 13}$\hspace{1mm}$_{- 15}$&-6.7& 0.2\\
$   936$&$10.03 $&$2.39$&$ 10.16\pm  1.02$&$  2.26\pm  1.01$&$ 34^{+ 13}$\hspace{1mm}$_{- 14}$&-3.2&-2.9\\
$   941$&$ 9.83 $&$2.79$&$ 10.10\pm  1.13$&$  2.52\pm  1.12$&$ 35^{+ 12}$\hspace{1mm}$_{- 16}$& 1.3& 1.9\\
$   951$&$10.30 $&$2.45$&$ 10.55\pm  0.99$&$  2.20\pm  0.98$&$ 30^{+ 15}$\hspace{1mm}$_{- 13}$& 2.7& 1.8\\
$   958$&$ 9.91 $&$2.66$&$ 10.15\pm  1.09$&$  2.42\pm  1.08$&$ 34^{+ 13}$\hspace{1mm}$_{- 15}$&-2.7&-1.4\\
$   973$&$11.47 $&$2.98$&$ 11.61\pm  1.28$&$  2.84\pm  1.26$&$ 17^{+ 21}$\hspace{1mm}$_{-  7}$&-0.8&-0.6\\
$   974$&$10.26 $&$2.63$&$ 10.51\pm  1.07$&$  2.38\pm  1.06$&$ 31^{+ 15}$\hspace{1mm}$_{- 14}$& 3.2& 0.7\\
$   982$&$10.14 $&$2.70^a$&$  9.96\pm  2.74$&$  2.87\pm  2.06$&$ 35^{+ 14}$\hspace{1mm}$_{- 20}$& 0.3&-0.5\\
$  1048$&$10.41 $&$2.45$&$ 10.56\pm  1.03$&$  2.30\pm  1.02$&$ 30^{+ 15}$\hspace{1mm}$_{- 13}$&-3.1& 0.7\\
$  1103$&$13.27 $&$2.67$&$ 13.22\pm  1.23$&$  2.72\pm  1.21$&$ 11^{+  8}$\hspace{1mm}$_{-  4}$&-0.9&-0.4\\
$  1104$&$10.44 $&$2.53$&$ 10.95\pm  0.91$&$  2.02\pm  0.90$&$ 23^{+ 20}$\hspace{1mm}$_{-  8}$& 2.7& 0.1\\
$  1123$&$10.22 $&$2.55$&$ 10.42\pm  1.06$&$  2.35\pm  1.05$&$ 32^{+ 14}$\hspace{1mm}$_{- 14}$& 2.4& 3.7\\
$  1134$&$12.99 $&$2.44$&$ 13.01\pm  1.11$&$  2.42\pm  1.08$&$ 11^{+  8}$\hspace{1mm}$_{-  3}$&-5.3&-1.7\\
$  1238$&$10.64 $&$2.62$&$ 10.31\pm  1.32$&$  2.95\pm  1.31$&$ 33^{+ 14}$\hspace{1mm}$_{- 16}$&-5.7&-2.0\\
$  1245$&$10.65 $&$2.44$&$ 10.79\pm  1.03$&$  2.30\pm  1.02$&$ 26^{+ 18}$\hspace{1mm}$_{- 11}$& 1.0&-0.5\\
$  1327$&$10.01 $&$3.12$&$ 10.30\pm  1.27$&$  2.83\pm  1.26$&$ 33^{+ 14}$\hspace{1mm}$_{- 15}$&-1.1&-0.7\\
$  1350$&$10.62 $&$2.55$&$ 10.88\pm  1.03$&$  2.28\pm  1.01$&$ 24^{+ 20}$\hspace{1mm}$_{-  9}$& 2.6& 3.6\\
$  1474$&$10.32 $&$2.79$&$ 10.48\pm  1.18$&$  2.63\pm  1.17$&$ 31^{+ 15}$\hspace{1mm}$_{- 14}$& 1.1& 1.0\\
$  1534$&$10.64 $&$2.73$&$ 10.81\pm  1.15$&$  2.55\pm  1.14$&$ 26^{+ 19}$\hspace{1mm}$_{- 11}$&-7.9&-0.0\\
$  1554$&$10.49 $&$2.81$&$ 10.65\pm  1.19$&$  2.65\pm  1.18$&$ 29^{+ 16}$\hspace{1mm}$_{- 13}$&-2.3&-0.3\\
$  1619$&$11.48 $&$2.64$&$ 11.58\pm  1.14$&$  2.53\pm  1.13$&$ 17^{+ 20}$\hspace{1mm}$_{-  6}$&-4.3& 1.4\\
$  1643$&$11.08 $&$2.52$&$ 11.32\pm  1.04$&$  2.28\pm  1.02$&$ 19^{+ 21}$\hspace{1mm}$_{-  7}$& 2.1& 0.3\\
$  1892$&$10.75 $&$2.76$&$ 11.00\pm  1.13$&$  2.51\pm  1.11$&$ 23^{+ 20}$\hspace{1mm}$_{-  9}$& 5.8& 1.6\\
$  1935$&$10.68 $&$2.77$&$ 10.65\pm  1.25$&$  2.80\pm  1.24$&$ 29^{+ 16}$\hspace{1mm}$_{- 14}$&-2.2& 1.4\\
$  2048$&$13.00 $&$2.64$&$...$&$...$&$...$&-3.8&-2.6\\
$  2233$&$11.11 $&$2.54$&$ 11.24\pm  1.08$&$  2.40\pm  1.07$&$ 20^{+ 21}$\hspace{1mm}$_{-  8}$&-3.1& 6.7\\
$  2314$&$11.68 $&$2.46$&$ 11.75\pm  1.07$&$  2.38\pm  1.06$&$ 17^{+ 18}$\hspace{1mm}$_{-  7}$&-0.3& 2.8\\
$  2420$&$11.16 $&$2.56$&$ 11.19\pm  1.14$&$  2.53\pm  1.12$&$ 20^{+ 21}$\hspace{1mm}$_{-  8}$& 5.5& 0.5\\
$  2446$&$11.35 $&$2.67$&$ 11.66\pm  1.06$&$  2.35\pm  1.04$&$ 17^{+ 19}$\hspace{1mm}$_{-  6}$& 3.2& 2.5\\
$  3308$&$12.09 $&$2.32$&$ 12.57\pm  0.83$&$  1.83\pm  0.82$&$ 12^{+  9}$\hspace{1mm}$_{-  3}$& 4.4&-2.1\\
$  3339$&$11.80 $&$2.63$&$ 14.00\pm  0.29$&$  0.42\pm  0.23$&$  7^{+  8}$\hspace{1mm}$_{-  1}$& 2.1& 4.4\\
$  3578$&$11.52 $&$2.69$&$ 11.18\pm  1.35$&$  3.03\pm  1.34$&$ 21^{+ 22}$\hspace{1mm}$_{-  9}$& 8.3&-1.2\\
$  3773$&$12.22 $&$2.70$&$...$&$...$&$...$& 0.0& 0.9\\
$ 11652$&$... $&$2.77$&$...$&$...$&$...$&-5.0&-0.1\\
\end{longtable}
\tablefoot{$K_S$ magnitudes taken from \cite{rainer10}, extinction corrected with the extinction map from \cite{rainer10}; extinction $A_{K_S}$ adopted from  the map of \cite{rainer10};  $K_{S,0}$ magnitude assuming an intrinsic colour of $(H-K_S)_0$=$-$0.1 for O/B stars; corresponding extinction $A_{K_S}$; stellar mass $M_*$ using isochrones with 3-8\,Myr age and solar metallicity. The last two columns denote the equivalent widths ($EW$) of CO and Na.$^{(a)}$ Beyond extinction map from \cite{rainer10}.}
\clearpage

\clearpage
\onecolumn
\normalsize
\begin{longtable}{rrccccc}
\caption{\label{tab:etkin}O/B stars III.}\\
\hline \hline 
Id&$p\hspace{1mm}$&$v_{RA}$&$v_{Dec}$&$v_{z}$&$v_{z}$&$h$\\
&$ \left[ ^{\arcsec} \right]$ &$ \left[ \text{km\;s}^{-1} \right]$&$ \left[ \text{km\;s}^{-1} \right]$&$ \left[ \text{km\;s}^{-1} \right]$&$\text{on\,disk}^g$\\
\hline
  \endfirsthead \caption{continued.}\\ \hline\hline
Id&$p\hspace{1mm}$&$v_{RA}$&$v_{Dec}$&$v_{z}$&$v_{z}$&$h$\\
&$ \left[ ^{\arcsec} \right]$ &$ \left[ \text{km\;s}^{-1} \right]$&$ \left[ \text{km\;s}^{-1} \right]$&$ \left[ \text{km\;s}^{-1} \right]$&$\text{on\,disk}^g$\\
\hline \\
\endhead 
\hline
\endfoot
    \hline
\endlastfoot  \\
$    64$&$    2.11$&$     -65\pm       4^b$&$     257\pm       4$&$     256\pm      12^c$&$1$&$  0.56$\\
$    96$&$    7.82$&$      69\pm       9^b$&$     250\pm       9$&$     136\pm      17^c$&$0$&$ -1.05$\\
$   109$&$    2.30$&$     350\pm       1^b$&$       6\pm       1$&$     149\pm      27^c$&$1$&$  0.63$\\
$   166$&$   10.62$&$      85\pm       5^a$&$      52\pm       4$&$     154\pm     102^c$&$0$&$  0.01$\\
$   205$&$    6.35$&$     -82\pm       8^b$&$     202\pm       6$&$      32\pm      16^c$&$1$&$  0.75$\\
$   209$&$    8.27$&$     -54\pm       3^a$&$     177\pm       6$&$      87\pm      31^c$&$0$&$ -0.51$\\
$   227$&$    6.57$&$     229\pm       2^b$&$      68\pm       3$&$     154\pm      28^c$&$1$&$  0.89$\\
$   230$&$    8.58$&$     -35\pm       7^b$&$     144\pm       6$&$    -114\pm      20^c$&$1$&$  0.15$\\
$   273$&$   10.26$&$     -79\pm       7^b$&$      59\pm       8$&$    -165\pm      25^c$&$1$&$  0.37$\\
$   294$&$    2.25$&$     137\pm       1^b$&$    -210\pm       1$&$     105\pm      61^c$&$1$&$  0.30$\\
$   331$&$    1.84$&$     100\pm       1^b$&$    -234\pm       1$&$     221\pm      45^c$&$1$&$ -0.01$\\
$   366$&$    9.34$&$      80\pm       4^a$&$      61\pm       4$&$     257\pm     135^c$&$1$&$  0.37$\\
$   372$&$    5.75$&$    -163\pm       1^b$&$     -72\pm       2$&$    -148\pm      23^c$&$0$&$ -0.54$\\
$   436$&$   19.09$&$...$&$...$&$      77\pm      33^c$&$...$&$...$\\
$   443$&$    2.09$&$     301\pm       1^b$&$     116\pm       1$&$     229\pm      36^c$&$1$&$  0.61$\\
$   445$&$    7.70$&$    -195\pm       7^b$&$     -71\pm       6$&$     -61\pm      21^c$&$1$&$  0.81$\\
$   483$&$   12.83$&$     -10\pm       4^a$&$       3\pm       4$&$    -180\pm      42^c$&$0$&$  0.04^h$\\
$   507$&$    3.19$&$     300\pm       1^b$&$     -49\pm       2$&$     -46\pm      57^c$&$...$&$  0.74$\\
$   508$&$    5.86$&$       1\pm      21^b$&$     229\pm       6$&$      24\pm      25^f$&$...$&$  0.80$\\
$   511$&$   10.60$&$      40\pm       5^a$&$    -127\pm       6$&$     231\pm     124^c$&$0$&$  0.54$\\
$   516$&$    5.04$&$     219\pm       8^b$&$     -87\pm       8$&$     -46\pm      64^c$&$...$&$ -0.76$\\
$   562$&$    0.94$&$    -520\pm       1^b$&$      66\pm       1$&$     110\pm      72^c$&$1$&$  0.72$\\
$   567$&$    6.25$&$     116\pm       6^b$&$    -228\pm       6$&$    -120\pm      77^c$&$0$&$ -0.92$\\
$   596$&$    7.35$&$      91\pm       5^a$&$     212\pm       4$&$     120\pm      52^c$&$1$&$  0.82$\\
$   610$&$    8.81$&$      14\pm       5^a$&$      50\pm       7$&$    -217\pm     211^c$&$0$&$  0.18$\\
$   617$&$   10.39$&$    -143\pm       3^a$&$     -60\pm       7$&$     202\pm     102^c$&$0$&$ -0.17$\\
$   663$&$   28.81$&$...$&$...$&$      -7\pm      24^c$&$...$&$...$\\
$   668$&$    1.18$&$     367\pm      13^a$&$     130\pm      12$&$    -221\pm     179^c$&$0$&$  0.50$\\
$   707$&$    3.30$&$     -27\pm       1^b$&$     164\pm       1$&$      53\pm      20^d$&$1$&$  0.15$\\
$   716$&$    7.37$&$      94\pm       8^b$&$     125\pm       9$&$     160\pm      50^e$&$1$&$  0.47$\\
$   718$&$    8.87$&$    -187\pm       5^a$&$      76\pm       7$&$       9\pm      88^c$&$...$&$ -0.87$\\
$   721$&$    8.74$&$     100\pm       6^a$&$     -98\pm       3$&$     -22\pm      44^c$&$...$&$ -0.59$\\
$   722$&$   12.45$&$     -21\pm       6^a$&$    -103\pm       8$&$     268\pm      30^c$&$0$&$  0.16$\\
$   725$&$    3.23$&$     220\pm       1^b$&$      80\pm       2$&$     117\pm      37^c$&$0$&$  0.20$\\
$   728$&$    8.58$&$       6\pm      10^b$&$      -5\pm       8$&$     277\pm     166^c$&$0$&$  0.03^h$\\
$   757$&$    9.49$&$    -124\pm       4^a$&$     -73\pm       3$&$     184\pm      89^c$&$0$&$  0.10$\\
$   762$&$    3.53$&$      -5\pm       1^b$&$     209\pm       1$&$     305\pm      70^e$&$1$&$  0.54$\\
$   785$&$    4.15$&$       5\pm       1^b$&$    -162\pm       2$&$     104\pm     170^c$&$...$&$ -0.08$\\
$   838$&$    7.96$&$    -138\pm      11^b$&$     -98\pm       9$&$      78\pm      64^c$&$0$&$  0.61$\\
$   847$&$    3.92$&$     -19\pm       1^b$&$    -143\pm       2$&$      52\pm      23^c$&$0$&$  0.24$\\
$   853$&$   11.38$&$     -14\pm       4^a$&$      -8\pm       6$&$     257\pm     124^c$&$1$&$ -0.07^h$\\
$   890$&$   11.34$&$     -64\pm       4^a$&$    -108\pm       6$&$      31\pm      45^c$&$...$&$ -0.42$\\
$   900$&$    1.62$&$     293\pm       1^b$&$     -99\pm       1$&$    -229\pm      35^c$&$1$&$  0.34$\\
$   936$&$    9.78$&$      31\pm       8^a$&$    -139\pm      10$&$...$&$...$&$  0.51$\\
$   941$&$    2.35$&$    -314\pm       1^b$&$      50\pm       1$&$     157\pm      55^c$&$1$&$  0.71$\\
$   951$&$   10.66$&$     -13\pm       8^b$&$    -161\pm       8$&$    -150\pm      40^e$&$0$&$ -0.67$\\
$   958$&$    9.01$&$     -13\pm       9^b$&$     -49\pm       8$&$      46\pm      68^c$&$...$&$  0.18$\\
$   973$&$    6.34$&$     190\pm       5^a$&$     221\pm      11$&$     133\pm      70^c$&$0$&$ -0.57$\\
$   974$&$    5.82$&$     -51\pm       1^b$&$    -134\pm       2$&$     140\pm      50^e$&$1$&$ -0.41$\\
$   982$&$   23.64$&$...$&$...$&$...$&$...$&$...$\\
$  1048$&$    9.57$&$      50\pm       3^a$&$     -35\pm       5$&$...$&$...$&$ -0.18$\\
$  1103$&$   11.10$&$     -68\pm      15^a$&$      -2\pm      16$&$      53\pm      92^c$&$...$&$ -0.08$\\
$  1104$&$   20.45$&$      57\pm      10^a$&$      66\pm       7$&$     123\pm      51^c$&$...$&$  0.10$\\
$  1123$&$    1.30$&$     188\pm       1^b$&$    -280\pm       1$&$    -364\pm      10^d$&$1$&$  0.53$\\
$  1134$&$    7.78$&$...$&$...$&$...$&$...$&$...$\\
$  1238$&$    1.64$&$     152\pm       1^b$&$    -182\pm       1$&$    -193\pm      41^d$&$1$&$  0.32$\\
$  1245$&$    9.04$&$     102\pm       7^a$&$     135\pm       6$&$...$&$...$&$  0.56$\\
$  1327$&$    5.62$&$     168\pm       8^b$&$      74\pm       8$&$     155\pm      50^e$&$0$&$ -0.12$\\
$  1350$&$    5.82$&$     -37\pm       1^b$&$    -143\pm       2$&$      10\pm      50^e$&$...$&$  0.03$\\
$  1474$&$    2.80$&$    -335\pm       1^b$&$      59\pm       1$&$     145\pm      72^c$&$0$&$  0.82$\\
$  1534$&$    2.25$&$     355\pm       1^b$&$    -126\pm       2$&$     -83\pm      42^c$&$1$&$  0.75$\\
$  1554$&$    7.58$&$    -149\pm      11^b$&$      18\pm      11$&$     222\pm      31^c$&$0$&$ -0.20$\\
$  1619$&$    1.20$&$...$&$...$&$     102\pm      94^c$&$0$&$...$\\
$  1643$&$    9.20$&$    -148\pm       7^b$&$    -130\pm       6$&$    -185\pm      50^e$&$1$&$  0.72$\\
$  1892$&$    3.86$&$     216\pm       2^b$&$     176\pm       2$&$    -153\pm      60^c$&$1$&$ -0.74$\\
$  1935$&$    8.87$&$      39\pm       4^a$&$     147\pm       3$&$...$&$...$&$ -0.64$\\
$  2048$&$   14.50$&$...$&$...$&$     130\pm      52^c$&$0$&$...$\\
$  2233$&$    0.82$&$      82\pm       1^b$&$     -35\pm       1$&$     -28\pm     104^c$&$...$&$  0.08$\\
$  2314$&$    0.11$&$    -415\pm      30^a$&$     748\pm     102$&$...$&$...$&$  0.20$\\
$  2420$&$    4.76$&$     113\pm       1^b$&$      60\pm       2$&$     -51\pm      65^f$&$...$&$  0.36$\\
$  2446$&$   19.46$&$      28\pm       6^a$&$      33\pm      11$&$...$&$...$&$  0.01$\\
$  3308$&$   16.67$&$       8\pm       3^a$&$     141\pm       5$&$    -104\pm      73^c$&$...$&$ -0.46$\\
$  3339$&$   19.70$&$...$&$...$&$      76\pm      60^c$&$...$&$...$\\
$  3578$&$    9.88$&$     -82\pm       6^a$&$     -36\pm       4$&$...$&$...$&$ -0.37$\\
$  3773$&$    8.60$&$     143\pm      18^a$&$      -1\pm      10$&$...$&$...$&$  0.20$\\
$ 11652$&$    7.31$&$     131\pm       9^a$&$     163\pm      17$&$...$&$...$&$ -0.66$\\
\end{longtable}
\tablefoot{The table lists: Projected distance to Sgr A* $p$ in arcseconds; proper motions $v_{RA}$ and $v_{Dec}$ are taken from previous studies; the radial velocity $v_z$ is adopted from this and previous studies; $h$ is the normalised projected angular momentum.
$^{(a)}$~Proper motions from \cite{Rainerpm09};
$^{(b)}$~Proper motions from \cite{yelda14};
$^{(c)}$~Radial velocity from this work;
$^{(d)}$~Radial velocity is error-weighted mean from \cite{bartko09} and \cite{yelda14};
$^{(e)}$~Radial velocity from \cite{bartko09};
$^{(f)}$~Radial velocity from \cite{yelda14};
$^{(g)}$~Disk membership excluded if 0, possible if 1 (based on $v_{z}$);
$^{(h)}$~High inclination (based on $v_{pm}/v_{tot}$).
}
\clearpage

\end{appendix}
\end{document}